\documentclass[aps,superscriptaddress,twocolumn,dvips]{revtex4}
\usepackage{feynmp}
\usepackage{amsmath}
\usepackage{amssymb}
\usepackage{epsfig}
\usepackage{graphicx}
\usepackage{subfigure}
\usepackage{hyperref}
\usepackage{bbm}
\usepackage{color}
\usepackage{longtable}
\usepackage{booktabs}
\usepackage{multirow}
\usepackage{array}

\newcommand{\Rmnum}[1]{\expandafter\@slowromancap\romannumeral #1@}
\allowdisplaybreaks[3]
\begin{document}

\title{Correlated interaction effects in three-dimensional semi-Dirac semimetal}

\author{Jing-Rong Wang}
\affiliation{High Magnetic Field Laboratory of Anhui Province,
Chinese Academy of Sciences, Hefei 230031, China}
\author{Wei Li}
\altaffiliation{Corresponding author: wliustc@theory.issp.ac.cn}
\affiliation{Key Laboratory of Materials Physics,
Institute of Solid State Physics, Chinese Academy
of Sciences, Hefei 230031, China }
\author{Chang-Jin Zhang}
\altaffiliation{Corresponding author: zhangcj@hmfl.ac.cn}
\affiliation{High Magnetic Field Laboratory of Anhui Province,
Chinese Academy of Sciences, Hefei 230031, China}
\affiliation{Institute of Physical Science and Information
Technology, Anhui University, Hefei 230601, China}

\begin{abstract}
Understanding the correlation effects in unconventional topological materials, in which the fermion excitations take
unusual dispersion, is an important topic in recent condensed matter physics. We study the influence
of short-range four-fermion interactions on three-dimensional semi-Dirac semimetal with an unusual fermion dispersion, that
is linear along two directions and quadratic along the third one.  Based on renormalization group theory, we
find all of 11 unstable fixed points including 5 quantum critical points, 5 bicritical points, and one tricritical point.
The physical essences of the quantum critical points are determined  by analyzing the
susceptibility exponents for all of the source terms in particle-hole and particle-particle channels. We also
verify phase diagrams of the system in the parameter space through numerically studying the flows
of the four-fermion coupling parameters and behaviors of the susceptibility exponents. These results are helpful
for us to understand the physical properties of candidate materials for three-dimensional semi-Dirac semimetal such as ZrTe$_{5}$.
\end{abstract}

\maketitle


\section{Introduction}

The past 15 years have witnessed that study about topological materials becomes one of the most important fields
in condensed matter physics \cite{Vafek14, Wehling14, Yan17, Hasan17, Armitage18, Kruthoff17, Tang19Wan,
Zhang19FangChen, VergnioryWangZ19}. Topological materials have wide potential implications as electronic devices due
to their  fascinating physical properties. In some topological materials, such as Dirac semimetal (DSM) including
Cd$_{3}$As$_{2}$ and Na$_{3}$Bi, and Weyl semimetal (WSM) including TaAs, TaP, NbAs, and NbP, the low-energy fermion excitations
are Dirac fermions or Weyl fermions which resemble the elementary particles in high energy physics. Thus, these
materials provide a platform to verify some important concepts in high energy physics.

\begin{figure}[htbp]
\center
\includegraphics[width=1.3in]{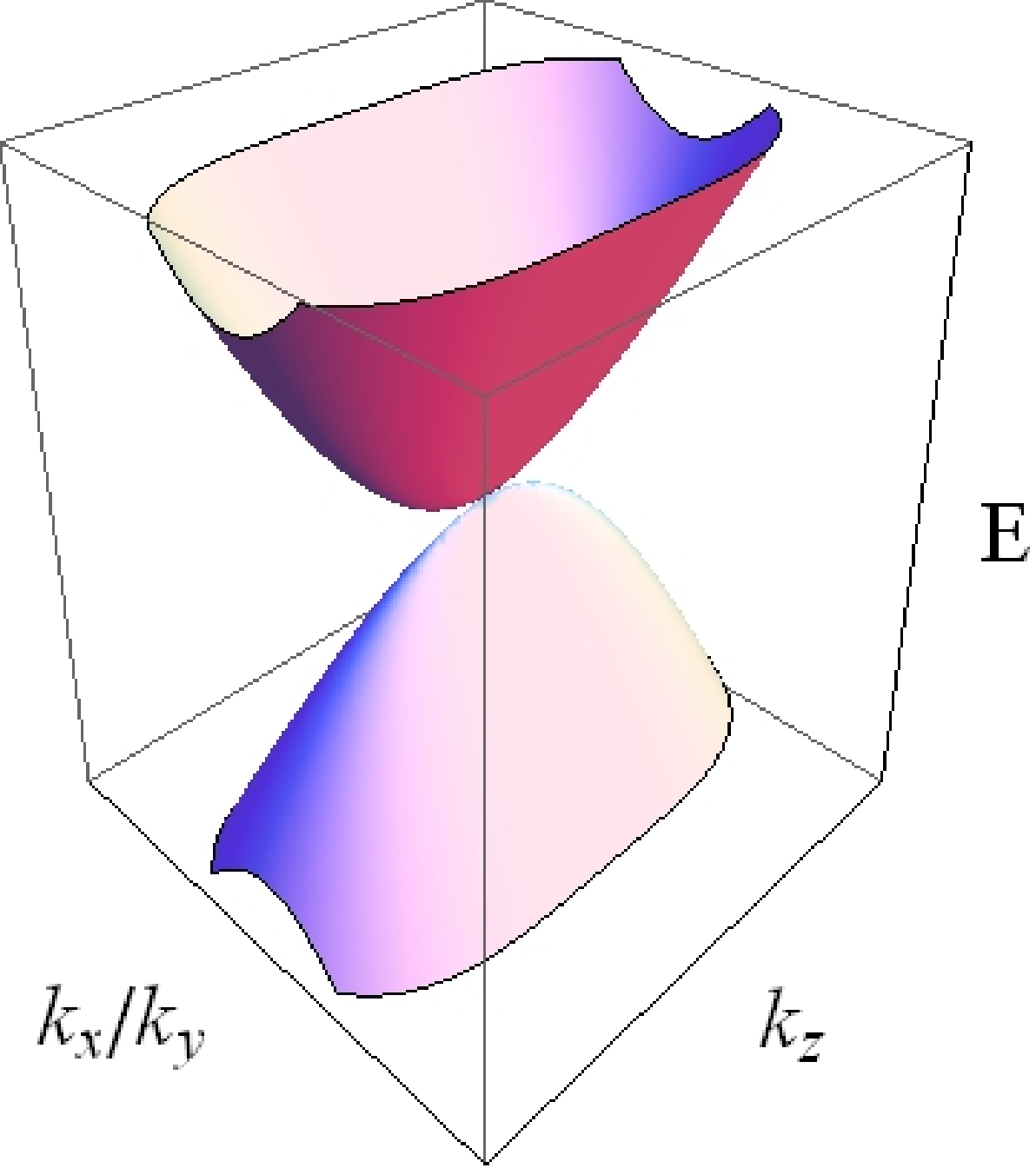}
\caption{Energy dispersion of fermions in 3D semi-DSM. \label{Fig:Dispersion}}
\end{figure}

Besides Dirac and Weyl fermions, there could be unconventional fermions with unusual dispersion in topological materials.
In double- (triple-) WSM, the fermion dispersion is quadratic (cubic) along two directions and linear along the third
one \cite{Xu11, Fang12}. Semi-DSM emerges at the topological quantum critical point (QCP) between DSM and band insulator\cite{Dietl08, Yang14}.
For two dimensional (2D) semi-DSM, the dispersion of fermion excitations is linear along one direction and quadratic along another one.
For three dimensional (3D) semi-DSM, the fermion dispersion is linear along two directions and quadratic along the third one as shown in Fig.~\ref{Fig:Dispersion}.
Higher spin fermions with multiband crossing have also attracted a lot of interest recently
\cite{Bradyln16, Tang17, Zhang18}. Spin-1 chiral fermions characterized by combination of a Dirac-like band and a flat band with three-bands crossing,
and spin-3/2 chiral fermions displaying a birefringent spectrum with two distinct fermion velocities have been observed recently \cite{Takane19, Rao19, Sanchez19, Schroter19, Schroter20}.

The correlation effects in Dirac and Weyl fermion systems are extensively studied, and are well understood relatively
\cite{Kotov12, WangLiu12, Hofmann14, Goswami11, Hosur12, Tang18, Leaw19, Herbut06, Herbut09, Maciejko14, Roy16, Szabo21}.
The influence of many-body interaction on unconventional fermion systems also attracted much interest and is an important topic.
There have been studies about influence of long-range Coulomb interaction \cite{YangNatPhys14, Abrikosov72, Isobe16, Cho16, WangLiuZhang17A,
Lai15, Jian15, WangLiuZhang17B, ZhangShiXin17, WangLiuZhang18, WangLiuZhang19, Han19, ZhangSX21, Roy18Birefringent, Kotov21},
short-range four-fermion interaction \cite{Vafek10, Roy17B, Roy18A, WangJing18, Szabo21B, Boettcher20}, and quantum fluctuation of order
parameter \cite{ Savary14, Uryszek19, Sur19, Uryszek20} on some unconventional fermion systems.
These studies revealed many novel behaviors, such as various quantum phase transitions,
non-Fermi liquid behaviors, anisotropic screening effect \emph{etc}. These studies also showed that the correlation
effects in unconventional fermion systems depend on the fermion dispersion subtly. For 2D semi-DSM, Isobe \emph{et al.}
showed that long-range Coulomb interaction results in non-Fermi liquid behaviors in a wide intermediate
energy range and marginal Fermi liquid behaviors in the lowest energy regime \cite{Isobe16}. However, for 3D semi-DSM,
it was revealed that long-range Coulomb becomes irrelevant in the lowest energy regime and the system exhibits Fermi liquid behaviors \cite{YangNatPhys14, Abrikosov72}.

There are still some important open questions about the correlation effects in unconventional fermion systems.
A insightful study about the influence of short-range four-fermion interactions on 2D semi-DSM was performed by Roy and Foster \cite{Roy18A}.
However, the effects of short-range four-fermion interactions in 3D semi-DSM is an urgent question, which is yet to be resoled.
In this article, we provide a comprehensive study for this question through renormalization group (RG) theory.

\section{Model}

The free action for 3D semi-DSM  can be written as
\begin{eqnarray}
S_{0}=\int\frac{d\omega}{2\pi}\frac{d^3\mathbf{k}}{(2\pi)^{3}}
\bar{\Psi}(\omega,\mathbf{k})\gamma_{0}\left[i\omega
+\mathcal{H}(\mathbf{k})\right]\Psi(\omega,\mathbf{k}),
\end{eqnarray}
where the Hamiltonian density $\mathcal{H}(\mathbf{k})$ is given by
\begin{eqnarray}
\mathcal{H}(\mathbf{k})=\gamma_{0}\left(iv\gamma_{1}k_{1}+ivk_{2}\gamma_{2}+iAk_{3}^{2}\gamma_{3}\right),
\end{eqnarray}
with $v$ and $A$ being model parameters. $\Psi$ is four component spinor, and
$\bar{\Psi}=\Psi^{\dag}\gamma_{0}$. The gamma matrices are defined as $\gamma_{0}=\tau_{3}\otimes\sigma_{0}$,
$\gamma_{1}=\tau_{2}\otimes\sigma_{1}$, $\gamma_{2}=\tau_{2}\otimes\sigma_{2}$, $\gamma_{3}=\tau_{2}\otimes\sigma_{3}$,
and $\gamma_{5}=\tau_{1}\otimes\sigma_{0}$, where $\tau_{1,2,3}$ and $\sigma_{1,2,3}$ are Pauli matrices. It is easy to
verify that $\gamma_{5}=\gamma_{0}\gamma_{1}\gamma_{2}\gamma_{3}$.
The gamma matrices satisfy the anticommutation relation $\left\{\gamma_{\mu},\gamma_{\nu}\right\}=2\delta_{\mu\nu}$ for $\mu,\nu=0,1,2,3,5$.
The gamma matrices have the properties as following
\begin{eqnarray}
\gamma_{\mu}^{\dag}&=&\gamma_{\mu},
\\
\gamma_{0,2,5}^{*}&=&\gamma_{0,2,5},\quad
\gamma_{1,3}^{*}=-\gamma_{1,3},
\\
\gamma_{0,2,5}^{T}&=&\gamma_{0,2,5},\quad
\gamma_{1,3}^{T}=-\gamma_{1,3}.
\end{eqnarray}
The energy dispersion of fermions takes the form $E(\mathbf{k})=\pm\sqrt{v^{2}k_{\bot}^{2}+A^{2}k_{3}^{4}}$ where
$k_{\bot}^{2}=k_{1}^{2}+k_{2}^{2}$. Density of states (DOS) is given by $\rho(\omega)\propto\omega^{3/2}/(v\sqrt{A})$,
which vanishes at the Fermi level.

The fermion action $S_{0}$ is invariant under the discrete transformations including parity ($\mathcal{P}$), time-reversal ($\mathcal{T}$), and
charge conjugation ($\mathcal{C}$). Under parity transformation, the fermion spinor fields satisfy
\begin{eqnarray}
\mathcal{P}\Psi_{\mathbf{k}}\mathcal{P}^{-1}&=&i\gamma_{1}\gamma_{2}\Psi_{-\mathbf{k}},
\\
\mathcal{P}\bar{\Psi}_{\mathbf{k}}\mathcal{P}^{-1}&=&
=-\bar{\Psi}_{-\mathbf{k}}i\gamma_{2}\gamma_{1}.
\end{eqnarray}
Utilizing time-reversal transformation, we have
\begin{eqnarray}
\mathcal{T}\Psi_{\mathbf{k}}\mathcal{T}^{-1}&=&-i\gamma_{1}\gamma_{5}\Psi_{-\mathbf{k}},
\\
\mathcal{T}\bar{\Psi}_{\mathbf{k}}\mathcal{T}^{-1}
&=&\bar{\Psi}_{-\mathbf{k}}i\gamma_{5}\gamma_{1}.
\end{eqnarray}
It should be notice that $\mathcal{T}i\mathcal{T}^{-1}=-i$.
The realization of charge conjugation on spinor fields reads as
\begin{eqnarray}
\mathcal{C}\Psi_{\mathbf{k}}\mathcal{C}^{-1}&=&-i\gamma_{0}\gamma_{1}\Psi_{\mathbf{k}}^{*}
=-\left(\bar{\Psi}_{\mathbf{k}}i\gamma_{1}\right)^{T},
\\
\mathcal{C}\bar{\Psi}_{\mathbf{k}}\mathcal{C}^{-1}&=&
-\left(i\gamma_{1}\Psi_{\mathbf{k}}\right)^{T}.
\end{eqnarray}

The fermion action $S_{0}$ remains  invariant under a continuous global $U(1)$ chiral rotation
\begin{eqnarray}
\Psi_{\mathbf{k}}&\rightarrow& e^{i\theta\gamma_{5}}\Psi_{\mathbf{k}},
\\
\bar{\Psi}_{\mathbf{k}}&\rightarrow&\bar{\Psi}_{\mathbf{k}}e^{i\theta\gamma_{5}}.
\end{eqnarray}
The fermion action $S_{0}$ is also symmetric under a discrete $Z_{2}$ chiral transformation
\begin{eqnarray}
\Psi_{\mathbf{k}}&\rightarrow&\gamma_{5}\Psi_{\mathbf{k}},
\\
\bar{\Psi}_{\mathbf{k}}&\rightarrow&-\bar{\Psi}_{\mathbf{k}}\gamma_{5}.
\end{eqnarray}

The $O(2)$ rotation about $z$ axis is generated by
\begin{eqnarray}
R_{z}(\phi)=e^{\frac{i\phi\Gamma_{03}}{2}},
\end{eqnarray}
where $\Gamma_{03}=\tau_{0}\otimes\sigma_{3}$. We notice that $\Gamma_{03}$ can be also expressed by
$\Gamma_{03}=i\gamma_{5}\gamma_{0}\gamma_{3}$.
Under the $O(2)$ transformation,
\begin{eqnarray}
R_{z}\left(\phi\right)\hat{h}(\mathbf{k})R_{z}^{-1}\left(\phi\right)
&=&\hat{h}(\mathbf{k}'),
\end{eqnarray}
where
\begin{eqnarray}
k_{1}'&=&k_{1}\cos(\phi)+k_{2}\sin(\phi),
\\
k_{2}'&=&-k_{1}\sin(\phi)+k_{2}\cos(\phi),
\\
k_{3}'&=&k_{3}.
\end{eqnarray}
Thus, $S_{0}$ is invariant under the $O(2)$ rotation.
For $\phi=\frac{\pi}{2}$,
\begin{eqnarray}
R_{z}\left(\frac{\pi}{2}\right)=e^{\frac{i\pi\Gamma_{03}}{4}},
\end{eqnarray}
which is just the $C_{4}$ rotation about $z$ axis.

If the four-fermion interaction is weak, it is irrelevant in 3D semi-DSM, due to the vanishing DOS. However, if the
four-fermion interaction is strong enough, the system could be driven to a new phase. As shown in the Appendix~\ref{App:FierzIdentity},
there are 12 kinds of four-fermion interactions. Due to the constraint by Fierz identity, five of
them are linearly independent. Here, we consider the interacting Lagrangian as following
\begin{eqnarray}
\mathcal{L}_{int}&=&g_{1}\left(\bar{\Psi}\gamma_{0}\Psi\right)^{2}+g_{2}\left(\bar{\Psi}\Psi\right)^{2}
+g_{4}\left(\bar{\Psi}\gamma_{0}\gamma_{5}\Psi\right)^{2}\nonumber
\\
&&+g_{5}\left(\bar{\Psi}i\gamma_{5}\Psi\right)^{2}
+g_{3z}\left(\bar{\Psi}\gamma_{0}\gamma_{3}\Psi\right)^{2}. \label{Eq:FFInteractionsLinIndependent}
\end{eqnarray}
In this article, we study the influence of four-fermion interactions on 3D semi-DSM through the RG method \cite{Shankar94}.

\section{Mean-field results}

In this section, taking the four-fermion interaction $g_{2}\left(\bar{\Psi}\Psi\right)^{2}$ as an example, we firstly
show the results of mean-field analysis.

\begin{figure}[htbp]
\center
\includegraphics[width=3.3in]{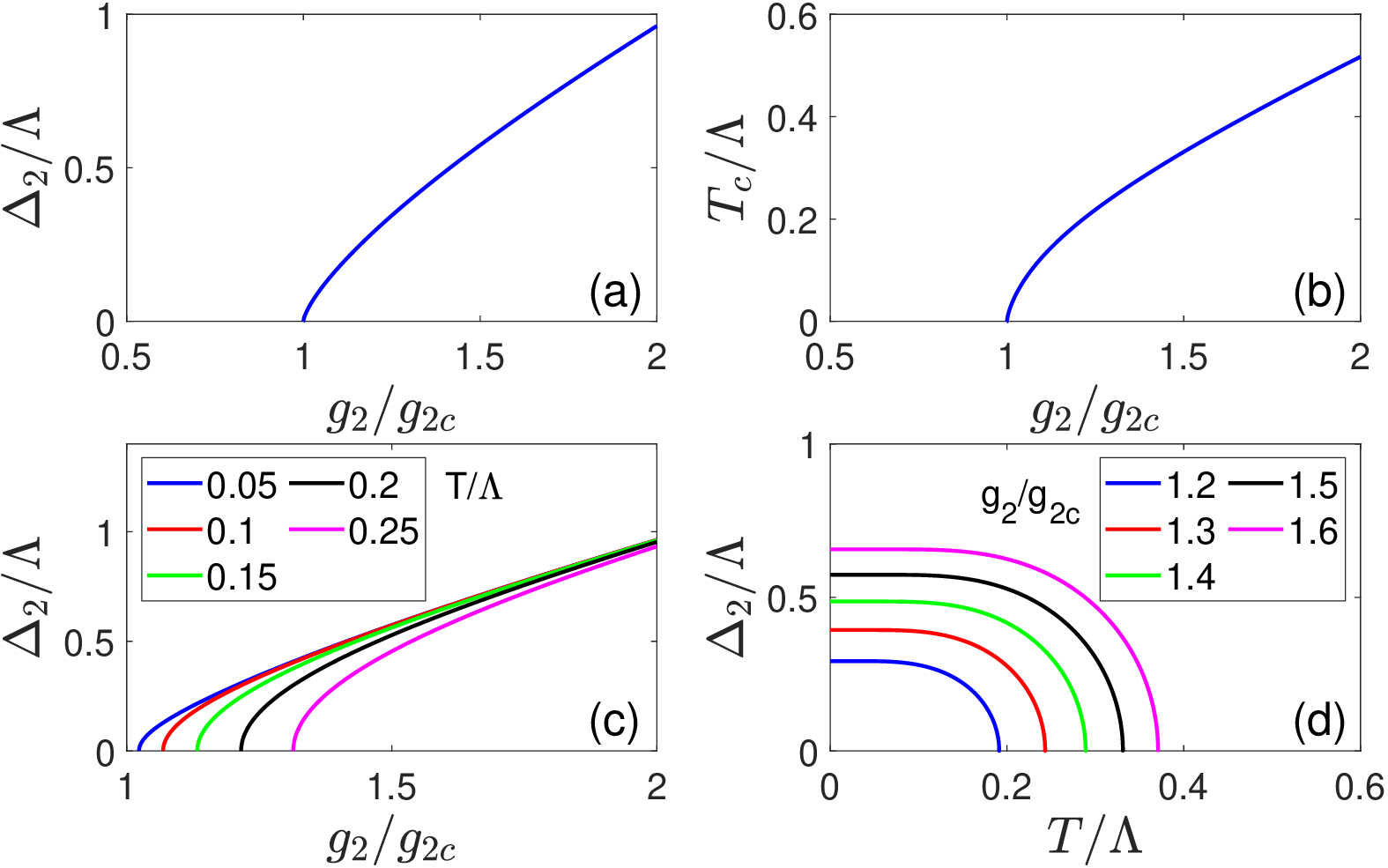}
\caption{Mean field results. (a) Dependence of $\Delta_{2}$ on $g_{2}$ at zero temperature; (b) Dependence of $T_{c}$ on $g_{2}$;
(c) Dependence of $\Delta_{2}$ on $g_{2}$ at different finite temperatures; (d) Dependence of $\Delta_{2}$ on temperature $T$ with
different values of $g_{2}$. \label{Fig:MeanFieldResults}}
\end{figure}

Under the influence of short-range four-fermion interaction $g_{2}\left(\bar{\Psi}\Psi\right)^{2}$,  the expectation value
\begin{eqnarray}
\Delta_{2}=\left<\bar{\Psi}\Psi\right>,
\end{eqnarray}
could become finite. According to the derivation shown in Appendix~\ref{App:MeanFieldAnalysis}, we obtain the free energy density
\begin{eqnarray}
f&=&-4T\int\frac{d^{3}\mathbf{k}}{(2\pi)^{3}}\ln\left[2\cosh\left(\frac{E_{\mathbf{k},\Delta_{2}}}{2T}\right)\right]
+\frac{\Delta_{2}^{2}}{2g_{2}},
\end{eqnarray}
where $E_{\mathbf{k},\Delta_{2}}=\sqrt{v^{2}k_{\bot}^{2}+A^{2}k_{3}^{4}+\Delta_{2}^{2}}$.

Through
\begin{eqnarray}
\frac{\partial f}{\partial \Delta_{2}}&=&0,
\end{eqnarray}
we get the self-consistent equation for $\Delta_{2}$ as following
\begin{eqnarray}
1&=&2g_{2}\int\frac{d^{3}\mathbf{k}}{(2\pi)^{3}}\tanh\left(\frac{E_{\mathbf{k},\Delta_{2}}}{2T}\right)
\frac{1}{E_{\mathbf{k},\Delta_{2}}}. \label{Eq:SelfConsistentEqDelta2FiniteT}
\end{eqnarray}
At zero temperature, the equation becomes
\begin{eqnarray}
1&=&2g_{2}\int\frac{d^{3}\mathbf{k}}{(2\pi)^{3}}
\frac{1}{E_{\mathbf{k},\Delta_{2}}}. \label{Eq:SelfConsistentEqDelta2ZeroT}
\end{eqnarray}

Based on analytical calculation for Eq.~(\ref{Eq:SelfConsistentEqDelta2ZeroT}), we find that $\Delta_{2}$ is given by
\begin{eqnarray}
\Delta_{2}&\approx&c_{1}\Lambda
\frac{\left(g_{2}-g_{2c}\right)^{\frac{2}{3}}}{g_{2}^{\frac{2}{3}}},
\end{eqnarray}
if $g_{2}$ is close to $g_{2c}$,
where
\begin{eqnarray}
g_{2c}&=&\frac{3\pi^{2}v^{2}\sqrt{A}}{2\Lambda^{\frac{3}{2}}},
\end{eqnarray}
and $c_{1}\approx0.662596$. Taking $\Delta_{2}=0$ for Eq.~(\ref{Eq:SelfConsistentEqDelta2FiniteT}), we notice that
the critical temperature $T_{c}$ satisfies
\begin{eqnarray}
T_{c}\approx c_{2}\Lambda\frac{\left(g_{2}-g_{2c}\right)^{\frac{2}{3}}}{g_{2}^{\frac{2}{3}}},
\end{eqnarray}
if $g_{2}$ is close to $g_{2c}$, where $c_{2}=1/\left(2\sqrt{2}a\right)^{\frac{2}{3}}\approx0.622863$.

Numerical results are shown in Figs.~\ref{Fig:MeanFieldResults}(a)-\ref{Fig:MeanFieldResults}(d). In \ref{Fig:MeanFieldResults}(a), dependence of
$\Delta_{2}$ on $g_{2}$ at zero temperature is depicted. Dependence of critical temperature $T_{c}$ on $g_{2}$ is  displayed in Fig.~\ref{Fig:MeanFieldResults}(b).
The behaviors of $\Delta_{2}$ at finite temperature are shown in Figs.~\ref{Fig:MeanFieldResults}(c) and \ref{Fig:MeanFieldResults}(d).

\section{RG results}

As shown in Appendix~\ref{App:DerivationRGEqs}, we firstly calculate all of the  corrections from the one-loop Feynman diagrams, by employing
a momentum shell $b\Lambda<\sqrt{v^2k_{\bot}^{2}+A^{2}k_{3}^{4}}<\Lambda$, where $b=e^{-\ell}$ with $\ell$ being the RG running
parameter. Then, we consider these corrections, and perform RG transformations to restore the original form of the
actions. Accordingly, we obtain the RG equations for $g_{a}$,  which can be written as
\begin{eqnarray}
\frac{dg_{a}}{d\ell}&=&-\frac{3}{2}g_{a}+F_{a}\left(g_{1},g_{2},g_{4},g_{5},g_{3z}\right), \label{Eq:RGEgaMainText}
\end{eqnarray}
where $a=1,2,4,5,3z$, The concrete expressions of $F_{a}$ can be found in Appendix~\ref{App:DerivationRGEqs}.

Solving the equations
\begin{eqnarray}
\left.\frac{dg_{a}}{d\ell}\right|_{(g_{1},g_{2},g_{4},g_{5},g_{3z})=(g_{1}^{*},g_{2}^{*},g_{4}^{*},g_{5}^{*},g_{3z}^{*})}=0,
\end{eqnarray}
we get 12 fixed points, including the trivial Gaussian fixed point
$(g_{1}^{*},g_{2}^{*},g_{4}^{*},g_{5}^{*},g_{3z}^{*})=(0,0,0,0,0)$ and 11 non-trivial fixed points
\begin{eqnarray}
\mathrm{FP}i:\quad (g_{1}^{*},g_{2}^{*},g_{4}^{*},g_{5}^{*},g_{3z}^{*})=(g_{1,i}^{*},g_{2,i}^{*},g_{4,i}^{*},g_{5,i}^{*},g_{3z,i}^{*}),
\end{eqnarray}
with $i=1,2,...,11$.

Expanding the RG equations of $g_{a}$ in the vicinity of a fixed point, we obtain
\begin{eqnarray}
\frac{d\delta g_{a}}{d\ell}&=&\sum_{b}M_{ab}\delta g_{b},
\end{eqnarray}
where $\delta g_{a}= g_{a}-g_{a}^{*}$. $M$ is five dimension square matrix, and the matrix elements
are expressions of $g_{1}^{*}$, $g_{2}^{*}$, $g_{4}^{*}$, $g_{5}^{*}$, $g_{3z}^{*}$.
From eigenvalues of $M$ at a fixed point $(g_{1}^{*}, g_{2}^{*}, g_{4}^{*}, g_{5}^{*},g_{3z}^{*})$, we can get the
properties of the fixed point. A negative (positive) eigenvalue is corresponding to a stable (unstable) eigendirection \cite{Maciejko14, Szabo21}.
There is one unstable direction for QCP, and there are two and three unstable
directions for bicritical point (BCP) and tricritical point (TCP) respectively.
Substituting the values of $g_{a}^{*}$ at each fixed point into the expression of $M$, we calculate the corresponding eigenvalues
of $M$. We find that FP1, FP2, FP3, FP4, and FP5 are QCPs, FP6, FP7, FP8, FP9, and FP10 are BCPs, and FP11 is a TCP.
The detailed calculations are presented in Appendix~\ref{App:NumResults}.

For a QCP, the correlation length exponent  $\nu$ is determined by the inverse of the corresponding positive
eigenvalue of $M$. For the five  QCPs, $\nu$ always satisfies
\begin{eqnarray}
\nu^{-1}=1.5.
\end{eqnarray}

\begin{table}[htbp]
\caption{ There are 5 QCPs among the 11 non-trivial unstable fixed points.  The corresponding order parameters
for the five QCPs are shown in the second rows.
\label{Table:FPSMeaning}}
\begin{center}
\begin{ruledtabular}
\begin{tabular}{lccccccccccc}
\toprule           & FP1 & FP2 & FP3 & FP4 & FP5
\\ \hline
\midrule Order parameter  & $\Delta_{2}$ & $\Delta_{5}$ & $\Delta_{2}/\Delta_{5}$ &  $\Delta_{7z}$ & $\Delta_{8z}$
\\
\bottomrule
\end{tabular}
\end{ruledtabular}
\end{center}
\end{table}

In order to determine the physical essences of the QCPs, we analyze the RG flows of all
the fermion bilinear source terms in particle-hole and particle-particle channels. The source terms
in particle-hole channels can be  written as
\begin{eqnarray}
S_{s}=\Delta_{X}\int\frac{d\omega}{2\pi}\frac{d^3\mathbf{k}}{(2\pi)^{3}}
\bar{\Psi}(\omega,\mathbf{k})\Gamma_{X}\Psi\left(\omega,\mathbf{k}\right). \label{Eq:SourceTermPH}
\end{eqnarray}
There are 12 choices for the matrix $\Gamma_{X}$, which corresponds to 12 different order parameters
in particle-hole channels.
The source terms in particle-particle channels take the form
\begin{eqnarray}
S_{s}=\Delta_{Y}\int\frac{d\omega}{2\pi}\frac{d^3\mathbf{k}}{(2\pi)^{3}}
\Psi^{\dag}(\omega,\mathbf{k})\Gamma_{Y}\Psi^{*}\left(\omega,\mathbf{k}\right).\label{Eq:SourceTermPP}
\end{eqnarray}
There are  $6$ choices for the matrix $\Gamma_{Y}$, which are corresponding to 6 different superconducting
pairings.

As presented in Appendix~\ref{App:SourceTerms},  we calculate the one-loop order corrections to the source terms as shown in Eqs.~(\ref{Eq:SourceTermPH})
and (\ref{Eq:SourceTermPP}) induced by the four-fermion interactions as shown in Eq.~(\ref{Eq:FFInteractionsLinIndependent}).
Then, we include these corrections and perform RG transformations to restore the original forms of the source terms.
Accordingly, through the RG transformations, we obtain the equations
\begin{eqnarray}
\bar{\beta}_{X,Y}=H_{X,Y}\left(g_{1},g_{2},g_{4},g_{5},g_{3z}\right), \label{Eq:BetaSourceTerm}
\end{eqnarray}
where
\begin{eqnarray}
\bar{\beta}_{X,Y}=\frac{d\ln\left(\Delta_{X,Y}\right)}{d\ell}-1.
\end{eqnarray}
$\bar{\beta}_{X,Y}$ are termed as susceptibility exponents or anomalous dimensions
for the fermion-bilinear source terms. $H_{X,Y}$ are functions of $g_{a}$ with $a=1,2,4,5,3z$.
The concrete expressions of $H_{X,Y}$ are shown in Appendix~\ref{App:SourceTerms}.
For a QCP, substituting the values of $g_{a}$ at the QCP into Eq.~(\ref{Eq:BetaSourceTerm}), and
finding the largest one among all of $\bar{\beta}_{X,Y}$, we can determine the physical meaning of the QCP.

For FP1,  $\bar{\beta}_{2}$ takes the largest value. It represents that this fixed point
is corresponding to the QCP to a state in which the order parameter
$\Delta_{2}=\langle\bar{\Psi}\Psi\rangle$ acquires finite value. The physical meaning of $\Delta_{2}$ is scalar mass. $\Delta_{2}$ breaks the continuous
$U(1)$ chiral symmetry,  but preserves $\mathcal{P}$, $\mathcal{T}$, $\mathcal{C}$ symmetries.
If $\Delta_{2}$ becomes finite, the fermion dispersion becomes $E_{\mathbf{k},\Delta_{2}}=\sqrt{v^{2}k_{\bot}^{2}+A^{2}k_{3}^{4}+\Delta_{2}^{2}}$,
which is gapped.

\begin{figure}[htbp]
\center
\includegraphics[width=3.3in]{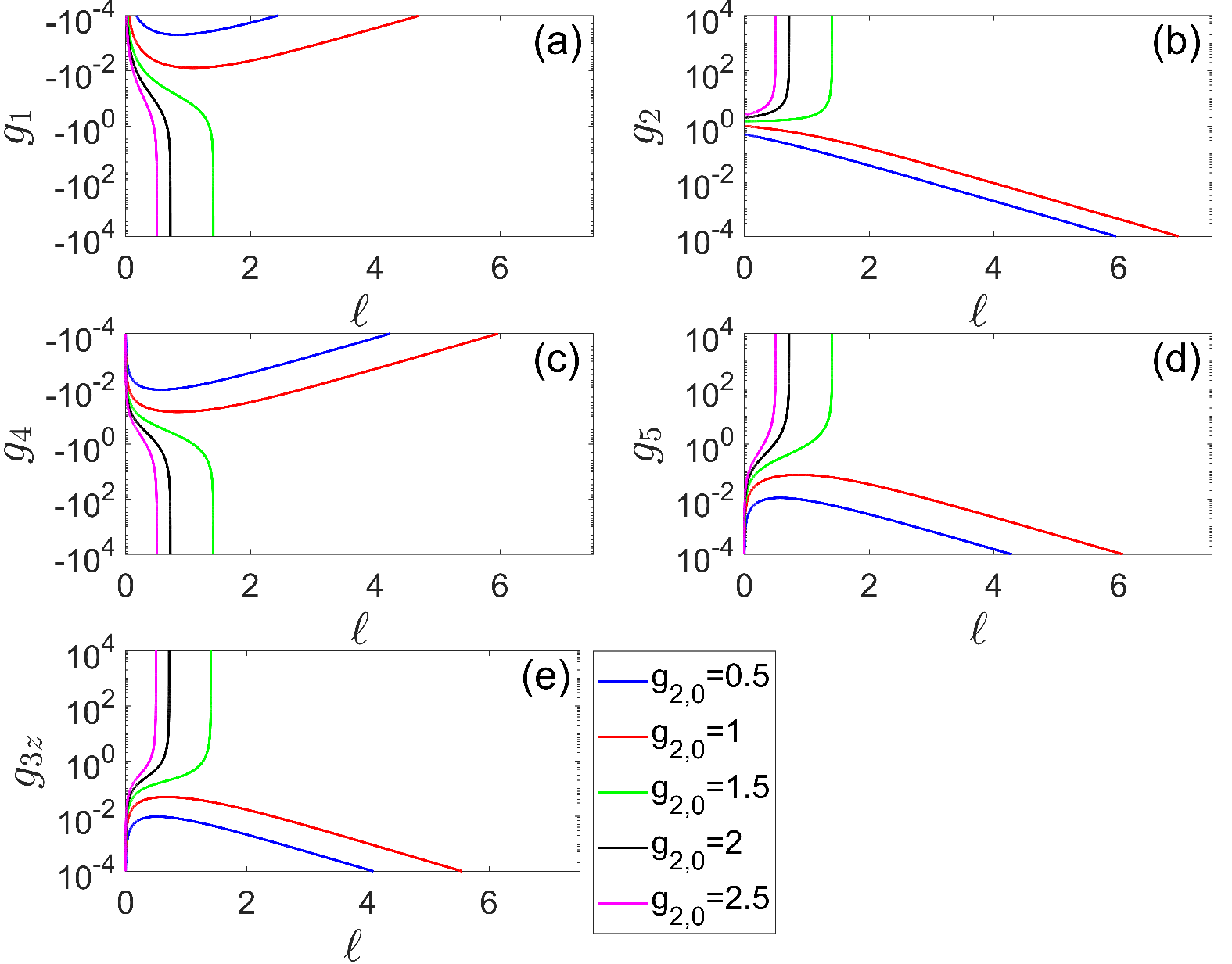}
\caption{(a)-(e): Flows of $g_{1}$, $g_{2}$, $g_{4}$, $g_{5}$, and $g_{3z}$ with different initial conditions.
 $g_{1,0}=0$, $g_{4,0}=0$, $g_{5,0}=0$, and $g_{3z,0}=0$ are taken. \label{Fig:gArray}}
\end{figure}

For FP2,  $\bar{\beta}_{5}$ is the largest one. It means that this fixed point
stands for the QCP to a state in which the order parameter
$\Delta_{5}=\langle\bar{\Psi}i\gamma_{5}\Psi\rangle$ becomes finite. The physical meaning of $\Delta_{5}$ corresponds to pseudoscalar mass. $\Delta_{5}$ breaks
continuous $U(1)$ chiral symmetry and $\mathcal{C}$ symmetry, but preserves $\mathcal{P}$ and $\mathcal{T}$ symmetries. Once
$\Delta_{5}$ becomes finite, the corresponding fermion dispersion can be written as $E_{\mathbf{k},\Delta_{5}}=\sqrt{v^{2}k_{\bot}^{2}+A^{2}k_{3}^{4}+\Delta_{5}^{2}}$.

For FP3, $\bar{\beta}_{2}$ and $\bar{\beta}_{5}$ are largest simultaneously. It indicates that the fixed point
corresponds to the QCP to a phase in which both of $\Delta_{2}$ and $\Delta_{5}$ become finite. The parameter
of this phase can be written as $\langle\bar{\Psi}\left(\cos(\theta)+i\gamma_{5}\sin(\theta)\right)\Psi\rangle$. In axionic insulating phase,
the continuous $U(1)$ chiral symmetry is broken. This
phase represents an axionic insulator \cite{Roy16}. In this case, the fermion dispersion takes the form
$E_{\mathbf{k},\Delta_{2}, \Delta_{5}}=\sqrt{v^{2}k_{\bot}^{2}+A^{2}k_{3}^{4}+\Delta_{2}^{2}+\Delta_{5}^{2}}$.

If the order parameters $\Delta_{2}$ and $\Delta_{5}$ are generated by four-fermion interactions, the continuous $U(1)$ chiral symmetry is broken.
There will be gapless Goldstone boson accompanied with breaking of continuous $U(1)$ chiral symmetry.
In real solid-state materials, some higher-order gradient terms such as $\bar{\Psi}k^{2}\mathbbm{1}_{4\times4}\Psi$, $\bar{\Psi}k^{4}\mathbbm{1}_{4\times4}\Psi$ \emph{etc.} could
appear in the action of free 3D semi-DSM. In this case, the action of free 3D semi-DSM breaks the continuous $U(1)$ chiral symmetry, but still satisfies
the discrete symmetries including $\mathcal{P}$, $\mathcal{T}$, and $\mathcal{C}$ symmetries.
Correspondingly, if $\Delta_{2}$ and $\Delta_{5}$ are generated by four-fermion
interactions, we notice that the discrete symmetry $\mathcal{C}$ is broken. Breaking of discrete symmetry will not lead gapless Goldstone mode.

\begin{figure}[htbp]
\center
\includegraphics[width=3.3in]{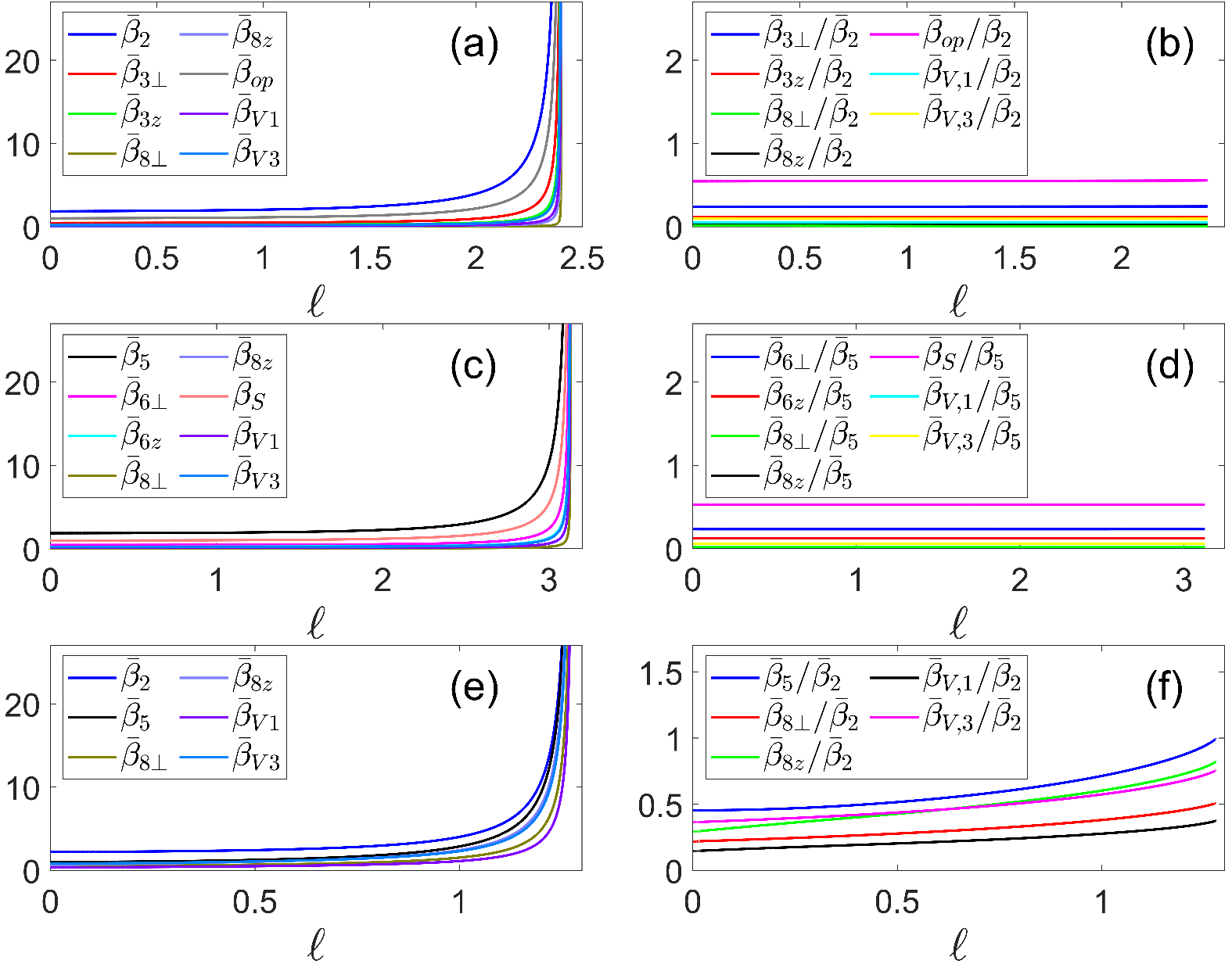}
\caption{Flows of $\bar{\beta}_{X,Y}$ which approach to positive infinity and  ratios between $\bar{\beta}_{X,Y}$. (a) and (b):
$g_{1,0}=0.15$, $g_{2,0}=1.3$, $g_{4,0}=0.46$, $g_{5,0}=-0.56$, and $g_{3z,0}=0.055$ are taken;
(c) and (d): $g_{2,0}=0.14$, $g_{2,0}=-0.59$, $g_{4,0}=0.42$, $g_{5,0}=1.36$, and $g_{3z,0}=0.07$ are taken;
(e) and (f): $g_{1,0}=0$, $g_{2,0}=1.4$, $g_{3,0}=0$, $g_{5,0}=0.2$, and $g_{3z,0}=0$ are taken. \label{Fig:SourceArray}}
\end{figure}

For FP4, $\bar{\beta}_{7z}$ takes the largest value. It signifies that this fixed point
is corresponding to the QCP to a state in which the order parameter
$\Delta_{7z}=\langle\bar{\Psi}i\gamma_{5}\gamma_{3}\Psi\rangle$ becomes finite.
$\Delta_{7z}$ stands for axial magnetization along $z$ axis. $\Delta_{7z}$ breaks $\mathcal{T}$ symmetry,
but preserves $\mathcal{P}$, $\mathcal{C}$, and $U(1)$ chiral symmetries.
If
$\Delta_{7z}>0$, the original fermion dispersion
becomes two dispersions $E_{\mathbf{k},\Delta_{2}}^{\pm}$ which can be written as
$E_{\mathbf{k},\Delta_{7z}}^{\pm}=\sqrt{v^{2}k_{\bot}^{2}+\left(Ak_{3}^{2}\pm\Delta_{7z}\right)^{2}}$.
It is easy to find that one dispersion $E_{\mathbf{k},\Delta_{7z}}^{+}$ is gapped, but another dispersion $E_{\mathbf{k},\Delta_{7z}}^{-}$
is gapless at two points $\mathbf{k}_{a}=(0,0,\sqrt{\frac{\Delta_{7z}}{A}})$ and $\mathbf{k}_{b}=(0,0,-\sqrt{\frac{\Delta_{7z}}{A}})$.
At these two gapless points, the fermions dispersion takes the form $E_{\mathbf{K},\Delta_{7z}}=\sqrt{v^2K_{\bot}^{2}+v_{z}^{2}K_{z}^{2}}$,
with $v_{z}=2\sqrt{A\Delta_{7z}}$ and $\mathbf{K}$ being the momentum relative to the point $\mathbf{k}_{a}$ or $\mathbf{k}_{b}$.
It is obvious that this fermion dispersion is linear within $xy$ plane and also linear along $z$ axis.

For FP5, $\bar{\beta}_{8z}$ is the largest one. It suggests that this fixed point
represents the QCP to a state with finite order parameter
$\Delta_{8z}=\langle\bar{\Psi}i\gamma_{3}\Psi\rangle$. $\Delta_{8z}$ does not break the symmetries of the free semi-DSM.
The physical meaning of $\Delta_{8z}$ is current along $z$ axis. If $\Delta_{8z}>0$, the fermion dispersion becomes
$E_{\mathbf{k},\Delta_{8z}}=\sqrt{v^{2}k_{\bot}^{2}+\left(Ak_{3}^{2}+\Delta_{8z}\right)^{2}}$, which is gapped.

For convenience, we summarize the corresponding order parameters for the five QCPs  in Table~\ref{Table:FPSMeaning}.

For general given initial conditions that is decided by $(g_{1,0}, g_{2,0}, g_{4,0}, g_{5,0}, g_{3z,0})$, we determine the corresponding phase
through the flows of four-fermion coupling parameters $g_{a}$ and flows of susceptibility exponents $\bar{\beta}_{X,Y}$. We show the flows of
$g_{1}$, $g_{2}$, $g_{4}$, $g_{5}$, and $g_{3z}$ under several initial conditions in Fig.~\ref{Fig:gArray}.
If $g_{a}$ with $a=1,2,4,5,3z$ approach to zero, it represents that the system is still in SM phase. If $|g_{a}|$  flow to infinity at
a finite running parameter $\ell_{c}$, the system becomes unstable to a new phase.

\begin{figure*}[htbp]
\center
\includegraphics[width=6.8in]{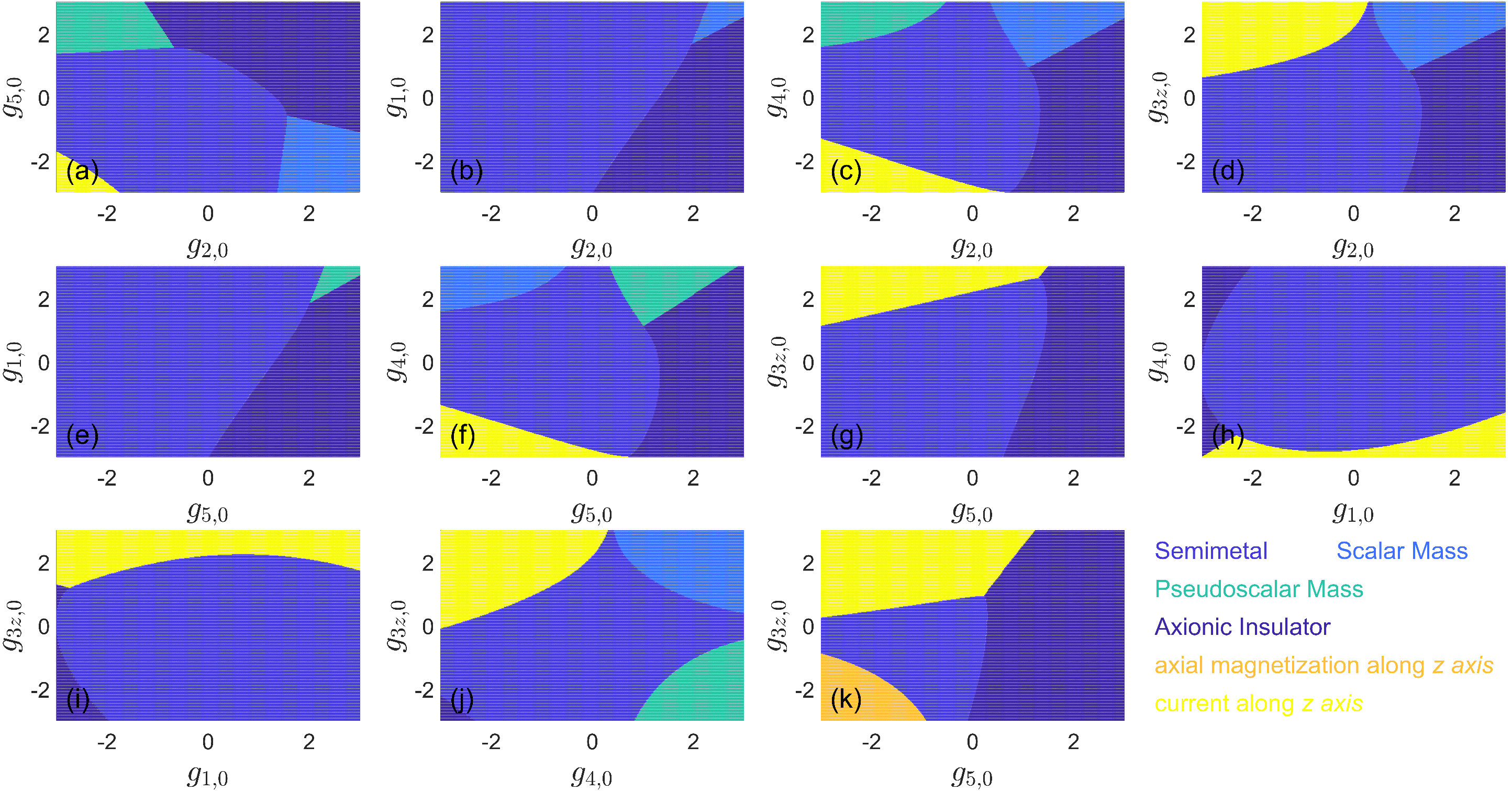}
\caption{Phase diagrams on the planes of two initial values of four-fermion coupling strength.
(a) $g_{2,0}$ and $g_{5,0}$; (b) $g_{2,0}$ and $g_{1,0}$; (c) $g_{2,0}$ and $g_{4,0}$; (d) $g_{2,0}$ and $g_{3z,0}$;
(e) $g_{5,0}$ and $g_{1,0}$; (f) $g_{5,0}$ and $g_{4,0}$; (g) $g_{5,0}$ and $g_{3z,0}$; (h) $g_{1,0}$ and $g_{4,0}$;
(i) $g_{1,0}$ and $g_{3z,0}$; (j) $g_{4,0}$ and $g_{3z,0}$; (k) $g_{5,0}$ and $g_{3z,0}$  In (a)-(j), the initial values of rest four-fermion coupling parameters
are taken as zero. For example, $g_{1,0}=0$, $g_{4,0}=0$ and $g_{3z,0}=0$ are taken in (a). In (k), $g_{1,0}=-2.3$, $g_{2,0}=0$, and $g_{4,0}=-0.61$ are taken.
\label{Fig:PhaseDiagramDiffPlane}}
\end{figure*}

In order to determine the physical essence of the new phase, we calculate the flows of the susceptibility exponents $\bar{\beta}_{X,Y}$ and compare them.
For three general initial conditions, the flows of $\bar{\beta}_{X,Y}$ and the ratio between them are presented in Figs.~\ref{Fig:SourceArray}(a)-\ref{Fig:SourceArray}(f).
Here we only show the susceptibility exponents that approach to positive infinity in Figs.~\ref{Fig:SourceArray}(a), \ref{Fig:SourceArray}(c),
and \ref{Fig:SourceArray}(e) respectively. For the initial condition corresponding to  Figs.~\ref{Fig:SourceArray}(a) and \ref{Fig:SourceArray}(b),
we find that $\bar{\beta_{2}}$ approaches to positive infinity most quickly. It means that scalar mass $\Delta_{2}$ is generated in the new phase.
For the initial condition corresponding to  Figs.~\ref{Fig:SourceArray}(c) and \ref{Fig:SourceArray}(d),
$\bar{\beta_{5}}$ flows to positive infinity with the largest speed. It indicates that pseudoscalar mass $\Delta_{5}$ becomes finite in the new phase.
For the initial condition corresponding to  Figs.~\ref{Fig:SourceArray}(e) and \ref{Fig:SourceArray}(f), we notice that $\bar{\beta}_{2}$ and $\bar{\beta}_{5}$
approach to positive infinity most quickly and $\bar{\beta}_{5}/\bar{\beta}_{2}\rightarrow1$.
It represents that the system becomes to a new phase that $\Delta_{2}$ and $\Delta_{5}$ acquire finite values
simultaneously. Namely, the system becomes to axionic insulating phase.

The phase diagrams on the planes composed by initial values of two four-fermion coupling parameters are shown in Fig.~\ref{Fig:PhaseDiagramDiffPlane}.
Different phases are marked by different colors. In Figs.~\ref{Fig:PhaseDiagramDiffPlane}(a)-\ref{Fig:PhaseDiagramDiffPlane}(j),
we show all the 10 phase diagrams on the planes composed
by initial values of two coupling parameters chosen from the five linearly independent coupling parameters.
The initial values of rest of three coupling parameters are taken as zero.
Taken Fig.~\ref{Fig:PhaseDiagramDiffPlane}(a) composed by $g_{2,0}$ and $g_{5,0}$ as an example,
we can notice that there are five phases, SM, insulator with scalar mass $\Delta_{2}$,
insulator characterized by pseudoscalar mass $\Delta_{5}$, axionic insulating phase, and a phase with current along $z$ axis $\Delta_{8z}$.
In Fig.~\ref{Fig:PhaseDiagramDiffPlane}(k), we present the phase diagram
composed by $g_{5,0}$ and $g_{3z,0}$. In this phase diagram, $g_{1,0}$, $g_{2,0}$
and $g_{4,0}$ are taken as proper values so that the phase with axial magnetization
along $z$ axis $\Delta_{7z}$ appears in the phase diagram.

The behaviors in the vicinity of a QCP is generally consistent with these indicated by Sur and Roy  \cite{Sur19}. In 3D semi-DSM, the Yukawa coupling between quantum fluctuation of order parameter and fermion
excitations becomes irrelevant in the low energy regime. Thus, the fermions should take Fermi liquid behaviors in the vicinity of a
QCP between SM phase and a symmetry breaking phase in 3D semi-DSM. Concretely, the residue of fermions $Z_{f}$ approaches to a finite constant value
in the lowest energy limit, and the Landau damping rate of fermions $\Gamma(\omega)$ satisfies
\begin{eqnarray}
\lim_{\omega\rightarrow0}\frac{\Gamma(\omega)}{\omega}\rightarrow0.
\end{eqnarray}
Additionally, under the influence of quantum fluctuation of order parameter, the observable quantities DOS $\rho$,
specific heat $C_{v}$, compressibility $\kappa$, optical conductivities within the $xy$ plane and along $z$ axis
$\sigma_{\bot\bot}$ and $\sigma_{zz}$ should respectively still take the behaviors
\begin{eqnarray}
&&\rho(\omega)\sim\omega^{3/2},\qquad C_{v}(T)\sim T^{5/2}, \qquad \kappa(T)\sim T^{3/2},\nonumber
\\
&&\sigma_{\bot\bot}(\omega)\sim\omega^{1/2},\qquad \sigma_{zz}(\omega)\sim\omega^{3/2},
\end{eqnarray}
which are qualitatively same as the ones for free fermions.

\section{Interplay with Coulomb interaction}

In 3D semi-DSM, the long-range Coulomb interaction becomes irrelevant
in the low energy regime \cite{Abrikosov72, YangNatPhys14}. Considering the interplay of short-range four-fermion interactions and long-range Coulomb
interaction, we find that the flow of Coulomb interaction is not changed and still becomes irrelevant in the low energy
regime. Whereas, the flows of four-fermion interactions are modified by Coulomb interaction. It is shown that Coulomb
interaction tends to enhance the instabilities in particle-hole channels. For the case that all the initial values of four-fermion coupling
strength vanish, if the Coulomb interaction is strong enough, the four-fermion interactions are generated and become divergent finally
driven by the Coulomb interaction. We notice that the system is driven into axionic insulating phase in this case.
The detailed derivation and numerical results are shown in Appendix~\ref{App:InterplayCoulomb}.

\section{Summary}

To conclude, we perform comprehensive studies about the influence of four-fermion interactions on
3D semi-DSM through RG theory. We find 11 unstable fixed points and show that five of them are QCPs,
five are BCPs, and the rest one is a TCP. The physical essence of the QCPs are determined
by analyzing the scalings of fermion bilinear source terms. The phase diagrams for general initial conditions
are also presented through detailed numerical calculations of flows of four-fermion couplings and susceptibility
exponents.

According to the theoretical study by Yang and Nagaosa \cite{Yang14}, 3D semi-DSM state can be realized at the topological QCP between 3D DSM and
band insulator. Through magneto-optics and magneto-transport, Yuan \emph{et al.} observed the evidence of 3D semi-DSM
phase in ZrTe$_{5}$ \cite{Yuan17}. The subsequent study about magnetotransport properties of ZrTe$_{5}$ under hydrostatic pressure
also supports the existence of 3D semi-DSM phase \cite{ZhangJingLei17}. Recent measurements of optical spectroscopy in ZrTe$_{5}$
are also consistent with 3D semi-DSM phase \cite{Martino19, SantoCottin20}. 3D semi-DSM state was also realized in pressured Cd$_{3}$As$_{2}$
\cite{ZhangCengXiuFaXian17}. Recently, Monhanta \emph{et al.} showed that nonmagnetic tetragonal perovskite oxides with $I4/mcm$ symmetry, e.g., SrNbO$_{3}$,
CaNbO$_{3}$, and SrMoO$_{3}$, host 3D semi-Dirac fermions which are protected by a nonsymmorphic symmetry \cite{Mohanta21}.
The present RG calculation results are helpful for understanding the physical properties of these candidate materials for 3D semi-DSM.

\section*{ACKNOWLEDGEMENTS}

J.R.W. is grateful to Prof. G.-Z. Liu for the valuable discussions.
We acknowledge the support from the National Natural Science
Foundation of China under Grants 11974356, 12274414, and U1832209. A portion of this work
was supported by the High Magnetic Field Laboratory of Anhui Province under
Grant AHHM-FX-2020-01.

\appendix

\section{Fierz Identity \label{App:FierzIdentity}}

\subsection{Fierz identity for 3D DSM}

The four-fermion interactions can be generally written as form $(\bar{\Psi}
M \Psi) (\bar{\Psi} N \Psi)$, where $M$ and $N$ are 4 dimensional matrices. There are 16 independent four-by-four matrices.
Accordingly, there are possible $16+16\times15/2=136$ kinds of four-fermion interactions.
However, the number of four-fermion interaction can be drastically reduced by the symmetries of the system \cite{Herbut09, Maciejko14, Roy16, Szabo21, Vafek10}.
The action of free fermions satisfies the symmetries including parity symmetry, time-reversal symmetry, charge conjugation symmetry,
rotation symmetry \emph{etc}. We could reduce the number of  four-fermion interactions by these symmetries of action of free fermions \cite{Herbut09, Maciejko14, Roy16, Szabo21, Vafek10}.
The four-fermion interactions  $(\bar{\Psi}M \Psi) (\bar{\Psi} N \Psi)$ with $M=N$ respect these symmetries, but the four-fermion interactions  $(\bar{\Psi}
M \Psi) (\bar{\Psi} N \Psi)$ with $M\neq N$ can not fulfill these symmetries. Therefore, we don't consider the four-fermion interactions $(\bar{\Psi}
M \Psi) (\bar{\Psi} N \Psi)$ with $M\neq N$ which are not allowed by the symmetries.

For 3D DSM, the interacting Lagrangian density can be written as \cite{Roy16}
\begin{eqnarray}
\mathcal{L}_{int}&=&g_{1}\left(\bar{\Psi}\gamma_{0}\Psi\right)^{2}
+g_{2}\left(\bar{\Psi}\Psi\right)^{2}+g_{3}\sum_{j=1}^{3}\left(\bar{\Psi}\gamma_{0}\gamma_{j}\Psi\right)^{2}\nonumber
\\
&&+g_{4}\left(\bar{\Psi}\gamma_{0}\gamma_{5}\right)^{2}
+g_{5}\left(\bar{\Psi}i\gamma_{5}\Psi\right)^{2}
\nonumber
\\
&&+g_{6}\sum_{\left<lk\right>}\left(\bar{\Psi}i\gamma_{l}\gamma_{k}\Psi\right)^{2}
+g_{7}\sum_{j=1}^{3}\left(\bar{\Psi}i\gamma_{5}\gamma_{j}\Psi\right)^{2}\nonumber
\\
&&+g_{8}\sum_{j=1}^{3}\left(\bar{\Psi}i\gamma_{j}\Psi\right)^{2}, \label{Eq:FourFermionAll3DDSM}
\end{eqnarray}
where
\begin{eqnarray}
\sum_{\left<lk\right>}\left(\bar{\Psi}i\gamma_{l}\gamma_{k}\Psi\right)^{2}
&=&\left[\left(\bar{\Psi}i\gamma_{2}\gamma_{3}\Psi\right)^{2}+\left(\bar{\Psi}i\gamma_{3}\gamma_{1}\Psi\right)^{2}\right.\nonumber
\\
&&\left.+\left(\bar{\Psi}i\gamma_{1}\gamma_{2}\Psi\right)^{2}\right].
\end{eqnarray}
There are eight kinds of four-fermion couplings. However, not all of them are linearly independent, due to
the constraint by Fierz identity \cite{Roy16, Herbut09, Szabo21}.

The Fierz identity indicates that \cite{Roy16, Herbut09, Szabo21}
\begin{eqnarray}
&&\left[\bar{\Psi}(x)M\Psi(x)\right]\left[\bar{\Psi}(y)N\Psi(y)\right]\nonumber
\\
&=&-\frac{1}{16}\mathrm{Tr}\left[M\Gamma_{a}N\Gamma_{b}\right]\left[\bar{\Psi}(x)\Gamma_{a}\Psi(y)\right]\nonumber
\\
&&\times\left[\bar{\Psi}(y)\Gamma_{b}\Psi(x)\right].
\label{Eq:FierzIdentityGeneral}
\end{eqnarray}
For local interaction, we have $x=y$. Thus,
\begin{eqnarray}
&&\left[\bar{\Psi}(x)M\Psi(x)\right]\left[\bar{\Psi}(x)N\Psi(x)\right]\nonumber
\\
&=&-\frac{1}{16}\mathrm{Tr}\left[M\Gamma_{a}N\Gamma_{b}\right]\left[\bar{\Psi}(x)\Gamma_{a}\Psi(x)\right]\nonumber
\\
&&\times
\left[\bar{\Psi}(x)\Gamma_{b}\Psi(x)\right].
\label{Eq:FierzIdentityGeneralLocal}
\end{eqnarray}
Repeat of indexes $a$ and $b$ in Eqs.~(\ref{Eq:FierzIdentityGeneral}) and (\ref{Eq:FierzIdentityGeneralLocal})
represents summation.
Substituting each four-fermion coupling in Eq.~(\ref{Eq:FourFermionAll3DDSM}) into Eq.~(\ref{Eq:FierzIdentityGeneralLocal}),
we could get eight equations, which can be compactly expressed by
\begin{eqnarray}
FX=0,
\end{eqnarray}
where
\begin{eqnarray}
X=\left(\begin{array}{c}
\left(\bar{\Psi}\gamma_{0}\Psi\right)^{2}
\\
\left(\bar{\Psi}\Psi\right)^{2}
\\
\sum\limits_{j=1}^{3}\left(\bar{\Psi}\gamma_{0}\gamma_{j}\Psi\right)^{2}
\\
\left(\bar{\Psi}\gamma_{0}\gamma_{5}\Psi\right)^{2}
\\
\left(\bar{\Psi}i\gamma_{5}\Psi\right)^{2}
\\
\sum\limits_{<lk>}\left(\bar{\Psi}i\gamma_{l}\gamma_{k}\Psi\right)^{2}
\\
\sum\limits_{j=1}^{3}\left(\bar{\Psi}i\gamma_{5}\gamma_{j}\Psi\right)^{2}
\\
\sum\limits_{j=1}^{3}\left(\bar{\Psi}i\gamma_{j}\Psi\right)^{2}
\end{array}\right),
\end{eqnarray}
and
\begin{eqnarray}
F=\left(
\begin{array}{rrrrrrrr}
5 & 1 & 1 & 1 & 1 & 1 & 1 & 1
\\
1 & 5 & -1 & -1 & -1 & 1 & 1 & -1
\\
3 & -3 & 3 &-3 & 3 & 1 & -1 & 1
\\
1 & -1 & -1 & 5 & -1 & -1 & 1 & 1
\\
1 & -1 & 1 & -1 & 5 & -1 & 1 & -1
\\
3 & 3 & 1 & -3 & -3 & 3 & -1 & 1
\\
3 & 3 & -1 & 3 & 3 & -1 & 3 & -1
\\
3 & -3 & 1 & 3 & -3 & 1 & -1 & 3
\end{array}\right).
\end{eqnarray}

It is easy to verify that rank of $F$ is 4, namely
\begin{eqnarray}
\mathrm{Rank}(F)=4.
\end{eqnarray}
Then, the number of linearly independent couplings is
\begin{eqnarray}
8-\mathrm{Rank}(F)=4.
\end{eqnarray}
For convenience, we take the four couplings
\begin{eqnarray}
\left(\bar{\Psi}\gamma_{0}\Psi\right)^{2},\quad\left(\bar{\Psi}\Psi\right)^{2},\quad
\left(\bar{\Psi}\gamma_{0}\gamma_{5}\Psi\right)^{2},\quad\left(\bar{\Psi}i\gamma_{5}\Psi\right)^{2}, \label{Eq:3DDSMLinearIndepenCouplings}
\end{eqnarray}
as linearly independent couplings. The other couplings
\begin{eqnarray}
&&\sum\limits_{j=1}^{3}\left(\bar{\Psi}\gamma_{0}\gamma_{j}\Psi\right)^{2},\qquad
\sum\limits_{<lk>}\left(\bar{\Psi}i\gamma_{l}\gamma_{k}\Psi\right)^{2},\nonumber
\\
&&\sum\limits_{j=1}^{3}\left(\bar{\Psi}i\gamma_{5}\gamma_{j}\Psi\right)^{2},\qquad
\sum\limits_{j=1}^{3}\left(\bar{\Psi}i\gamma_{j}\Psi\right)^{2},
\end{eqnarray}
can be expressed by the four independent couplings shown in Eq.~(\ref{Eq:3DDSMLinearIndepenCouplings}).
In order to obtain the concrete expressions for other couplings, we define
\begin{eqnarray}
\tilde{X}=\left(\begin{array}{c}
\sum\limits_{j=1}^{3}\left(\bar{\Psi}\gamma_{0}\gamma_{j}\Psi\right)^{2}
\\
\sum\limits_{<lk>}\left(\bar{\Psi}i\gamma_{l}\gamma_{k}\Psi\right)^{2}
\\
\sum\limits_{j=1}^{3}\left(\bar{\Psi}i\gamma_{5}\gamma_{j}\Psi\right)^{2}
\\
\sum\limits_{j=1}^{3}\left(\bar{\Psi}i\gamma_{j}\Psi\right)^{2}
\\
\left(\bar{\Psi}\gamma_{0}\Psi\right)^{2}
\\
\left(\bar{\Psi}\Psi\right)^{2}
\\
\left(\bar{\Psi}\gamma_{0}\gamma_{5}\Psi\right)^{2}
\\
\left(\bar{\Psi}i\gamma_{5}\Psi\right)^{2}
\end{array}\right).
\end{eqnarray}
It is easy to find that
\begin{eqnarray}
\tilde{F}\tilde{X}=0,
\end{eqnarray}
where
\begin{eqnarray}
\tilde{F}=\left(
\begin{array}{rrrrrrrr}
1 & 1 & 1 & 1 & 5 & 1 & 1 & 1
\\
-1 & 1 & 1 & -1 & 1 & 5 & -1 & -1
\\
3 & 1 & -1 & 1 & 3 & -3 & -3 & 3
\\
-1 & -1 & 1 & 1 & 1 & -1 & 5 & -1
\\
1 & -1 & 1 & -1 & 1 & -1 & -1 & 5
\\
1 & 3 & -1 & 1 & 3 & 3 & -3 & -3
\\
-1 & -1 & 3 & -1 & 3 & 3 & 3 & 3
\\
1 & 1 & -1 & 3 & 3 & -3 & 3 & -3
\end{array}\right).
\end{eqnarray}
Performing a series of similarity transformations for $\tilde{F}$,
\begin{eqnarray}
\tilde{F}\rightarrow\tilde{F}',
\end{eqnarray}
we obtain
\begin{eqnarray}
\tilde{F}'\tilde{X}=0, \label{Eq:EchelonMatrix3DDSM}
\end{eqnarray}
where
\begin{eqnarray}
\tilde{F}'=\left(
\begin{array}{rrrrrrrr}
1 & 0 & 0 & 0 & 1 & -1 & -1 & 2
\\
0 & 1 & 0 & 0 & 1 & 2 & -1 & -1
\\
0 & 0 & 1 & 0 & 2 & 1 & 1 & 1
\\
0 & 0 & 0 & 1 & 1 & -1 & 2 & -1
\\
0 & 0 & 0 & 0 & 0 & 0 & 0 & 0
\\
0 & 0 & 0 & 0 & 0 & 0 & 0 & 0
\\
0 & 0 & 0 & 0 & 0 & 0 & 0 & 0
\\
0 & 0 & 0 & 0 & 0 & 0 & 0 & 0
\end{array}\right).
\end{eqnarray}
Eq.~(\ref{Eq:EchelonMatrix3DDSM}) can be equivalently expressed by
\begin{eqnarray}
\sum\limits_{j=1}^{3}\left(\bar{\Psi}\gamma_{0}\gamma_{j}\Psi\right)^{2}
&=&-\left(\bar{\Psi}\gamma_{0}\Psi\right)^{2}
+\left(\bar{\Psi}\Psi\right)^{2}+\left(\bar{\Psi}\gamma_{0}\gamma_{5}\Psi\right)^{2}\nonumber
\\
&&
-2\left(\bar{\Psi}i\gamma_{5}\Psi\right)^{2},
\\
\sum\limits_{<lk>}\left(\bar{\Psi}i\gamma_{l}\gamma_{k}\Psi\right)^{2}
&=&-\left(\bar{\Psi}\gamma_{0}\Psi\right)^{2}
-2\left(\bar{\Psi}\Psi\right)^{2}\nonumber
\\
&&+\left(\bar{\Psi}\gamma_{0}\gamma_{5}\Psi\right)^{2}
+\left(\bar{\Psi}i\gamma_{5}\Psi\right)^{2},
\\
\sum\limits_{j=1}^{3}\left(\bar{\Psi}i\gamma_{5}\gamma_{j}\Psi\right)^{2}
&=&-2\left(\bar{\Psi}\gamma_{0}\Psi\right)^{2}
-\left(\bar{\Psi}\Psi\right)^{2}\nonumber
\\
&&-\left(\bar{\Psi}\gamma_{0}\gamma_{5}\Psi\right)^{2}
-\left(\bar{\Psi}i\gamma_{5}\Psi\right)^{2},
\\
\sum\limits_{j=1}^{3}\left(\bar{\Psi}i\gamma_{j}\Psi\right)^{2}
&=&-\left(\bar{\Psi}\gamma_{0}\Psi\right)^{2}
+\left(\bar{\Psi}\Psi\right)^{2}\nonumber
\\
&&-2\left(\bar{\Psi}\gamma_{0}\gamma_{5}\Psi\right)^{2}
+\left(\bar{\Psi}i\gamma_{5}\Psi\right)^{2}.
\end{eqnarray}

\begin{widetext}

\subsection{Fierz identity for 3D semi-DSM}

\begin{table*}[htbp]
\caption{ Transformation properties of various fermion bilinears under parity ($\mathcal{P}$), time-reversal ($\mathcal{T}$),  charge conjugation ($\mathcal{C}$),
$Z_{2}$ chiral, $U(1)$ chiral, and $O(2)$ ration transformations. Notice  that $j=1,2$, and $lk=23,31$. Here, $+$ ($-$) represents that the fermion bilinear is even (odd) under  a transformation.
In the sixth column, $\checkmark$ represents that the bilinear is a scalar under the $U(1)$ chiral transformation. Fermion bilinears which transform as components of a chiral $U(1)$
vector under the $U(1)$ chiral transformation are marked by  circles with same color. The colors red, green, blue correspond to three different chiral $U(1)$ vectors. In the seventh column,
0 (1) stands for that the bilinear transforms as a scalar (vector) under the $O(2)$ rotation about $z$ axis.
\label{Table:FermionBilinearCPTTransformations}}
\begin{center}
\begin{ruledtabular}
\begin{tabular}{lllllll}
\toprule   bilinear  &      $\mathcal{P}$ &  $\mathcal{T}$ & $\mathcal{C}$ & $Z_{2}$ & $U(1)$ & $O(2)$
\\ \hline
\midrule
$\bar{\Psi}\gamma_{0}\Psi$    & $+$  & $+$ & $-$ & $+$ & $\checkmark$ & 0
\\
$\bar{\Psi}\Psi$    & $+$  & $+$ & $+$ & $-$ & \textcolor[rgb]{1.00,0.00,0.00}{$\bullet$} & 0
\\
$\bar{\Psi}\gamma_{0}\gamma_{j}\Psi$    & $-$  & $+$ & $-$ & $-$ & \textcolor[rgb]{0.00,1.00,0.00}{$\bullet$} & 1
\\
$\bar{\Psi}\gamma_{0}\gamma_{3}\Psi$    & $+$  & $-$ & $+$ & $-$ & \textcolor[rgb]{0.00,0.00,1.00}{$\bullet$} & 0
\\
$\bar{\Psi}\gamma_{0}\gamma_{5}\Psi$    & $+$  & $-$ & $-$ & $+$ & $\checkmark$ & 0
\\
$\bar{\Psi}i\gamma_{5}\Psi$    & $+$  & $+$ & $-$ & $-$ & \textcolor[rgb]{1.00,0.00,0.00}{$\bullet$} & 0
\\
$\bar{\Psi}i\gamma_{l}\gamma_{k}\Psi$    & $-$  & $+$ & $+$ & $-$ & \textcolor[rgb]{0.00,1.00,0.00}{$\bullet$} & 1
\\
$\bar{\Psi}i\gamma_{1}\gamma_{2}\Psi$    & $+$  & $-$ & $-$ & $-$ & \textcolor[rgb]{0.00,0.00,1.00}{$\bullet$} & 0
\\
$\bar{\Psi}i\gamma_{5}\gamma_{j}\Psi$    & $-$  & $+$ & $-$ & $+$ & $\checkmark$ & 1
\\
$\bar{\Psi}i\gamma_{5}\gamma_{3}\Psi$    & $+$  & $-$ & $+$ & $+$ & $\checkmark$ & 0
\\
$\bar{\Psi}i\gamma_{j}\Psi$    & $-$  & $-$ & $-$ & $+$ & $\checkmark$ & 1
\\
$\bar{\Psi}i\gamma_{3}\Psi$    & $+$  & $+$ & $+$ & $+$ & $\checkmark$ & 0
\\
\bottomrule
\end{tabular}
\end{ruledtabular}
\end{center}
\end{table*}

For 3D semi-DSM, the interacting Lagrangian density is described by
\begin{eqnarray}
\mathcal{L}_{\mathrm{int}}&=&g_{1}\left(\bar{\Psi}\gamma_{0}\Psi\right)^{2}
+g_{2}\left(\bar{\Psi}\Psi\right)^{2}+g_{3\bot}\sum_{j=1}^{2}\left(\bar{\Psi}\gamma_{0}\gamma_{j}\Psi\right)^{2}
+g_{3z}\left(\bar{\Psi}\gamma_{0}\gamma_{3}\Psi\right)^{2}
+g_{4}\left(\bar{\Psi}\gamma_{0}\gamma_{5}\Psi\right)^{2}+g_{5}\left(\bar{\Psi}i\gamma_{5}\Psi\right)^{2}\nonumber
\\
&&
+g_{6\bot}\sum_{<<lk>>}\left(\bar{\Psi}i\gamma_{l}\gamma_{k}\Psi\right)^{2}
+g_{6z}\left(\bar{\Psi}i\gamma_{1}\gamma_{2}\Psi\right)^{2}
+g_{7\bot}\sum_{j=1}^{2}\left(\bar{\Psi}i\gamma_{5}\gamma_{j}\Psi\right)^{2}
+g_{7z}\left(\bar{\Psi}i\gamma_{5}\gamma_{3}\Psi\right)^{2}\nonumber
\\
&&+g_{8\bot}\sum_{j=1}^{2}\left(\bar{\Psi}i\gamma_{j}\Psi\right)^{2}
+g_{8z}\left(\bar{\Psi}i\gamma_{3}\Psi\right)^{2},\label{Eq:FourFermionAll3DsemiDSM}
\end{eqnarray}
where
\begin{eqnarray}
\sum_{<<lk>>}\left(\bar{\Psi}i\gamma_{l}\gamma_{k}\Psi\right)^{2}=\left[\left(\bar{\Psi}i\gamma_{2}\gamma_{3}\Psi\right)^{2}
+\left(\bar{\Psi}i\gamma_{3}\gamma_{1}\Psi\right)^{2}\right].
\end{eqnarray}
As shown in Eq.~(\ref{Eq:FourFermionAll3DDSM}), there are 8 four-fermion couplings for 3D DSM. However, we consider 12 kinds of
four-fermion couplings as shown in Eq.~(\ref{Eq:FourFermionAll3DsemiDSM}) for 3D semi-DSM, due to the anisotropy of the fermion dispersion.

After careful derivation, as shown in Table~\ref{Table:FermionBilinearCPTTransformations}, we obtain the properties of each fermion biliners under the parity ($\mathcal{P}$), time-reversal ($\mathcal{T}$),
charge conjugation ($\mathcal{C}$), and $Z_{2}$ chiral, $U(1)$ chiral, and $O(2)$ rotation transformations. We reduce 136 possible four-fermion interactions
$(\bar{\Psi}M \Psi) (\bar{\Psi} N \Psi)$ by imposing discrete transformations including $\mathcal{P}$, $\mathcal{T}$, $\mathcal{C}$, and $Z_{2}$ chiral symmetries.
It is easy to find that both of $(\bar{\Psi}M \Psi)$ and $(\bar{\Psi} N \Psi)$ should be either even or odd under $\mathcal{P}$, $\mathcal{T}$,
$C$, and $Z_{2}$ transformations, such that the four-fermion interaction is invariant under all these four individual discrete symmetries.
We can find that there are no two identical rows under these four symmetry transformations in Table~\ref{Table:FermionBilinearCPTTransformations}. Therefore,
there exists no interaction term $(\bar{\Psi}M \Psi) (\bar{\Psi} N \Psi)$ with $M\neq N$ that mixes any two different fermion bilinears.

Substituting each four-fermion coupling in Eq.~(\ref{Eq:FourFermionAll3DsemiDSM}) into Eq.~(\ref{Eq:FierzIdentityGeneralLocal}),
we could get 12 equations, which can be compactly expressed by
\begin{eqnarray}
FX=0,
\end{eqnarray}
where
\begin{eqnarray}
X=\left(\begin{array}{c}
\left(\bar{\Psi}\gamma_{0}\Psi\right)^{2}
\\
\left(\bar{\Psi}\Psi\right)^{2}
\\
\sum\limits_{j=1}^{2}\left(\bar{\Psi}\gamma_{0}\gamma_{j}\Psi\right)^{2}
\\
\left(\bar{\Psi}\gamma_{0}\gamma_{3}\Psi\right)^{2}
\\
\left(\bar{\Psi}\gamma_{0}\gamma_{5}\Psi\right)^{2}
\\
\left(\bar{\Psi}i\gamma_{5}\Psi\right)^{2}
\\
\sum\limits_{<<lk>>}\left(\bar{\Psi}i\gamma_{l}\gamma_{k}\Psi\right)^{2}
\\
\left(\bar{\Psi}i\gamma_{1}\gamma_{2}\Psi\right)^{2}
\\
\sum\limits_{j=1}^{2}\left(\bar{\Psi}i\gamma_{5}\gamma_{j}\Psi\right)^{2}
\\
\left(\bar{\Psi}i\gamma_{5}\gamma_{3}\Psi\right)^{2}
\\
\sum\limits_{j=1}^{2}\left(\bar{\Psi}i\gamma_{j}\Psi\right)^{2}
\\
\left(\bar{\Psi}i\gamma_{3}\Psi\right)^{2}
\end{array}\right),
\end{eqnarray}
and
\begin{eqnarray}
F=\left(
\begin{array}{rrrrrrrrrrrr}
5 & 1 & 1 & 1 & 1 & 1 & 1 & 1 & 1 & 1 & 1 & 1
\\
1 & 5 & -1 & -1 & -1 & -1 & 1 & 1 & 1 & 1 & -1 & -1
\\
1 & -1 & 2 & -1 & -1 & 1 & 0 & 1 & 0 & -1 & 0 & 1
\\
1 & -1 & -1 & 5 & -1 & 1 & 1 & -1 & -1 & 1 & 1 & -1
\\
1 & -1 & -1 & -1 & 5 & -1 & -1 & -1 & 1 & 1 & 1 & 1
\\
1 & -1 & 1 & 1 & -1 & 5 & -1 & -1 & 1 & 1 & -1 & -1
\\
1 & 1 & 0 & 1 & -1 & -1 & 2 & -1 & 0 & -1 & 0 & 1
\\
1 & 1 & 1 & -1 & -1 & -1 & -1 & 5 & -1 & 1 & 1 & -1
\\
1 & 1 & 0 & -1 & 1 & 1 & 0 & -1 & 2 & -1 & 0 & -1
\\
1 & 1 & -1 & 1 & 1 & 1 & -1 & 1 & -1 & 5 & -1 & 1
\\
1 & -1 & 0 & 1 & 1 & -1 & 0 & 1 & 0 & -1 & 2 & -1
\\
1 & -1 & 1 & -1 & 1 & -1 & 1 & -1 & -1 & 1 & -1 & 5
\end{array}\right).
\end{eqnarray}

It is easy to find that
\begin{eqnarray}
\mathrm{Rank}(F)=7.
\end{eqnarray}
Then the number of linearly independent couplings is
\begin{eqnarray}
12-\mathrm{Rank}(F)=5.
\end{eqnarray}
For convenience, we take the five couplings
\begin{eqnarray}
\left(\bar{\Psi}\gamma_{0}\Psi\right)^{2},\qquad\left(\bar{\Psi}\Psi\right)^{2},\qquad
\left(\bar{\Psi}\gamma_{0}\gamma_{5}\Psi\right)^{2},\qquad\left(\bar{\Psi}i\gamma_{5}\Psi\right)^{2},
\qquad\left(\bar{\Psi}\gamma_{0}\gamma_{3}\Psi\right)^{2}, \label{Eq:LinIndep3DSemiDSM}
\end{eqnarray}
as linearly independent couplings. The other couplings
\begin{eqnarray}
&&\sum\limits_{j=1}^{2}\left(\bar{\Psi}\gamma_{0}\gamma_{j}\Psi\right)^{2},\qquad
\sum\limits_{<<lk>>}\left(\bar{\Psi}i\gamma_{l}\gamma_{k}\Psi\right)^{2},\qquad
\left(\bar{\Psi}i\gamma_{1}\gamma_{2}\Psi\right)^{2},\qquad
\sum\limits_{j=1}^{2}\left(\bar{\Psi}i\gamma_{5}\gamma_{j}\Psi\right)^{2},\qquad
\left(\bar{\Psi}i\gamma_{5}\gamma_{3}\Psi\right)^{2},\nonumber
\\
&&
\sum\limits_{j=1}^{2}\left(\bar{\Psi}i\gamma_{j}\Psi\right)^{2},\qquad
\left(\bar{\Psi}i\gamma_{3}\Psi\right)^{2},
\end{eqnarray}
can be expressed by the five independent couplings shown in Eq.~(\ref{Eq:LinIndep3DSemiDSM}).
In order to obtain the concrete expressions for other couplings, we define
\begin{eqnarray}
\tilde{X}=\left(\begin{array}{c}
\sum\limits_{j=1}^{2}\left(\bar{\Psi}\gamma_{0}\gamma_{j}\Psi\right)^{2}
\\
\sum\limits_{<<lk>>}\left(\bar{\Psi}i\gamma_{l}\gamma_{k}\Psi\right)^{2}
\\
\left(\bar{\Psi}i\gamma_{1}\gamma_{2}\Psi\right)^{2}
\\
\sum\limits_{j=1}^{2}\left(\bar{\Psi}i\gamma_{5}\gamma_{j}\Psi\right)^{2}
\\
\left(\bar{\Psi}i\gamma_{5}\gamma_{3}\Psi\right)^{2}
\\
\sum\limits_{j=1}^{2}\left(\bar{\Psi}i\gamma_{j}\Psi\right)^{2}
\\
\left(\bar{\Psi}i\gamma_{3}\Psi\right)^{2}
\\
\left(\bar{\Psi}\gamma_{0}\Psi\right)^{2}
\\
\left(\bar{\Psi}\Psi\right)^{2}
\\
\left(\bar{\Psi}\gamma_{0}\gamma_{5}\Psi\right)^{2}
\\
\left(\bar{\Psi}i\gamma_{5}\Psi\right)^{2}
\\
\left(\bar{\Psi}\gamma_{0}\gamma_{3}\Psi\right)^{2}
\end{array}\right).
\end{eqnarray}
It is easy to get that
\begin{eqnarray}
\tilde{F}\tilde{X}=0,
\end{eqnarray}
where
\begin{eqnarray}
\tilde{F}&=&\left(
\begin{array}{rrrrrrrrrrrr}
1 & 1 & 1 & 1 & 1 & 1 & 1 & 5 & 1 & 1 & 1 & 1
\\
-1 & 1 & 1 & 1 & 1 & -1 & -1 & 1 & 5 & -1 & -1 & -1
\\
2 & 0 & 1 & 0 & -1 & 0 & 1 & 1 & -1 & -1 & 1 & -1
\\
-1 & 1 & -1 & -1 & 1 & 1 & -1 & 1 & -1 & -1 & 1 & 5
\\
-1 & -1 & -1 & 1 & 1 & 1 & 1 & 1 & -1 & 5 & -1 & -1
\\
1 & -1 & -1 & 1 & 1 & -1 & -1 & 1 & -1 & -1 & 5 & 1
\\
0 & 2 & -1 & 0 & -1 & 0 & 1 & 1 & 1 & -1 & -1 & 1
\\
1 & -1 & 5 & -1 & 1 & 1 & -1 & 1 & 1 & -1 & -1 & -1
\\
0 & 0 & -1 & 2 & -1 & 0 & -1 & 1 & 1 & 1 & 1 & -1
\\
-1 & -1 & 1 & -1 & 5 & -1 & 1 & 1 & 1 & 1 & 1 & 1
\\
0 & 0 & 1 & 0 & -1 & 2 & -1 & 1 & -1 & 1 & -1 & 1
\\
1 & 1 & -1 & -1 & 1 & -1 & 5 & 1 & -1 & 1 & -1 & -1
\end{array}\right).
\end{eqnarray}
Carrying out a series of similarity transformations for $\tilde{F}$,
\begin{eqnarray}
\tilde{F}\rightarrow\tilde{F}',
\end{eqnarray}
we arrive
\begin{eqnarray}
\tilde{F}'\tilde{X}=0, \label{Eq:EchelonMatrix3DsemiDSM}
\end{eqnarray}
where
\begin{eqnarray}
\tilde{F}'&=&\left(
\begin{array}{rrrrrrrrrrrr}
1 & 0 & 0 & 0 & 0 & 0 & 0 & 1 & -1 & -1 & 2 & 1
\\
0 & 1 & 0 & 0 & 0 & 0 & 0 & 1 & 1 & -1 & 0 & 1
\\
0 & 0 & 1 & 0 & 0 & 0 & 0 & 0 & 1 & 0 & -1 & -1
\\
0 & 0 & 0 & 1 & 0 & 0 & 0 & 1 & 1 & 1 & 0 & -1
\\
0 & 0 & 0 & 0 & 1 & 0 & 0 & 1 & 0 & 0 & 1 & 1
\\
0 & 0 & 0 & 0 & 0 & 1 & 0 & 1 & -1 & 1 & 0 & 1
\\
0 & 0 & 0 & 0 & 0 & 0 & 1 & 0 & 0 & 1 & -1 & -1
\\
0 & 0 & 0 & 0 & 0 & 0 & 0 & 0 & 0 & 0 & 0 & 0
\\
0 & 0 & 0 & 0 & 0 & 0 & 0 & 0 & 0 & 0 & 0 & 0
\\
0 & 0 & 0 & 0 & 0 & 0 & 0 & 0 & 0 & 0 & 0 & 0
\\
0 & 0 & 0 & 0 & 0 & 0 & 0 & 0 & 0 & 0 & 0 & 0
\\
0 & 0 & 0 & 0 & 0 & 0 & 0 & 0 & 0 & 0 & 0 & 0
\end{array}\right).
\end{eqnarray}
Eq.~(\ref{Eq:EchelonMatrix3DsemiDSM}) can be also written as
\begin{eqnarray}
\sum_{j=1}^{2}\left(\bar{\Psi}\gamma_{0}\gamma_{j}\Psi\right)^{2}
&=&-\left(\bar{\Psi}\gamma_{0}\Psi\right)^{2}
+\left(\bar{\Psi}\Psi\right)^{2}
+\left(\bar{\Psi}\gamma_{0}\gamma_{5}\Psi\right)^{2}
-2\left(\bar{\Psi}i\gamma_{5}\Psi\right)^{2}
-\left(\bar{\Psi}\gamma_{0}\gamma_{3}\Psi\right)^{2}, \label{Eq:semiDSMFFCouplingRelation1}
\\
\sum\limits_{<<lk>>}\left(\bar{\Psi}i\gamma_{l}\gamma_{k}\Psi\right)^{2}
&=&-\left(\bar{\Psi}\gamma_{0}\Psi\right)^{2}
-\left(\bar{\Psi}\Psi\right)^{2}
+\left(\bar{\Psi}\gamma_{0}\gamma_{5}\Psi\right)^{2}
-\left(\bar{\Psi}\gamma_{0}\gamma_{3}\Psi\right)^{2},\label{Eq:semiDSMFFCouplingRelation2}
\\
\left(\bar{\Psi}i\gamma_{1}\gamma_{2}\Psi\right)^{2}
&=&-\left(\bar{\Psi}\Psi\right)^{2}
+\left(\bar{\Psi}i\gamma_{5}\Psi\right)^{2}
+\left(\bar{\Psi}\gamma_{0}\gamma_{3}\Psi\right)^{2}, \label{Eq:semiDSMFFCouplingRelation3}
\\
\sum\limits_{j=1}^{2}\left(\bar{\Psi}i\gamma_{5}\gamma_{j}\Psi\right)^{2}
&=&-\left(\bar{\Psi}\gamma_{0}\Psi\right)^{2}
-\left(\bar{\Psi}\Psi\right)^{2}
-\left(\bar{\Psi}\gamma_{0}\gamma_{5}\Psi\right)^{2}
+\left(\bar{\Psi}\gamma_{0}\gamma_{3}\Psi\right)^{2}, \label{Eq:semiDSMFFCouplingRelation4}
\\
\left(\bar{\Psi}i\gamma_{5}\gamma_{3}\Psi\right)^{2}
&=&-\left(\bar{\Psi}\gamma_{0}\Psi\right)^{2}
-\left(\bar{\Psi}i\gamma_{5}\Psi\right)^{2}
-\left(\bar{\Psi}\gamma_{0}\gamma_{3}\Psi\right)^{2}, \label{Eq:semiDSMFFCouplingRelation5}
\\
\sum\limits_{j=1}^{2}\left(\bar{\Psi}i\gamma_{j}\Psi\right)^{2}
&=&-\left(\bar{\Psi}\gamma_{0}\Psi\right)^{2}
+\left(\bar{\Psi}\Psi\right)^{2}
-\left(\bar{\Psi}\gamma_{0}\gamma_{5}\Psi\right)^{2}
-\left(\bar{\Psi}\gamma_{0}\gamma_{3}\Psi\right)^{2}, \label{Eq:semiDSMFFCouplingRelation6}
\\
\left(\bar{\Psi}i\gamma_{3}\Psi\right)^{2}
&=&-\left(\bar{\Psi}\gamma_{0}\gamma_{5}\Psi\right)^{2}
+\left(\bar{\Psi}i\gamma_{5}\Psi\right)^{2}
+\left(\bar{\Psi}\gamma_{0}\gamma_{3}\Psi\right)^{2}. \label{Eq:semiDSMFFCouplingRelation7}
\end{eqnarray}

\end{widetext}

\section{Mean-filed analysis \label{App:MeanFieldAnalysis}}

Here taking the short-range four-fermion interaction $g_{2}\left(\bar{\Psi}\Psi\right)^{2}$ as an example, we
give the details of the derivation and calculation for the mean-field analysis. Analysis for other kinds of
four-fermion interactions could be carried out similarly.

\subsection{Derivation of the self-consistent equation \label{Subsec:DerivationSelfConsistentEq}}

Under the influence of short-range four-fermion interaction $g_{2}\left(\bar{\Psi}\Psi\right)^{2}$,  the expectation value
\begin{eqnarray}
\Delta_{2}=\left<\bar{\Psi}\Psi\right>,
\end{eqnarray}
could become finite. Considering the order parameter $\Delta_{2}$, the fermion propagator can be written as
\begin{eqnarray}
G(i\omega,\mathbf{k},\Delta_{2})&=&\left[i\omega\gamma_{0}+iv\left(k_{1}\gamma_{1}+k_{2}\gamma_{2}\right)
+iAk_{3}^{2}\gamma_{3}\right.\nonumber
\\
&&\left.+\Delta_{2}\right]^{-1}.
\end{eqnarray}
For finite temperature, we employ the propagator in Matsubara formalism as following
\begin{eqnarray}
G(i\omega_{n},\mathbf{k},\Delta_{2})&=&\left[i\omega_{n}\gamma_{0}+iv\left(k_{1}\gamma_{1}+k_{2}\gamma_{2}\right)
+iAk_{3}^{2}\gamma_{3}\right.\nonumber
\\
&&\left.+\Delta_{2}\right]^{-1},
\end{eqnarray}
where $\omega_{n}=(2n+1)\pi T$ with $n$ being integers and $T$ the temperature.

The partition functions is given by
\begin{eqnarray}
Z&=&\int\mathcal{D}\bar{\Psi}\mathcal{D}\Psi e^{S}\nonumber
\\
&=&\prod_{\omega_{n}}\prod_{\mathbf{k}}\int\mathcal{D}\bar{\Psi}\mathcal{D}\Psi e^{ \bar{\Psi}_{\omega_{n},\mathbf{k}}
\beta G^{-1}(i\omega_{n},\mathbf{k},\Delta_{2})\Psi_{\omega_{n},\mathbf{k}}}\nonumber
\\
&&\times e^{-\int d\tau \int d^{3}\mathbf{x}\frac{\Delta_{2}^{2}}{2g_{2}}},
\end{eqnarray}
where $\beta=\frac{1}{T}$.
Using the functional integral formula
\begin{eqnarray}
\int\mathcal{D}\bar{\eta}\mathcal{D}\eta e^{\bar{\eta} K\eta}=\det{K},
\end{eqnarray}
we get
\begin{eqnarray}
Z&=&\prod_{\omega_{n}}\prod_{\mathbf{k}}\beta^{4}\det \left[G^{-1}(i\omega_{n},\mathbf{k},\Delta_{2})\right]\nonumber
\\
&&\times e^{-\int d\tau \int d^{3}\mathbf{x}\frac{\Delta_{2}^{2}}{2g_{2}}},
\end{eqnarray}
which leads to
\begin{eqnarray}
\ln Z
&=&\sum_{\omega_{n}}\sum_{\mathbf{k}}\ln\left\{\beta^{4}\det \left[G^{-1}(i\omega_{n},\mathbf{k},\Delta_{2})\right]\right\}\nonumber
\\
&&-\int d\tau \int d^{3}\mathbf{x}\frac{\Delta_{2}^{2}}{2g_{2}}.
\end{eqnarray}
It is easy to obtain
\begin{eqnarray}
\mathrm{det}\left[G^{-1}(i\omega_{n},\mathbf{k},\Delta_{2})\right]
&=&\left(\omega_{n}^{2}+E_{\mathbf{k},\Delta_{2}}^{2}\right)^{2},
\end{eqnarray}
where
\begin{eqnarray}
E_{\mathbf{k},\Delta_{2}}&=&\sqrt{v^{2}k_{\bot}^{2}+A^{2}k_{3}^{4}+\Delta_{2}^{2}}.
\end{eqnarray}
Thus, we arrive
\begin{eqnarray}
\ln Z
&=&\sum_{\omega_{n}}\sum_{\mathbf{k}}\ln\left[\beta^{4}\left(\omega_{n}^{2}+E_{\mathbf{k},\Delta_{2}}^{2}\right)^{2}\right]\nonumber
\\
&&-\int d\tau \int d^{3}\mathbf{x}\frac{\Delta_{2}^{2}}{2g_{2}}.
\end{eqnarray}
Carrying out the summarization of frequency, we get
\begin{eqnarray}
\ln Z
&=&4\sum_{\mathbf{k}}\ln\left[2\cosh\left(\frac{E_{\mathbf{k},\Delta_{2}}}{2T}\right)\right]
-\beta\mathcal{V}\frac{\Delta_{2}^{2}}{2g_{2}},
\end{eqnarray}
where $\mathcal{V}$ is volume of sample.

The free energy density $f$ and free energy $F$ are defined as
\begin{eqnarray}
f&=&\frac{F}{\mathcal{V}}=-\frac{1}{\beta}\ln Z\nonumber
\\
&=&-4T\frac{1}{\mathcal{V}}\sum_{\mathbf{k}}\ln\left[2\cosh\left(\frac{E_{\mathbf{k},\Delta_{2}}}{2T}\right)\right]
+\frac{\Delta_{2}^{2}}{2g_{2}}.
\end{eqnarray}
Taking the  continuous limit by using the replacement
\begin{eqnarray}
\frac{1}{\mathcal{V}}\sum_{\mathbf{k}}\rightarrow\int\frac{d^3\mathbf{k}}{(2\pi)^{3}},
\end{eqnarray}
we obtain
\begin{eqnarray}
f&=&-4T\int\frac{d^{3}\mathbf{k}}{(2\pi)^{3}}\ln\left[2\cosh\left(\frac{E_{\mathbf{k},\Delta_{2}}}{2T}\right)\right]
+\frac{\Delta_{2}^{2}}{2g_{2}}.
\end{eqnarray}

The self-consistent  equation for $\Delta_{2}$ is  determined by
\begin{eqnarray}
\frac{\partial f}{\partial \Delta_{2}}&=&0.
\end{eqnarray}
Concretely, the self-consistent equation is given by
\begin{eqnarray}
1&=&2g_{2}\int\frac{d^{3}\mathbf{k}}{(2\pi)^{3}}\tanh\left(\frac{E_{\mathbf{k},\Delta_{2}}}{2T}\right)
\frac{1}{E_{\mathbf{k},\Delta_{2}}}.
\end{eqnarray}
At zero temperature, the equation becomes
\begin{eqnarray}
1&=&2g_{2}\int\frac{d^{3}\mathbf{k}}{(2\pi)^{3}}
\frac{1}{E_{\mathbf{k},\Delta_{2}}}.
\end{eqnarray}

\subsection{Solving the self-consistent equation \label{Subsec:SolveSelfConsistentEq}}

\subsubsection{Zero temperature}

At zero temperature, the self-consistent equation  can be written as
\begin{eqnarray}
1&=&2g_{2}\int\frac{d^3\mathbf{k}}{(2\pi)^{3}}
\frac{1}{\sqrt{v^2k_{\bot}^{2}+A^{2}k_{3}^{4}+\Delta_{2}^{2}}}\nonumber
\\
&=&\frac{g_{2}}{\pi^{2}}\int dk_{\bot}d|k_{3}|k_{\bot}
\frac{1}{\sqrt{v^2k_{\bot}^{2}+A^{2}k_{3}^{4}+\Delta_{2}^{2}}}.
\end{eqnarray}

We employing the transformations
\begin{eqnarray}
E=\sqrt{v^2q_{\bot}^{2}+A^{2}q_{3}^{4}},\qquad
\delta =\frac{Aq_{3}^{2}}{vq_{\bot}}, \label{Eq:TransformationA}
\end{eqnarray}
which are equivalent to
\begin{eqnarray}
q_{\bot}=\frac{E}{v\sqrt{1+\delta^2}},\qquad
|q_{3}|=\frac{\sqrt{\delta}\sqrt{E}}{\sqrt{A}\left(1+\delta^2\right)^{\frac{1}{4}}}. \label{Eq:TransformationB}
\end{eqnarray}
The integration measures satisfy the relation
\begin{eqnarray}
dq_{\bot}d|q_{3}|
&=&\frac{\sqrt{E}}{2v\sqrt{A}\sqrt{\delta}\left(1+\delta^2\right)^{\frac{3}{4}}}dEd\delta. \label{Eq:TransformationC}
\end{eqnarray}

Utilizing the transformations Eqs.~(\ref{Eq:TransformationA})-(\ref{Eq:TransformationC}), we obtain
\begin{eqnarray}
1&=&\frac{g_{2}}{2\pi^{2}v^{2}\sqrt{A}}\int_{0}^{\Lambda} dE \frac{E^{\frac{3}{2}}}
{\sqrt{E^{2}+\Delta_{2}^{2}}}\int_{0}^{+\infty}d\delta\frac{1}{\sqrt{\delta}\left(1+\delta^2\right)^{\frac{5}{4}}}\nonumber
\\
&=&\frac{g_{2}}{\pi^{2}v^{2}\sqrt{A}}\int_{0}^{\Lambda} dE \frac{E^{\frac{3}{2}}}
{\sqrt{E^{2}+\Delta_{2}^{2}}}.
\end{eqnarray}
It can be further written as
\begin{eqnarray}
1&=&\frac{g_{2}\Lambda^{\frac{3}{2}}}{\pi^{2}v^{2}\sqrt{A}}\Bigg[\int_{0}^{1} dx \bigg(\frac{x^{\frac{3}{2}}}
{\sqrt{x^{2}+\left(\frac{\Delta_{2}}{\Lambda}\right)^{2}}}-x^{\frac{1}{2}}\bigg)\nonumber
\\
&&+\frac{2}{3}\Bigg].
\end{eqnarray}
Taking $\Delta_{2}=0$, we can get the critical coupling strength $g_{2c}$, which satisfies
\begin{eqnarray}
g_{2c}&=&\frac{3\pi^{2}v^{2}\sqrt{A}}{2\Lambda^{\frac{3}{2}}}.
\end{eqnarray}
In the limit $\Delta_{2}\ll\Lambda$, we have
\begin{eqnarray}
1&\approx&\frac{2g_{2}\Lambda^{\frac{3}{2}}}{3\pi^{2}v^{2}\sqrt{A}}\left[1-\frac{\Gamma\left(\frac{1}{4}\right)\Gamma\left(\frac{5}{4}\right)}{\sqrt{\pi}}
\left(\frac{\Delta_{2}}{\Lambda}\right)^{\frac{3}{2}}\right]\nonumber
\\
&=&\frac{g_{2}}{g_{2c}}\left[1-\frac{\Gamma\left(\frac{1}{4}\right)\Gamma\left(\frac{5}{4}\right)}{\sqrt{\pi}}\left(\frac{\Delta_{2}}{\Lambda}\right)^{\frac{3}{2}}\right].
\end{eqnarray}
Thus, $\Delta_{2}$ is given by
\begin{eqnarray}
\Delta_{2}&\approx&c_{1}\Lambda
\frac{\left(g_{2}-g_{2c}\right)^{\frac{2}{3}}}{g_{2}^{\frac{2}{3}}},
\end{eqnarray}
where
\begin{eqnarray}
c_{1}&=&\left(\frac{\sqrt{\pi}}{\Gamma\left(\frac{1}{4}\right)\Gamma\left(\frac{5}{4}\right)}\right)^{\frac{2}{3}}\approx0.662596.
\end{eqnarray}

\subsubsection{Finite temperature}

At finite temperature, the self-consistent equation can be written as
\begin{eqnarray}
\frac{1}{g_{2}}
&=&\frac{1}{\pi^{2}}\int dk_{\bot}d|k_{3}|k_{\bot}\frac{1}{\sqrt{v^{2}k_{\bot}^{2}
+A^{2}k_{3}^{4}+\Delta_{2}^{2}}}\nonumber
\\
&&\times\tanh\left( \frac{\sqrt{v^{2}k_{\bot}^{2}
+A^{2}k_{3}^{4}+\Delta_{2}^{2}}}{2T}\right).
\end{eqnarray}
Employing the transformations Eqs.~(\ref{Eq:TransformationA})-(\ref{Eq:TransformationC}) and carrying out the
integration of $\delta$, we obtain
\begin{eqnarray}
\frac{1}{g_{2}}
&=&\frac{1}{\pi^{2}v^{2}\sqrt{A}}\int_{0}^{\Lambda} dE
\frac{E^{\frac{3}{2}}}{\sqrt{E^2+\Delta_{2}^{2}}}\nonumber
\\
&&\times\tanh\left( \frac{\sqrt{E^{2}+\Delta_{2}^{2}}}{2T}\right).
\end{eqnarray}

\begin{table*}[htbp]
\caption{ Energy dispersions of fermion excitations considering different order parameters.
\label{Table:EnergyDispersionOrderParameter}}
\begin{center}
\begin{ruledtabular}
\begin{tabular}{lll}
\toprule     Order Parameter        & Expectation Value&  Energy Dispersion
\\ \hline
\midrule $\Delta_{1}$       & $\left<\bar{\Psi}\gamma_{0}\Psi\right>$ & $E_{\mathbf{k},\Delta_{1}}^{\pm}=\sqrt{v^{2}k_{\bot}^{2}+A^{2}k_{3}^{4}}\pm\Delta_{1}$
\\
\midrule $\Delta_{2}$       & $\left<\bar{\Psi}\Psi\right>$ & $E_{\mathbf{k},\Delta_{2}}=\sqrt{v^{2}k_{\bot}^{2}+A^{2}k_{3}^{4}+\Delta_{2}^{2}}$
\\
\midrule $\Delta_{3\bot}$       & $\sum_{j=1,2}\left<\bar{\Psi}\gamma_{0}\gamma_{j}\Psi\right>$ & $E_{\mathbf{k}\Delta_{3\bot}}^{\pm}=\sqrt{\frac{1}{2}v^{2}\left(k_{1}+k_{2}\right)^{2}+2\left[\Delta_{3\bot}\pm\frac{1}{2}\sqrt{v^{2}\left(k_{2}-k_{1}\right)^{2}
+2A^{2}k_{3}^{4}}\right]^{2}}$
\\
\midrule $\Delta_{3z}$       & $\left<\bar{\Psi}\gamma_{0}\gamma_{3}\Psi\right>$ & $E_{\mathbf{k},\Delta_{3z}}^{\pm}=\sqrt{\left(vk_{\bot}\pm\Delta_{3z}\right)^{2}+A^{2}k_{3}^{4}}$
\\
\midrule $\Delta_{4}$       & $\left<\bar{\Psi}\gamma_{0}\gamma_{5}\Psi\right>$  & $E_{\mathbf{k},\Delta_{4}}^{\pm}=\sqrt{v^{2}k_{\bot}^{2}+A^{2}k_{3}^{4}}\pm\Delta_{4}$
\\
\midrule $\Delta_{5}$       & $\left<\bar{\Psi}i\gamma_{5}\Psi\right>$ & $E_{\mathbf{k},\Delta_{5}}=\sqrt{v^{2}k_{\bot}^{2}+A^{2}k_{3}^{4}+\Delta_{5}^{2}}$
\\
\midrule $\Delta_{6\bot}$       & $\left<\bar{\Psi}\left(i\gamma_{2}\gamma_{3}+i\gamma_{3}\gamma_{1}\right)\Psi\right>$ & $E_{\mathbf{k}\Delta_{6\bot}}^{\pm}
=\sqrt{\frac{1}{2}v^{2}\left(k_{1}+k_{2}\right)^{2}+2\left[\Delta_{6\bot}\pm\frac{1}{2}\sqrt{v^{2}\left(k_{2}-k_{1}\right)^{2}
+2A^{2}k_{3}^{4}}\right]^{2}}$
\\
\midrule $\Delta_{6z}$       & $\left<\bar{\Psi}i\gamma_{1}\gamma_{2}\Psi\right>$ & $E_{\mathbf{k},\Delta_{6z}}^{\pm}=\sqrt{\left(vk_{\bot}\pm\Delta_{6z}\right)^{2}+A^{2}k_{3}^{4}}$
\\
\midrule $\Delta_{7\bot}$       & $\sum_{j=1,2}\left<\bar{\Psi}i\gamma_{5}\gamma_{j}\Psi\right>$ & $E_{\mathbf{k},\Delta_{7\bot}}^{\pm}=\sqrt{\frac{1}{2}\left(k_{1}-k_{2}\right)^{2}+A^{2}k_{3}^{4}+\frac{1}{2}\left(k_{1}+k_{2}\pm2\Delta_{7\bot}\right)^{2}}$
\\
\midrule $\Delta_{7z}$       & $\left<\bar{\Psi}i\gamma_{5}\gamma_{3}\Psi\right>$ & $E_{\mathbf{k},\Delta_{7z}}^{\pm}=\sqrt{v^{2}k_{\bot}^{2}+\left(Ak_{3}^{2}\pm\Delta_{7z}\right)^{2}}$
\\
\midrule $\Delta_{8\bot}$   & $\sum_{j=1,2}\left<\bar{\Psi}i\gamma_{j}\Psi\right>$    & $E_{\mathbf{k},\Delta_{8\bot}}=\sqrt{\left(vk_{1}+\Delta_{8\bot}\right)^{2}+\left(vk_{2}+\Delta_{8\bot}\right)^{2}+A^{2}k_{3}^{4}}$
\\
\midrule $\Delta_{8z}$       & $\left<\bar{\Psi}i\gamma_{3}\Psi\right>$ & $E_{\mathbf{k},\Delta_{8z}}=\sqrt{v^{2}k_{\bot}^{2}+\left(Ak_{3}^{2}+\Delta_{8z}\right)^{2}}$
\\
\bottomrule
\end{tabular}
\end{ruledtabular}
\end{center}
\end{table*}

$T_{c}$ is determined by
\begin{eqnarray}
\frac{1}{g_{2}}
&=&\frac{1}{\pi^{2}v^{2}\sqrt{A}}\int_{0}^{\Lambda} dE\sqrt{E}
\tanh\left( \frac{E}{2T_{c}}\right)\nonumber
\\
&=&\frac{2\sqrt{2}T_{c}^{\frac{3}{2}}}{\pi^{2}v^{2}\sqrt{A}}\left[\frac{2}{3}\left(\frac{\Lambda}{2T_{c}}\right)^{\frac{3}{2}}
\tanh\left( \frac{\Lambda}{2T_{c}}\right)\right.\nonumber
\\
&&\left.-\frac{2}{3}\int_{0}^{\frac{\Lambda}{2T_{c}}} dxx^{\frac{3}{2}}
\frac{1}{\cosh^{2}\left( x\right)}\right].
\end{eqnarray}
If $T_{c}\ll \Lambda$, the equation can be approximated by
\begin{eqnarray}
\frac{1}{g_{2}}
&\approx&\frac{2\sqrt{2}T_{c}^{\frac{3}{2}}}{\pi^{2}v^{2}\sqrt{A}}\left[\frac{2}{3}\left(\frac{\Lambda}{2T_{c}}\right)^{\frac{3}{2}}-\frac{2}{3}\int_{0}^{+\infty} dxx^{\frac{3}{2}}
\frac{1}{\cosh^{2}\left( x\right)}\right]\nonumber
\\
&=&\frac{1}{g_{2c}}-2\sqrt{2}a\frac{1}{g_{2c}}\left(\frac{T_{c}}{\Lambda}\right)^{\frac{3}{2}},
\end{eqnarray}
where
\begin{eqnarray}
a=\int_{0}^{+\infty} dxx^{\frac{3}{2}}
\frac{1}{\cosh^{2}\left( x\right)}\approx0.719227.
\end{eqnarray}
Then $T_{c}$ satisfies
\begin{eqnarray}
T_{c}\approx c_{2}\Lambda\frac{\left(g_{2}-g_{2c}\right)^{\frac{2}{3}}}{g_{2}^{\frac{2}{3}}},
\end{eqnarray}
where $c_{2}=1/\left(2\sqrt{2}a\right)^{\frac{2}{3}}\approx0.622863$.

\subsection{Dispersion of fermions with finite order parameter}

Mean-field analysis for other four-fermion couplings can be performed through similar procedures as subsections~\ref{Subsec:DerivationSelfConsistentEq} and
\ref{Subsec:SolveSelfConsistentEq}. For convenience, we show the fermion dispersions with various finite order parameters in Table~\ref{Table:EnergyDispersionOrderParameter}.

If $\Delta_{1}>0$, the original fermion dispersion $E_{\mathbf{k}}=\sqrt{v^{2}k_{\bot}^{2}+A^{2}k_{3}^{2}}$ becomes two dispersions $E_{\mathbf{k},\Delta_{1}}^{+}$ and $E_{\mathbf{k},\Delta_{1}}^{-1}$.
$E_{\mathbf{k},\Delta_{1}}^{+}$ is gapped. Whereas, $E_{\mathbf{k},\Delta_{1}}^{-}$ is gapless when $\sqrt{v^{2}k_{\bot}^{2}+A^{2}k_{3}^{2}}=\Delta_{1}$. It indicates that the gapless nodal point becomes gapless on a surface.
If $\Delta_{2}>0$, the fermion dispersion becomes to $E_{\mathbf{k},\Delta_{2}}$ which is gapped. If $\Delta_{3\bot}>0$, there are two fermion dispersions $E_{\mathbf{k},\Delta_{3\bot}}^{+}$ and $E_{\mathbf{k},\Delta_{3\bot}}^{-}$. We find that $E_{\mathbf{k},\Delta_{3\bot}}^{+}$ is gapped, but $E_{\mathbf{k},\Delta_{3}\bot}^{-}$ is gapless along a nodal line which is determined by
\begin{eqnarray}
k_{1}+k_{2}&=&0,
\\
\frac{1}{2}\sqrt{v^{2}\left(k_{2}-k_{1}\right)^{2}
+2A^{2}k_{3}^{4}}&=&\Delta_{3\bot}.
\end{eqnarray}
If $\Delta_{3z}>0$, one dispersion $E_{\mathbf{k},\Delta_{3z}}^{+}$ is gapped, but another dispersion $E_{\mathbf{k},\Delta_{3z}}^{-}$ is gapless along a nodal line which is decided by $vk_{\bot}=\Delta_{3z}$
and $k_{3}=0$. If $\Delta_{4}>0$, the dispersion $E_{\mathbf{k},\Delta_{4}}^{+}$ is gapped, whereas $E_{\mathbf{k},\Delta_{4}}^{-}$ is gapless on the surface which satisfies $\sqrt{v^{2}k_{\bot}^{2}+A^{2}k_{3}^{2}}=\Delta_{4}$.
If $\Delta_{5}>0$, the corresponding fermion dispersion $E_{\mathbf{k},\Delta_{5}}$ is gapped. If $\Delta_{6\bot}>0$, the dispersion $E_{\mathbf{k},\Delta_{6\bot}}^{+}$ is gapped, but the dispersion
$E_{\mathbf{k},\Delta_{6\bot}}^{-}$ is gapless along a nodal line which satisfies
\begin{eqnarray}
k_{1}+k_{2}&=&0,
\\
\frac{1}{2}\sqrt{v^{2}\left(k_{2}-k_{1}\right)^{2}
+2A^{2}k_{3}^{4}}&=&\Delta_{6\bot}.
\end{eqnarray}
If $\Delta_{6z}>0$, we can find that one dispersion $E_{\mathbf{k},\Delta_{6z}}^{+}$ is gapped, but another dispersion $E_{\mathbf{k},\Delta_{6z}}^{-}$ is gapless along a nodal line which is determined by $vk_{\bot}=\Delta_{6z}$
and $k_{3}=0$. If $\Delta_{7\bot}>0$, there are two fermions dispersions $E_{\mathbf{k},\Delta_{7\bot}}^{+}$ and $E_{\mathbf{k},\Delta_{7\bot}}^{-}$.  It is easy to verify that $E_{\mathbf{k},\Delta_{7\bot}}^{+}$ is gapless at
the point $(-\Delta_{7\bot},-\Delta_{7\bot},0)$ and $E_{\mathbf{k},\Delta_{7\bot}}^{-}$ is gapless at  $(\Delta_{7\bot},\Delta_{7\bot},0)$. At these two gapless points, the fermion dispersions are still linear within $xy$ plane and quadratic along the $z$ axis. If $\Delta_{7z}>0$, we can find that one dispersion $E_{\mathbf{k},\Delta_{7z}}^{+}$ is gapped, but another dispersion $E_{\mathbf{k},\Delta_{7z}}^{-}$ is gapless at two points
\begin{eqnarray}
\mathbf{k}_{a}&=&(0,0,\sqrt{\frac{\Delta_{7z}}{A}}),\quad
\mathbf{k}_{b}=(0,0,-\sqrt{\frac{\Delta_{7z}}{A}}).
\end{eqnarray}
At these two gapless points, the fermions dispersion can be written as
\begin{eqnarray}
E_{\mathbf{K},\Delta_{7z}}=\sqrt{v^2K_{\bot}^{2}+v_{z}^{2}K_{z}^{2}},
\end{eqnarray}
with $v_{z}=2\sqrt{A\Delta_{7z}}$ and $\mathbf{K}$ being the momentum relative to the point $\mathbf{k}_{a}$ or $\mathbf{k}_{b}$. It is clear that this fermion dispersion is linear within $xy$ plane and also linear along $z$ axis.
If $\Delta_{8\bot}>0$, the fermions dispersion $E_{\mathbf{k},\Delta_{8\bot}}$ is gapless at the point $(-\Delta_{8\bot}/v,-\Delta_{8\bot}/v,0)$. If $\Delta_{8z}>0$, the fermion dispersion $E_{\mathbf{k},\Delta_{8z}}>0$ is gapped.

\section{Derivation of the RG equations for the  strength of four-fermion couplings \label{App:DerivationRGEqs}}

\subsection{Self-energy of the fermions}

\begin{figure}[htbp]
\center
\includegraphics[width=3.1in]{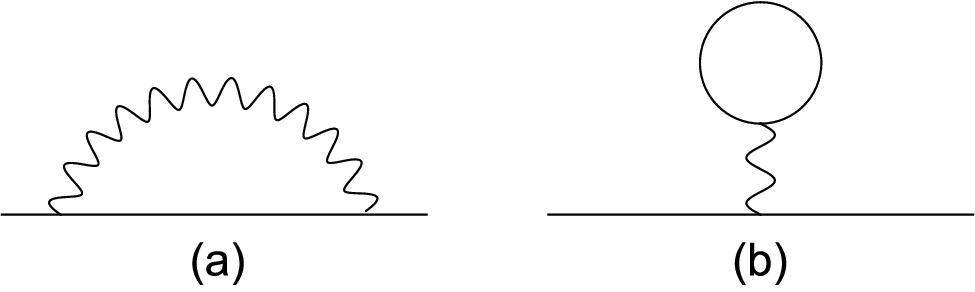}
\caption{Feynman diagrams for the self-energies  of fermions induced by four-fermion interactions. Solid
line represents the fermion propagator, and wavy line stands for the four-fermion interaction.\label{Fig:FermionSelfEnergy}}
\end{figure}

The fermion propagator reads as
\begin{eqnarray}
G_{0}(i\omega,\mathbf{k})&=&-\frac{i\omega\gamma_{0}+iv\left(k_{1}\gamma_{1}+k_{2}\gamma_{2}\right)
+iAk_{3}^{2}\gamma_{3}}{\omega^2+E_{\mathbf{k}}^{2}}, \label{Eq:FermionPropagator}
\end{eqnarray}
where $E_{\mathbf{k}}=\sqrt{v^2k_{\bot}^{2}+A^{2}k_{3}^{4}}$ with $k_{\bot}^{2}=k_{1}^{2}+k_{2}^{2}$.
The self-energy of fermions resulting from Fig.~\ref{Fig:FermionSelfEnergy}(a) takes the form
\begin{eqnarray}
\Sigma_{1}=\sum_{a}g_{a}\int_{-\infty}^{+\infty}\frac{d\omega}{2\pi}
\int'\frac{d^3\mathbf{k}}{(2\pi)^{3}}\Gamma_{a}G_{0}(\omega,\mathbf{k})\Gamma_{a}, \label{Eq:SelfEnergyA}
\end{eqnarray}
where
\begin{eqnarray}
\sum_{a}\equiv\sum_{a=1,2,4,5,3z}.
\end{eqnarray}
 $\int'$ represents that a momentum shell will be properly taken.
Figure~\ref{Fig:FermionSelfEnergy}(b) induces the self-energy of fermions as following
\begin{eqnarray}
\Sigma_{2}=\sum_{a}g_{a}\int_{-\infty}^{+\infty}\frac{d\omega}{2\pi}
\int'\frac{d^3\mathbf{k}}{(2\pi)^{3}}\mathrm{Tr}\left[G_{0}(\omega,\mathbf{k})\Gamma_{a}\right]. \label{Eq:SelfEnergyB}
\end{eqnarray}
Substituting Eq.~(\ref{Eq:FermionPropagator}) into Eqs.~(\ref{Eq:SelfEnergyA}) and
(\ref{Eq:SelfEnergyB}), we obtain
\begin{eqnarray}
\Sigma_{1}&=&0, \label{Eq:Sigma1Result}
\\
\Sigma_{2}&=&0. \label{Eq:Sigma2Result}
\end{eqnarray}
It should be notice that a generated constant term in $\Sigma_{1}$ has been discarded.
The generated constant term in self-energy is also discarded in the study about long-range
Coulomb interaction in 3D semi-DSM \cite{Abrikosov72}.
According to Eqs.~(\ref{Eq:Sigma1Result}) and (\ref{Eq:Sigma2Result}),  the fermion propagator is not renormalized by the
four-fermion interactions to one-loop order.

\begin{figure}[htbp]
\center
\includegraphics[width=3.1in]{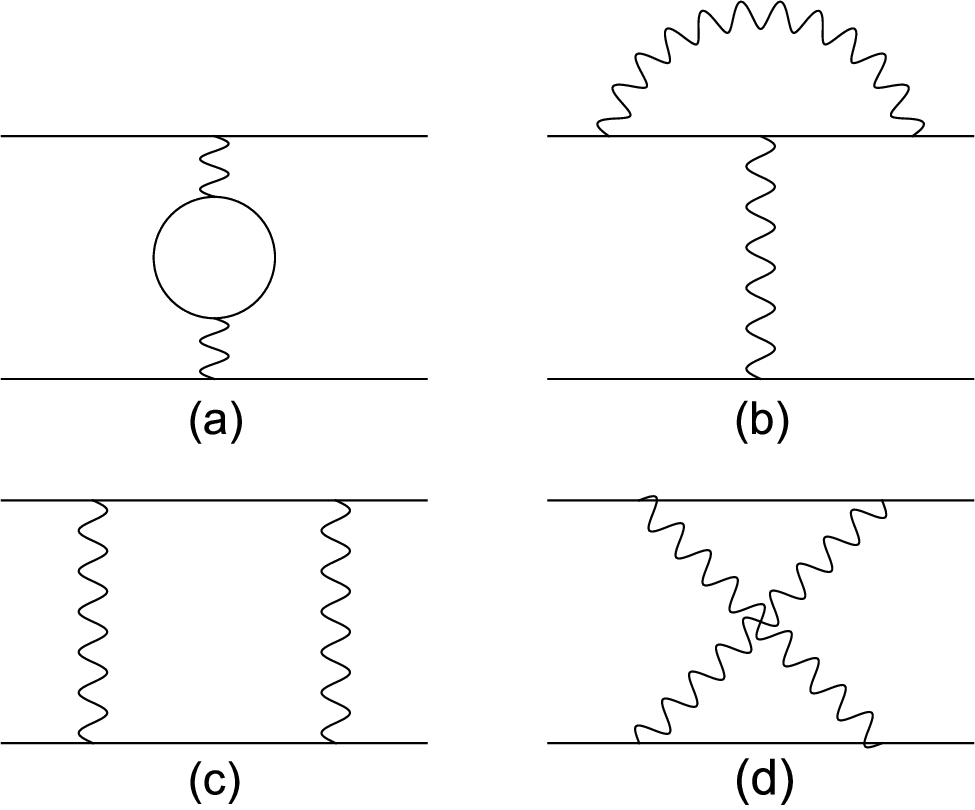}
\caption{One-loop Feynman diagrams for the corrections to the
four-fermion couplings. \label{Fig:VertexCorrection}}
\end{figure}

For the five independent four-fermion interactions shown in Eq.~(\ref{Eq:FFInteractionsLinIndependent}),
there is not a constant term in $\Sigma_{2}$, and $\Sigma_{2}$ always equals to zero. If we consider the four-fermion interaction $\left(\bar{\Psi}i\gamma_{3}\Psi\right)^{2}$,
we can find that there is a constant term in $\Sigma_{2}$. This constant term is actually a correction for the chemical potential $\mu$.
This constant term could modify the chemical potential $\mu$ from zero to finite and thus drives the Fermi level away from the node.
In this case, we assume that the system parameters (for examples, gate voltage, pressure \emph{etc.}) are fine-tuned in such a way that effective chemical potential is zero.
This way we can study the influence of interactions on 3D semi-DSM with zero chemical potential.

\subsection{One-loop corrections for the four-fermion couplings}

Fig.~\ref{Fig:VertexCorrection}(a) leads to the correction
\begin{eqnarray}
V_{a}^{(1)}&=&-2g_{a}^{2}\left(\bar{\Psi}\Gamma_{a}\Psi\right)^{2}\int_{-\infty}^{+\infty}
\frac{d\omega}{2\pi}\int'\frac{d^{3}\mathbf{k}}{(2\pi)^{3}}\mathrm{Tr}\left[\Gamma_{a}
\right.\nonumber
\\
&&\left.\times G_{0}(i\omega,\mathbf{k})\Gamma_{a}G_{0}(i\omega,\mathbf{k})\right]. \label{Eq:FirstDiagramExpression}
\end{eqnarray}
Fig.~\ref{Fig:VertexCorrection}(b) results in the correction
\begin{eqnarray}
V_{a}^{(2)}=\sum_{b}V_{ab}^{(2)}, \label{Eq:SecondDiagramExpressionA}
\end{eqnarray}
where
\begin{eqnarray}
V_{ab}^{(2)}&=&4g_{a}g_{b}\left(\bar{\Psi}\Gamma_{a}\bar{\Psi}\right)\int_{-\infty}^{+\infty}\frac{d\omega}{2\pi}
\int'\frac{d^3\mathbf{k}}{(2\pi)^{3}}\left(\bar{\Psi}\Gamma_{b}\right.\nonumber
\\
&&\left.\times G_{0}(i\omega,\mathbf{k})\Gamma_{a}
G_{0}(i\omega,\mathbf{k})\Gamma_{b}\Psi\right). \label{Eq:SecondDiagramExpressionB}
\end{eqnarray}
The Figs.~\ref{Fig:VertexCorrection}(c) and \ref{Fig:VertexCorrection}(d) induce the correction
\begin{eqnarray}
V^{(3)+(4)}&=&\sum_{a}\sum_{a\le b}V_{ab}^{(3)+(4)},
\end{eqnarray}
where
\begin{eqnarray}
V_{ab}^{(3)+(4)}&=&4g_{a}g_{b}\int_{-\infty}^{+\infty}\frac{d\omega}{2\pi}\int'\frac{d^{3}\mathbf{k}}{(2\pi)^{3}}
\nonumber
\\
&&\times\left(\bar{\Psi}\Gamma_{a}G_{0}(i\omega,\mathbf{k})\Gamma_{b}\Psi\right)\bar{\Psi}\left\{\left[\Gamma_{b}G_{0}(i\omega,\mathbf{k})\Gamma_{a}\right.\right.\nonumber
\\
&&\left.\left.+\Gamma_{a}G_{0}(-i\omega,-\mathbf{k})\Gamma_{b}\right]\right\}\Psi. \label{Eq:ThirdFourthDiagramExpression}
\end{eqnarray}

Substituting Eq.~(\ref{Eq:FermionPropagator}) into Eq.~(\ref{Eq:FirstDiagramExpression}), we obtain
\begin{eqnarray}
V_{a}^{(1)}&=&\delta g_{a}^{(1)}\left(\bar{\Psi}\Gamma_{a}\Psi\right)^{2},
\end{eqnarray}
where
\begin{eqnarray}
\delta g_{1}^{(1)}&=&0,
\\
\delta g_{2}^{(1)}&=& g_{2}^{2}\frac{2\Lambda^{\frac{3}{2}}}{\pi^{2}v^{2}\sqrt{A}}\ell,
\\
\delta g_{4}^{(1)}&=& 0,
\\
\delta g_{5}^{(1)}&=&g_{5}^{2}\frac{2\Lambda^{\frac{3}{2}}}{\pi^{2}v^{2}\sqrt{A}}\ell,
\\
\delta g_{3z}^{(1)}&=&g_{3z}^{2}
\frac{2\Lambda^{\frac{3}{2}}}{5\pi^{2}v^{2}\sqrt{A}}\ell.
\end{eqnarray}

Substituting Eq.~(\ref{Eq:FermionPropagator}) into Eqs.~(\ref{Eq:SecondDiagramExpressionA}) and (\ref{Eq:SecondDiagramExpressionB}), we find that
the contribution from  Fig.~\ref{Fig:VertexCorrection}(b) can be written as
\begin{eqnarray}
V_{a}^{(2)}&=&\delta g_{a}^{(2)}\left(\bar{\Psi}\Gamma_{a}\Psi\right)^{2},
\end{eqnarray}
where
\begin{eqnarray}
\delta g_{1}^{(2)}&=&0,
\\
\delta g_{2}^{(2)}&=&\left(-g_{2}g_{1}-g_{2}^{2}+g_{2}g_{4}+g_{2}g_{5}+g_{2}g_{3z}\right)\nonumber
\\
&&\times\frac{\Lambda^{\frac{3}{2}}}{\pi^{2}v^{2}\sqrt{A}}\ell,
\\
\delta g_{4}^{(2)}&=&0,
\\
\delta g_{5}^{(2)}&=&\left(-g_{5}g_{1}+g_{5}g_{2}+g_{5}g_{4}-g_{5}^{2}-g_{5}g_{3z}\right)\nonumber
\\
&&\times\frac{\Lambda^{\frac{3}{2}}}{\pi^{2}v^{2}\sqrt{A}}\ell,
\\
\delta g_{3z}^{(2)}&=&\left(-g_{3z}g_{1}+g_{3z}g_{2}+g_{3z}g_{4}-g_{3z}g_{5}-g_{3z}^{2}\right)\nonumber
\\
&&\times\frac{\Lambda^{\frac{3}{2}}}{5\pi^{2}v^{2}\sqrt{A}}\ell.
\end{eqnarray}

Substituting Eq.~(\ref{Eq:FermionPropagator}) into Eq.~(\ref{Eq:ThirdFourthDiagramExpression}),
the contribution from Figs.~\ref{Fig:VertexCorrection}(c) and \ref{Fig:VertexCorrection}(d) can be written as
\begin{eqnarray}
V_{1,1}^{(3)+(4)}
&=&g_{1}^{2}\frac{\Lambda^{\frac{3}{2}}}{5\pi^{2}v^{2}\sqrt{A}}\ell
\left(\bar{\Psi}i\gamma_{3}\Psi\right)^{2},
\\
V_{2,2}^{(3)+(4)}
&=&g_{2}^{2}\frac{\Lambda^{\frac{3}{2}}}{5\pi^{2}v^{2}\sqrt{A}}\ell
\left(\bar{\Psi}i\gamma_{3}\Psi\right)^{2},
\\
V_{4,4}^{(3)+(4)}
&=&g_{4}^{2}\frac{\Lambda^{\frac{3}{2}}}{5\pi^{2}v^{2}\sqrt{A}}\ell
\left(\bar{\Psi}i\gamma_{3}\Psi\right)^{2},
\\
V_{5,5}^{(3)+(4)}
&=&g_{5}^{2}\frac{\Lambda^{\frac{3}{2}}}{5\pi^{2}v^{2}\sqrt{A}}\ell
\left(\bar{\Psi}i\gamma_{3}\Psi\right)^{2},
\\
V_{3z,3z}^{(3)+(4)}
&=&g_{3z}^{2}\frac{\Lambda^{\frac{3}{2}}}{5\pi^{2}v^{2}\sqrt{A}}\ell
\left(\bar{\Psi}i\gamma_{3}\Psi\right)^{2},
\\
V_{1,2}^{(3)+(4)}
&=&g_{1}g_{2}\frac{2\Lambda^{\frac{3}{2}}}{5\pi^{2}v^{2}\sqrt{A}}\ell
\sum_{j=1}^{2}\left(\bar{\Psi}\gamma_{0}\gamma_{j}\Psi\right)^{2},
\\
V_{1,4}^{(3)+(4)}
&=&g_{1}g_{4}\frac{\Lambda^{\frac{3}{2}}}{5\pi^{2}v^{2}\sqrt{A}}\ell
\left(\bar{\Psi}i\gamma_{5}\gamma_{3}\Psi\right)^{2},
\\
V_{1,5}^{(3)+(4)}
&=&g_{1}g_{5}\frac{2\Lambda^{\frac{3}{2}}}{5\pi^{2}v^{2}\sqrt{A}}\ell
\sum_{<<lk>>}\left(\bar{\Psi}i\gamma_{l}\gamma_{k}\Psi\right)^{2},
\\
V_{1,3z}^{(3)+(4)}&=&0,
\\
V_{2,4}^{(3)+(4)}
&=&-g_{2}g_{4}\frac{\Lambda^{\frac{3}{2}}}{\pi^{2}v^{2}\sqrt{A}}\ell\left(\bar{\Psi}i\gamma_{5}\Psi\right)^{2}\nonumber
\\
&&+g_{2}g_{4}\frac{\Lambda^{\frac{3}{2}}}{5\pi^{2}v^{2}\sqrt{A}}\ell\left(\bar{\Psi}i\gamma_{1}\gamma_{2}\Psi\right)^{2},
\\
V_{2,5}^{(3)+(4)}
&=&-g_{2}g_{5}\frac{\Lambda^{\frac{3}{2}}}{\pi^{2}v^{2}\sqrt{A}}\ell\left(\bar{\Psi}\gamma_{0}\gamma_{5}\Psi\right)^{2}\nonumber
\\
&&+g_{2}g_{5}\sum_{j=1}^{2}\frac{2\Lambda^{\frac{3}{2}}}{5\pi^{2}v^{2}\sqrt{A}}\ell\left(\bar{\Psi}i\gamma_{5}\gamma_{j}\Psi\right)^{2},
\\
V_{2,3z}^{(3)+(4)}
&=&-g_{2}g_{3z}\frac{\Lambda^{\frac{3}{2}}}{\pi^{2}v^{2}\sqrt{A}}\ell\left(\bar{\Psi}i\gamma_{3}\Psi\right)^{2},
\\
V_{4,5}^{(3)+(4)}
&=&-g_{4}g_{5}\frac{\Lambda^{\frac{3}{2}}}{\pi^{2}v^{2}\sqrt{A}}\ell\left(\bar{\Psi}\Psi\right)^{2}\nonumber
\\
&&+g_{4}g_{5}\frac{\Lambda^{\frac{3}{2}}}{5\pi^{2}v^{2}\sqrt{A}}\ell\left(\bar{\Psi}\gamma_{0}\gamma_{3}\Psi\right)^{2},
\\
V_{4,3z}^{(3)+(4)}
&=&-g_{4}g_{3z}
\frac{\Lambda^{\frac{3}{2}}}{\pi^{2}v^{2}\sqrt{A}}\ell
\left(\bar{\Psi}i\gamma_{1}\gamma_{2}\Psi\right)^{2}\nonumber
\\
&&+g_{4}g_{3z}\sum_{j=1}^{2}
\frac{2\Lambda^{\frac{3}{2}}}{5\pi^{2}v^{2}\sqrt{A}}\ell
\left(\bar{\Psi}\gamma_{0}\gamma_{j}\Psi\right)^{2}\nonumber
\\
&&+g_{4}g_{3z}
\frac{\Lambda^{\frac{3}{2}}}{5\pi^{2}v^{2}\sqrt{A}}\ell
\left(\bar{\Psi}i\gamma_{5}\Psi\right)^{2},
\\
V_{5,3z}^{(3)+(4)}
&=&g_{5}g_{3z}\sum_{j=1}^{2}
\frac{2\Lambda^{\frac{3}{2}}}{5\pi^{2}v^{2}\sqrt{A}}\ell\left(\bar{\Psi}i\gamma_{j}\Psi\right)^{2}\nonumber
\\
&&+g_{5}g_{3z}
\frac{\Lambda^{\frac{3}{2}}}{5\pi^{2}v^{2}\sqrt{A}}\ell\left(\bar{\Psi}\gamma_{0}\gamma_{5}\Psi\right)^{2}.
\end{eqnarray}
Using the relations shown in Eqs.~(\ref{Eq:semiDSMFFCouplingRelation1})-(\ref{Eq:semiDSMFFCouplingRelation7}), we further get
\begin{eqnarray}
V_{1,1}^{(3)+(4)}
&=&g_{0}^{2}\frac{\Lambda^{\frac{3}{2}}}{5\pi^{2}v^{2}\sqrt{A}}\ell
\left[-\left(\bar{\Psi}\gamma_{0}\gamma_{5}\Psi\right)^{2}
+\left(\bar{\Psi}i\gamma_{5}\Psi\right)^{2}\right.\nonumber
\\
&&\left.+\left(\bar{\Psi}\gamma_{0}\gamma_{3}\Psi\right)^{2}\right],
\\
V_{2,2}^{(3)+(4)}
&=&g_{2}^{2}\frac{\Lambda^{\frac{3}{2}}}{5\pi^{2}v^{2}\sqrt{A}}\ell
\left[-\left(\bar{\Psi}\gamma_{0}\gamma_{5}\Psi\right)^{2}
+\left(\bar{\Psi}i\gamma_{5}\Psi\right)^{2}\right.\nonumber
\\
&&\left.+\left(\bar{\Psi}\gamma_{0}\gamma_{3}\Psi\right)^{2}\right],
\\
V_{4,4}^{(3)+(4)}
&=&g_{4}^{2}
\frac{\Lambda^{\frac{3}{2}}}{5\pi^{2}v^{2}\sqrt{A}}\ell
\left[-\left(\bar{\Psi}\gamma_{0}\gamma_{5}\Psi\right)^{2}
+\left(\bar{\Psi}i\gamma_{5}\Psi\right)^{2}\right.\nonumber
\\
&&\left.+\left(\bar{\Psi}\gamma_{0}\gamma_{3}\Psi\right)^{2}\right],
\\
V_{5,5}^{(3)+(4)}
&=&g_{5}^{2}\frac{\Lambda^{\frac{3}{2}}}{5\pi^{2}v^{2}\sqrt{A}}\ell
\left[-\left(\bar{\Psi}\gamma_{0}\gamma_{5}\Psi\right)^{2}
+\left(\bar{\Psi}i\gamma_{5}\Psi\right)^{2}\right.\nonumber
\\
&&\left.+\left(\bar{\Psi}\gamma_{0}\gamma_{3}\Psi\right)^{2}\right],
\\
V_{3z,3z}^{(3)+(4)}
&=&g_{3z}^{2}\frac{\Lambda^{\frac{3}{2}}}{5\pi^{2}v^{2}\sqrt{A}}\ell
\left[-\left(\bar{\Psi}\gamma_{0}\gamma_{5}\Psi\right)^{2}
+\left(\bar{\Psi}i\gamma_{5}\Psi\right)^{2}\right.\nonumber
\\
&&\left.+\left(\bar{\Psi}\gamma_{0}\gamma_{3}\Psi\right)^{2}\right],
\\
V_{1,2}^{(3)+(4)}
&=&g_{1}g_{2}\frac{2\Lambda^{\frac{3}{2}}}{5\pi^{2}v^{2}\sqrt{A}}\ell
\left[-\left(\bar{\Psi}\gamma_{0}\Psi\right)^{2}+\left(\bar{\Psi}\Psi\right)^{2}\right.\nonumber
\\
&&+\left(\bar{\Psi}\gamma_{0}\gamma_{5}\Psi\right)^{2}
-2\left(\bar{\Psi}i\gamma_{5}\Psi\right)^{2}\nonumber
\\
&&\left.-\left(\bar{\Psi}\gamma_{0}\gamma_{3}\Psi\right)^{2}\right],
\\
V_{1,4}^{(3)+(4)}
&=&g_{1}g_{4}\frac{\Lambda^{\frac{3}{2}}}{5\pi^{2}v^{2}\sqrt{A}}\ell
\left[-\left(\bar{\Psi}\gamma_{0}\Psi\right)^{2}
-\left(\bar{\Psi}i\gamma_{5}\Psi\right)^{2}\right.\nonumber
\\
&&\left.-\left(\bar{\Psi}\gamma_{0}\gamma_{3}\Psi\right)^{2}\right],
\\
V_{1,5}^{(3)+(4)}
&=&g_{1}g_{5}\frac{2\Lambda^{\frac{3}{2}}}{5\pi^{2}v^{2}\sqrt{A}}\ell
\left[-\left(\bar{\Psi}\gamma_{0}\Psi\right)^{2}
-\left(\bar{\Psi}\Psi\right)^{2}\right.\nonumber
\\
&&\left.+\left(\bar{\Psi}\gamma_{0}\gamma_{5}\Psi\right)^{2}
-\left(\bar{\Psi}\gamma_{0}\gamma_{3}\Psi\right)^{2}\right],
\\
V_{1,3z}^{(3)+(4)}&=&0,
\\
V_{2,4}^{(3)+(4)}
&=&g_{2}g_{4}\frac{\Lambda^{\frac{3}{2}}}{5\pi^{2}v^{2}\sqrt{A}}\ell
\left[-\left(\bar{\Psi}\Psi\right)^{2}
-4\left(\bar{\Psi}i\gamma_{5}\Psi\right)^{2}\right.\nonumber
\\
&&\left.+\left(\bar{\Psi}\gamma_{0}\gamma_{3}\Psi\right)^{2}\right],
\\
V_{2,5}^{(3)+(4)}
&=&g_{2}g_{5}\frac{2\Lambda^{\frac{3}{2}}}{5\pi^{2}v^{2}\sqrt{A}}\ell
\left[-\left(\bar{\Psi}\gamma_{0}\Psi\right)^{2}
-\left(\bar{\Psi}\Psi\right)^{2}\right.\nonumber
\\
&&\left.-\frac{7}{2}\left(\bar{\Psi}\gamma_{0}\gamma_{5}\Psi\right)^{2}
+\left(\bar{\Psi}\gamma_{0}\gamma_{3}\Psi\right)^{2}\right],
\\
V_{2,3z}^{(3)+(4)}
&=&-g_{2}g_{3z}\frac{\Lambda^{\frac{3}{2}}}{\pi^{2}v^{2}\sqrt{A}}\ell
\left[-\left(\bar{\Psi}\gamma_{0}\gamma_{5}\Psi\right)^{2}
+\left(\bar{\Psi}i\gamma_{5}\Psi\right)^{2}\right.\nonumber
\\
&&\left.+\left(\bar{\Psi}\gamma_{0}\gamma_{3}\Psi\right)^{2}\right],
\\
V_{4,5}^{(3)+(4)}
&=&g_{4}g_{5}\frac{\Lambda^{\frac{3}{2}}}{5\pi^{2}v^{2}\sqrt{A}}\ell\left[-5\left(\bar{\Psi}\Psi\right)^{2}\right.\nonumber
\\
&&\left.+\left(\bar{\Psi}\gamma_{0}\gamma_{3}\Psi\right)^{2}\right],
\\
V_{4,3z}^{(3)+(4)}
&=&g_{4}g_{3z}\frac{2\Lambda^{\frac{3}{2}}}{5\pi^{2}v^{2}\sqrt{A}}\ell
\left[-\left(\bar{\Psi}\gamma_{0}\Psi\right)^{2}
+\frac{7}{2}\left(\bar{\Psi}\Psi\right)^{2}\right.\nonumber
\\
&&+\left(\bar{\Psi}\gamma_{0}\gamma_{5}\Psi\right)^{2}
-4\left(\bar{\Psi}i\gamma_{5}\Psi\right)^{2}\nonumber
\\
&&\left.-\frac{7}{2}\left(\bar{\Psi}\gamma_{0}\gamma_{3}\Psi\right)^{2}\right],
\\
V_{5,3z}^{(3)+(4)}
&=&g_{5}g_{3z}\frac{2\Lambda^{\frac{3}{2}}}{5\pi^{2}v^{2}\sqrt{A}}\ell
\left[-\left(\bar{\Psi}\gamma_{0}\Psi\right)^{2}
+\left(\bar{\Psi}\Psi\right)^{2}\right.\nonumber
\\
&&\left.-\frac{1}{2}\left(\bar{\Psi}\gamma_{0}\gamma_{5}\Psi\right)^{2}
-\left(\bar{\Psi}\gamma_{0}\gamma_{3}\Psi\right)^{2}\right].
\end{eqnarray}
Thus, the contribution from Figs.~\ref{Fig:VertexCorrection}(c) and \ref{Fig:VertexCorrection}(d) is given by
\begin{eqnarray}
V^{(3)+(4)}&=&\sum_{a=1,2,4,5,3z}\sum_{a\leq b}V_{ab}^{(3)+(4)}\nonumber
\\
&=&\sum_{a=1,2,4,5,3z}\delta g_{a}^{(3)+(4)}\left(\bar{\Psi}\Gamma_{a}\Psi\right)^{2},
\end{eqnarray}
where
\begin{eqnarray}
\delta g_{1}^{(3)+(4)}&=&\bigg(-g_{1}g_{2}-\frac{1}{2}g_{1}g_{4}-g_{1}g_{5}
-g_{2}g_{5}-g_{4}g_{3z}\nonumber
\\
&&-g_{5}g_{3z}\bigg)\frac{2\Lambda^{\frac{3}{2}}}{5\pi^{2}v^{2}\sqrt{A}}\ell,
\\
\delta g_{2}^{(3)+(4)}&=&\left(g_{1}g_{2}-g_{1}g_{5}-\frac{1}{2}g_{2}g_{4}
-g_{2}g_{5}-\frac{5}{2}g_{4}g_{5}\right.\nonumber
\\
&&\left.+\frac{7}{2}g_{4}g_{3z}+g_{5}g_{3z}\right)
\frac{2\Lambda^{\frac{3}{2}}}{5\pi^{2}v^{2}\sqrt{A}}\ell,
\\
\delta g_{4}^{(3)+(4)}&=&\left(-\frac{1}{2}g_{1}^{2}-\frac{1}{2}g_{2}^{2}-\frac{1}{2}g_{4}^{2}
-\frac{1}{2}g_{5}^{2}-\frac{1}{2}g_{3z}^{2}\right.\nonumber
\\
&&\left.+g_{1}g_{2}+g_{1}g_{5}-\frac{7}{2}g_{2}g_{5}
+\frac{5}{2}g_{2}g_{3z}+g_{4}g_{3z}\right.\nonumber
\\
&&\left.-\frac{1}{2}g_{5}g_{3z}\right)\frac{2\Lambda^{\frac{3}{2}}}{5\pi^{2}v^{2}\sqrt{A}}\ell,
\\
\delta g_{5}^{(3)+(4)}&=&\left(\frac{1}{2}g_{1}^{2}+\frac{1}{2}g_{2}^{2}+\frac{1}{2}g_{4}^{2}
+\frac{1}{2}g_{5}^{2}+\frac{1}{2}g_{3z}^{2}-2g_{1}g_{2}\right.\nonumber
\\
&&\left.-\frac{1}{2}g_{1}g_{4}-2g_{2}g_{4}
-\frac{5}{2}g_{2}g_{3z}-4g_{4}g_{3z}\right)\nonumber
\\
&&\times\frac{2\Lambda^{\frac{3}{2}}}{5\pi^{2}v^{2}\sqrt{A}}\ell,
\\
\delta g_{3z}^{(3)+(4)}&=&\left(\frac{1}{2}g_{1}^{2}+\frac{1}{2}g_{2}^{2}+\frac{1}{2}g_{4}^{2}
+\frac{1}{2}g_{5}^{2}+\frac{1}{2}g_{3z}^{2}-g_{1}g_{2}\right.\nonumber
\\
&&-\frac{1}{2}g_{1}g_{4}-g_{1}g_{5}
+\frac{1}{2}g_{2}g_{4}+g_{2}g_{5}-\frac{5}{2}g_{2}g_{3z}\nonumber
\\
&&\left.+\frac{1}{2}g_{4}g_{5}-\frac{7}{2}g_{4}g_{3z}-g_{5}g_{3z}\right)\nonumber
\\
&&\times\frac{2\Lambda^{\frac{3}{2}}}{5\pi^{2}v^{2}\sqrt{A}}\ell.
\end{eqnarray}

As shown in above, the one-loop corrections are proportional to $\Lambda^{\frac{3}{2}}$. This characteristic  actually is
easy to see from the expressions of the one-loop corrections to four-fermion interactions. From Eqs.~(\ref{Eq:FirstDiagramExpression}) to
(\ref{Eq:ThirdFourthDiagramExpression}), we can find that
the one-loop corrections should be proportional to
\begin{eqnarray}
\frac{\Lambda^{\frac{2}{z_{\bot}}+\frac{1}{z_{3}}}}{\Lambda}=\Lambda^{\frac{3}{2}},
\end{eqnarray}
where $z_{\bot}=1$ and $z_{3}=2$. The numerator $\Lambda^{\frac{2}{z_{\bot}}+\frac{1}{z_{3}}}=\Lambda^{\frac{5}{2}}$ comes from the integral measure $\int'd^{3}\mathbf{k}$.
The denominator $\Lambda$ results from the expression of integrand after the integration of energy $\omega$ is carried out.

From the above results, we obtain
\begin{eqnarray}
\delta g_{a}&=&\delta g_{a}^{(1)}+\delta g_{a}^{(2)}+\delta g_{a}^{(3)+(4)}.
\end{eqnarray}
Concretely,
\begin{eqnarray}
\delta g_{1}
&=&\left(-g_{1}g_{2}-\frac{1}{2}g_{1}g_{4}-g_{1}g_{5}
-g_{2}g_{5}-g_{4}g_{3z}\right.\nonumber
\\
&&\left.-g_{5}g_{3z}\right)\frac{2\Lambda^{\frac{3}{2}}}{5\pi^{2}v^{2}\sqrt{A}}\ell, \label{Eq:Deltag1}
\\
\delta g_{2}
&=&\left(\frac{5}{2}g_{2}^{2}-\frac{3}{2}g_{1}g_{2}-g_{1}g_{5}+2g_{2}g_{4}
+\frac{3}{2}g_{2}g_{5}\right.\nonumber
\\
&&\left.+\frac{5}{2}g_{2}g_{3z}-\frac{5}{2}g_{4}g_{5}+\frac{7}{2}g_{4}g_{3z}+g_{5}g_{3z}\right)\nonumber
\\
&&\times\frac{2\Lambda^{\frac{3}{2}}}{5\pi^{2}v^{2}\sqrt{A}}\ell, \label{Eq:Deltag2}
\\
\delta g_{4}
&=&\left(-\frac{1}{2}g_{1}^{2}-\frac{1}{2}g_{2}^{2}-\frac{1}{2}g_{4}^{2}
-\frac{1}{2}g_{5}^{2}-\frac{1}{2}g_{3z}^{2}+g_{1}g_{2}\right.\nonumber
\\
&&\left.+g_{1}g_{5}-\frac{7}{2}g_{2}g_{5}
+\frac{5}{2}g_{2}g_{3z}+g_{4}g_{3z}-\frac{1}{2}g_{5}g_{3z}\right)\nonumber
\\
&&\times\frac{2\Lambda^{\frac{3}{2}}}{5\pi^{2}v^{2}\sqrt{A}}\ell, \label{Eq:Deltag4}
\\
\delta g_{5}
&=&\left(\frac{1}{2}g_{1}^{2}+\frac{1}{2}g_{2}^{2}+\frac{1}{2}g_{4}^{2}
+3g_{5}^{2}+\frac{1}{2}g_{3z}^{2}-2g_{1}g_{2}\right.\nonumber
\\
&&-\frac{1}{2}g_{1}g_{4}
-\frac{5}{2}g_{1}g_{5}-2g_{2}g_{4}+\frac{5}{2}g_{2}g_{5}
-\frac{5}{2}g_{2}g_{3z}\nonumber
\\
&&\left.+\frac{5}{2}g_{4}g_{5}-4g_{4}g_{3z}-\frac{5}{2}g_{5}g_{3z}\right)
\frac{2\Lambda^{\frac{3}{2}}}{5\pi^{2}v^{2}\sqrt{A}}\ell, \label{Eq:Deltag5}
\\
\delta g_{3z}
&=&\left(\frac{1}{2}g_{1}^{2}+\frac{1}{2}g_{2}^{2}+\frac{1}{2}g_{4}^{2}
+\frac{1}{2}g_{5}^{2}+g_{3z}^{2}-g_{1}g_{2}-\frac{1}{2}g_{1}g_{4}\right.\nonumber
\\
&&-g_{1}g_{5}-\frac{1}{2}g_{1}g_{3z}
+\frac{1}{2}g_{2}g_{4}+g_{2}g_{5}-2g_{2}g_{3z}\nonumber
\\
&&\left.+\frac{1}{2}g_{4}g_{5}-3g_{4}g_{3z}-\frac{3}{2}g_{5}g_{3z}\right)
\frac{2\Lambda^{\frac{3}{2}}}{5\pi^{2}v^{2}\sqrt{A}}\ell. \label{Eq:Deltag3z}
\end{eqnarray}

\subsection{Scaling transformations}

The free action of fermions is
\begin{eqnarray}
S_{\Psi}&=&\int\frac{d\omega}{2\pi}\frac{d^{3}\mathbf{k}}{(2\pi)^{3}}\bar{\Psi}(\omega,\mathbf{k})
\left(i\omega\gamma_{0}+ivk_{1}\gamma_{1}+ivk_{2}\gamma_{2}\right.\nonumber
\\
&&\left.+iAk_{3}^{2}\gamma_{3}
\right)\Psi(\omega,\mathbf{k}).
\end{eqnarray}
The fermion self-energy induced by four-fermion interactions to one-loop order vanishes. Thus, the form of
action $S_{\Psi}$ is not changed. Employing the transformations
\begin{eqnarray}
\omega&=&\omega'e^{-\ell}, \label{Eq:Scalingomega}
\\
k_{1}&=&k_{1}'e^{-\ell}, \label{Eq:Scalingk1}
\\
k_{2}&=&k_{2}'e^{-\ell}, \label{Eq:Scalingk2}
\\
k_{3}&=&k_{3}'e^{-\frac{\ell}{2}}, \label{Eq:Scalingk3}
\\
v&=&v', \label{Eq:Scalingv}
\\
A&=&A', \label{Eq:ScalingA}
\\
\Psi&=&\Psi' e^{\frac{9}{4}\ell}, \label{Eq:ScalingPsi}
\end{eqnarray}
the action becomes
\begin{eqnarray}
S_{\Psi'}&=&\int\frac{d\omega'}{2\pi}\frac{d^{3}\mathbf{k}'}{(2\pi)^{3}}\bar{\Psi}'(\omega',\mathbf{k}')
\left(i\omega'\gamma_{0}+iv'k_{1}'\gamma_{1}+iv'k_{2}'\gamma_{2}\right.\nonumber
\\
&&\left.+iA'k_{3}'^{2}\gamma_{3}
\right)\Psi'(\omega',\mathbf{k}'),
\end{eqnarray}
which has the same form as the original action.

The original action of four-fermion interactions takes the form
\begin{eqnarray}
S_{\Psi^{4}}&=&\sum_{a=1,2,4,5,3z}g_{a}\int\frac{d\omega_{1}}{2\pi}\frac{d^{3}\mathbf{k}_{1}}{(2\pi)^{3}}
\frac{d\omega_{2}}{2\pi}\frac{d^{3}\mathbf{k}_{2}}{(2\pi)^{3}}
\frac{d\omega_{3}}{2\pi}\frac{d^{3}\mathbf{k}_{3}}{(2\pi)^{3}}\nonumber
\\
&&\times\bar{\Psi}(\omega_{1},\mathbf{k}_{1})
\Gamma_{a}\Psi(\omega_{2},\mathbf{k}_{2})\bar{\Psi}(\omega_{3},k_{3})\Gamma_{a}\nonumber
\\
&&\times\Psi(\omega_{1}-\omega_{2}+\omega_{3},
\mathbf{k}_{1}-\mathbf{k}_{2}+\mathbf{k}_{3}).
\end{eqnarray}
Including the one-loop order correction, the action becomes
\begin{eqnarray}
S_{\Psi^{4}}&=&\sum_{a=1,2,4,5,3z}\left(g_{a}+\delta g_{a}\right)\int\frac{d\omega_{1}}{2\pi}\frac{d^{3}\mathbf{k}_{1}}{(2\pi)^{3}}
\frac{d\omega_{2}}{2\pi}\frac{d^{3}\mathbf{k}_{2}}{(2\pi)^{3}}\nonumber
\\
&&\times\frac{d\omega_{3}}{2\pi}\frac{d^{3}\mathbf{k}_{3}}{(2\pi)^{3}}\bar{\Psi}(\omega_{1},\mathbf{k}_{1})
\Gamma_{a}\Psi(\omega_{2},\mathbf{k}_{2})\bar{\Psi}(\omega_{3},k_{3})\Gamma_{a}\nonumber
\\
&&\times\Psi(\omega_{1}-\omega_{2}+\omega_{3},
\mathbf{k}_{1}-\mathbf{k}_{2}+\mathbf{k}_{3}).
\end{eqnarray}
Utilizing the transformations Eqs.~(\ref{Eq:Scalingomega})-(\ref{Eq:Scalingk3}) and (\ref{Eq:ScalingPsi}),
we get
\begin{eqnarray}
S_{\Psi'^{4}}
&=&\sum_{a=1,2,4,5,3z}\left(g_{a}+\delta g_{a}\right)e^{-\frac{3}{2}\ell}\int\frac{d\omega_{1}'}{2\pi}\frac{d^{3}\mathbf{k}_{1}'}{(2\pi)^{3}}
\frac{d\omega_{2}'}{2\pi}\nonumber
\\
&&\times\frac{d^{3}\mathbf{k}_{2}'}{(2\pi)^{3}}
\frac{d\omega_{3}'}{2\pi}\frac{d^{3}\mathbf{k}_{3}'}{(2\pi)^{3}}\bar{\Psi}'(\omega_{1}',\mathbf{k}_{1}')
\nonumber
\\
&&\times\Gamma_{a}\Psi'(\omega_{2}',\mathbf{k}_{2}')\bar{\Psi}'(\omega_{3}',k_{3}')\Gamma_{a}\nonumber
\\
&&\times\Psi'(\omega_{1}'-\omega_{2}'+\omega_{3}',
\mathbf{k}_{1}'-\mathbf{k}_{2}'+\mathbf{k}_{3}').
\end{eqnarray}
Let
\begin{eqnarray}
g_{a}'&=&\left(g_{a}+\delta g_{a}\right)e^{-\frac{3}{2}\ell}\approx g_{a}-\frac{3}{2}g_{a}\ell+\delta g_{a},\label{Eq:GeneralRescalingga}
\end{eqnarray}
we obtain
\begin{eqnarray}
S_{\Psi'^{4}}
&=&\sum_{a=1,2,4,5,3z}g_{a}'\int\frac{d\omega_{1}'}{2\pi}\frac{d^{3}\mathbf{k}_{1}'}{(2\pi)^{3}}
\frac{d\omega_{2}'}{2\pi}\frac{d^{3}\mathbf{k}_{2}'}{(2\pi)^{3}}
\frac{d\omega_{3}'}{2\pi}\frac{d^{3}\mathbf{k}_{3}'}{(2\pi)^{3}}\nonumber
\\
&&\times\bar{\Psi}'(\omega_{1}',\mathbf{k}_{1}')
\Gamma_{a}\Psi'(\omega_{2}',\mathbf{k}_{2}')\bar{\Psi}'(\omega_{3}',k_{3}')\Gamma_{a}\nonumber
\\
&&\times\Psi'(\omega_{1}'-\omega_{2}'+\omega_{3}',
\mathbf{k}_{1}'-\mathbf{k}_{2}'+\mathbf{k}_{3}'),
\end{eqnarray}
which recovers the original form of the action.

From Eq.~(\ref{Eq:GeneralRescalingga}), we get the RG equation for $g_{a}$ as following
\begin{eqnarray}
\frac{dg_{a}}{d\ell}&=&-\frac{3}{2}g_{a}+\frac{d\delta g_{a}}{d\ell}. \label{Eq:RGEGeneralga}
\end{eqnarray}
Substituting Eqs.~(\ref{Eq:Deltag1})-(\ref{Eq:Deltag3z}) into Eq.~(\ref{Eq:RGEGeneralga}), we find
\begin{eqnarray}
\frac{d g_{1}}{d\ell}
&=&-\frac{3}{2}g_{1}-\frac{2}{5}g_{1}\left(g_{2}+\frac{1}{2}g_{4}+g_{5}\right)
-\frac{2}{5}\left(g_{2}g_{5}\right.\nonumber
\\
&&\left.+g_{4}g_{3z}+g_{5}g_{3z}\right), \label{Eq:RGEg1}
\\
\frac{d g_{2}}{d\ell}
&=&-\frac{3}{2}g_{2}+g_{2}^{2}+g_{2}\left(-\frac{3}{5}g_{1}+\frac{4}{5}g_{4}+\frac{3}{5}g_{5}+g_{3z}\right)\nonumber
\\
&&-\frac{2}{5}g_{1}g_{5}+g_{4}\left(-g_{5}+\frac{7}{5}g_{3z}\right)+\frac{2}{5}g_{5}g_{3z}, \label{Eq:RGEg2}
\\
\frac{d g_{4}}{d\ell}
&=&-\frac{3}{2}g_{4}-\frac{1}{5}g_{4}^{2}-\frac{1}{5}\left(g_{1}^{2}+g_{2}^{2}+g_{5}^{2}+g_{3z}^{2}\right)\nonumber
\\
&&+\frac{2}{5}g_{4}g_{3z}+\frac{2}{5}g_{1}\left(g_{2}+g_{5}\right)+g_{2}\left(-\frac{7}{5}g_{5}+g_{3z}\right)\nonumber
\\
&&-\frac{1}{5}g_{5}g_{3z}, \label{Eq:RGEg4}
\\
\frac{d g_{5}}{d\ell}
&=&-\frac{3}{2}g_{5}+\frac{6}{5}g_{5}^{2}+\frac{1}{5}\left(g_{1}^{2}+g_{2}^{2}+g_{4}^{2}+g_{3z}^{2}\right)\nonumber
\\
&&+g_{5}\left(-g_{1}+g_{2}+g_{4}-g_{3z}\right)-\frac{2}{5}g_{1}\left(2g_{2}+\frac{1}{2}g_{4}\right)\nonumber
\\
&&-g_{2}\left(\frac{4}{5}g_{4}+g_{3z}\right)-\frac{8}{5}g_{4}g_{3z}, \label{Eq:RGEg5}
\\
\frac{d g_{3z}}{d\ell}
&=&-\frac{3}{2}g_{3z}+\frac{2}{5}g_{3z}^{2}+\frac{1}{5}\left(g_{1}^{2}+g_{2}^{2}+g_{4}^{2}+g_{5}^{2}\right)\nonumber
\\
&&-\frac{2}{5}g_{3z}\left(\frac{1}{2}g_{1}+2g_{2}+3g_{4}+\frac{3}{2}g_{5}\right)\nonumber
\\
&&-\frac{2}{5}g_{1}\left(g_{2}+\frac{1}{2}g_{4}+g_{5}\right)
+\frac{2}{5}g_{2}\left(\frac{1}{2}g_{4}+g_{5}\right)\nonumber
\\
&&+\frac{1}{5}g_{4}g_{5}. \label{Eq:RGEg3z}
\end{eqnarray}
The redefinition
\begin{eqnarray}
\frac{\Lambda^{\frac{3}{2}}g_{a}}{\pi^{2}v^{2}\sqrt{A}}\rightarrow g_{a},
\end{eqnarray}
has been employed.

\section{Susceptibility of source terms \label{App:SourceTerms}}

We consider the Lagrangian for the source terms as following
\begin{eqnarray}
\mathcal{L}_{s}&=&\Delta_{1}\bar{\Psi}\gamma_{0}\Psi+\Delta_{2}\bar{\Psi}\Psi
+\Delta_{3\bot}\sum_{j=1}^{2}\bar{\Psi}\gamma_{0}\gamma_{j}\Psi\nonumber
\\
&&+\Delta_{3z}\bar{\Psi}\gamma_{0}\gamma_{3}\Psi+\Delta_{4}\bar{\Psi}\gamma_{0}\gamma_{5}\Psi
+\Delta_{5}\bar{\Psi}i\gamma_{5}\Psi\nonumber
\\
&&+\Delta_{6\bot}\sum_{<<lk>>}\left(\bar{\Psi}i\gamma_{l}\gamma_{k}\Psi\right)
+\Delta_{6z}\bar{\Psi}i\gamma_{1}\gamma_{2}\Psi\nonumber
\\
&&+\Delta_{7\bot}\sum_{j=1}^{2}\bar{\Psi}i\gamma_{5}\gamma_{j}\Psi+\Delta_{7z}\bar{\Psi}i\gamma_{5}\gamma_{3}\Psi
\nonumber
\\
&&+\Delta_{8\bot}\sum_{j=1}^{2}\bar{\Psi}i\gamma_{j}\Psi
+\Delta_{8z}\bar{\Psi}i\gamma_{3}\Psi+\Delta_{S}\Psi^{\dag}i\gamma_{0}\gamma_{5}\gamma_{2}\Psi^{*}
\nonumber
\\
&&
+\Delta_{op}\Psi^{\dag}i\gamma_{0}\gamma_{2}\Psi^{*}+\Delta_{V,1}\Psi^{\dag}\gamma_{3}\Psi^{*}
+\Delta_{V,2}\Psi^{\dag}i\gamma_{0}\gamma_{5}\Psi^{*}\nonumber
\\
&&+\Delta_{V,3}\Psi^{\dag}\gamma_{1}\Psi^{*}+\Delta_{V,0}\Psi^{\dag}i\gamma_{0}\gamma_{2}\gamma_{3}\Psi^{*}.
\end{eqnarray}

\begin{figure}[htbp]
\center
\includegraphics[width=3.3in]{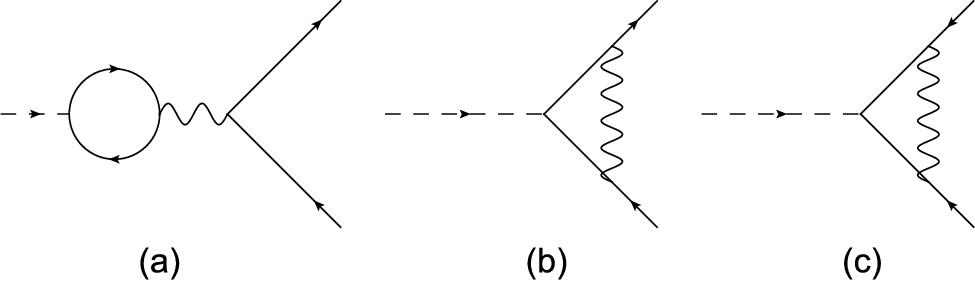}
\caption{(a) and (b): One-loop Feynman diagrams for the corrections to the
source terms in particle-hole channels. (c): One-loop Feynman diagram for the corrections to the
source terms in particle-particle channels.\label{Fig:SourceTerm}}
\end{figure}

\subsection{One-loop order corrections for source terms in particle-hole channels}

There are two one-loop Feynman diagrams lead to the corrections for source terms in particle-hole
channels. The one-loop correction for the source term $\Delta_{X}$ from  Fig.~\ref{Fig:SourceTerm}(a) is given by
\begin{eqnarray}
W_{\Delta_{X}}^{(1)}&=&-2\Delta_{X}g_{X}\left(\bar{\Psi}\Gamma_{X}\Psi\right)\sum_{a=1,2,4,5,3z}g_{a}\int_{-\infty}^{+\infty}\frac{d\omega}{2\pi}\nonumber
\\
&&\times\int'\frac{d^3\mathbf{k}}{(2\pi)^{3}}
\mathrm{Tr}\left[\Gamma_{X}G_{0}(i\omega,\mathbf{k})\Gamma_{a}\right.\nonumber
\\
&&\left.\times G_{0}(i\omega,\mathbf{k})\right]. \label{Eq:SourceTermCorrectionA}
\end{eqnarray}
The one-loop correction for the source term $\Delta_{X}$ resulting from  Fig.~\ref{Fig:SourceTerm}(b) can be written as
\begin{eqnarray}
W_{\Delta_{X}}^{(2)}&=&2\Delta_{X}\sum_{a=1,2,4,5,3z}g_{a}
\int_{-\infty}^{+\infty}\frac{d\omega}{2\pi}
\int'\frac{d^3\mathbf{k}}{(2\pi)^{3}}\nonumber
\\
&&\times\left(\bar{\Psi}
\Gamma_{a}G_{0}(i\omega,\mathbf{k})\Gamma_{X}G_{0}(i\omega,\mathbf{k})\Gamma_{a}\Psi\right). \label{Eq:SourceTermCorrectionB}
\end{eqnarray}

Substituting Eq.~(\ref{Eq:FermionPropagator}) into Eq.~(\ref{Eq:SourceTermCorrectionA}), we find
\begin{eqnarray}
W_{\Delta_{1}}^{(1)}
&=&0,
\\
W_{\Delta_{2}}^{(1)}
&=&\Delta_{2}g_{2}
\frac{2\Lambda^{\frac{3}{2}}}{\pi^{2}v^{2}\sqrt{A}}\ell
\left(\bar{\Psi}\Psi\right),
\\
W_{\Delta_{3\bot}}^{(1)}&=&0,
\\
W_{3z}^{(1)}&=&\Delta_{3z}g_{3z}
\frac{2\Lambda^{\frac{3}{2}}}{5\pi^{2}v^{2}\sqrt{A}}\ell
\left(\bar{\Psi}\gamma_{0}\gamma_{3}\Psi\right),
\\
W_{\Delta_{4}}^{(1)}&=&0,
\\
W_{\Delta_{5}}^{(1)}
&=&\Delta_{5}g_{5}\frac{2\Lambda^{\frac{3}{2}}}{\pi^{2}v^{2}\sqrt{A}}\ell
\left(\bar{\Psi}i\gamma_{5}\Psi\right),
\\
W_{\Delta_{6\bot}}^{(1)}&=&0,
\\
W_{\Delta_{6z}}^{(1)}&=&0,
\\
W_{\Delta_{7\bot}}^{(1)}&=&0,
\\
W_{\Delta_{7z}}^{(1)}&=&0,
\\
W_{\Delta_{8\bot}}^{(1)}&=&0,
\\
W_{\Delta_{8z}}^{(1)}&=&0.
\end{eqnarray}
Substituting Eq.~(\ref{Eq:FermionPropagator}) into Eq.~(\ref{Eq:SourceTermCorrectionB}), we obtain
\begin{eqnarray}
W_{\Delta_{1}}^{(2)}
&=&0,
\\
W_{\Delta_{2}}^{(2)}
&=&\frac{1}{2}\Delta_{2}\left(-g_{1}-g_{2}+g_{4}+g_{5}+g_{3z}\right)\nonumber
\\
&&\times\frac{\Lambda^{\frac{3}{2}}}{\pi^{2}v^{2}\sqrt{A}}\ell\left(\bar{\Psi}\Psi\right),
\\
W_{\Delta_{3\bot}}^{(2)}
&=&\Delta_{3\bot}\left(-g_{1}+g_{2}+g_{4}-g_{5}+g_{3z}\right)\nonumber
\\
&&\times\frac{\Lambda^{\frac{3}{2}}}{5\pi^{2}v^{2}\sqrt{A}}\ell\sum_{j=1}^{2}\left(\bar{\Psi}
\gamma_{0}\gamma_{j}\Psi\right),
\\
W_{\Delta_{3z}}^{(2)}
&=&\Delta_{3z}\left(-g_{1}+g_{2}+g_{4}-g_{5}-g_{3z}\right)\nonumber
\\
&&\times\frac{\Lambda^{\frac{3}{2}}}{10\pi^{2}v^{2}\sqrt{A}}\ell\left(\bar{\Psi}\gamma_{0}\gamma_{3}\Psi\right),
\\
W_{\Delta_{4}}^{(2)}&=&0,
\\
W_{\Delta_{5}}^{(2)}
&=&\Delta_{5}\left(-g_{1}+g_{2}+g_{4}-g_{5}-g_{3z}\right)\nonumber
\\
&&\times\frac{\Lambda^{\frac{3}{2}}}{2\pi^{2}v^{2}\sqrt{A}}\ell\left(\bar{\Psi}i\gamma_{5}
\Psi\right),
\\
W_{\Delta_{6\bot}}^{(2)}
&=&\Delta_{6\bot}\left(-g_{1}-g_{2}+g_{4}+g_{5}-g_{3z}\right)\nonumber
\\
&&\times\frac{\Lambda^{\frac{3}{2}}}{5\pi^{2}v^{2}\sqrt{A}}\ell\sum_{<<lk>>}\left(\bar{\Psi}i\gamma_{l}\gamma_{k}\Psi\right),
\\
W_{\Delta_{6z}}^{(2)}
&=&\Delta_{6z}\left(-g_{1}-g_{2}+g_{4}+g_{5}+g_{3z}\right)\nonumber
\\
&&\times\frac{\Lambda^{\frac{3}{2}}}{10\pi^{2}v^{2}\sqrt{A}}\ell\left(\bar{\Psi}i\gamma_{1}\gamma_{2}
\Psi\right),
\\
W_{\Delta_{7\bot}}^{(2)}
&=&\Delta_{7\bot}\left(-g_{1}-g_{2}-g_{4}-g_{5}+g_{3z}\right)\nonumber
\\
&&\times\frac{3\Lambda^{\frac{3}{2}}}{10\pi^{2}v^{2}\sqrt{A}}\ell\sum_{j=1}^{2}\left(\bar{\Psi}i\gamma_{5}\gamma_{j}\Psi\right),
\\
W_{\Delta_{7z}}^{(2)}
&=&-\Delta_{7z}\left(g_{1}+g_{2}+g_{4}+g_{5}+g_{3z}\right)\nonumber
\\
&&\frac{2\Lambda^{\frac{3}{2}}}{5\pi^{2}v^{2}\sqrt{A}}\ell\left(\bar{\Psi}i\gamma_{5}\gamma_{3}\Psi\right),
\\
W_{\Delta_{8\bot}}^{(2)}
&=&\Delta_{8\bot}\left(-g_{1}+g_{2}-g_{4}+g_{5}-g_{3z}\right)\nonumber
\\
&&\times\frac{3\Lambda^{\frac{3}{2}}}{10\pi^{2}v^{2}\sqrt{A}}\ell\sum_{j=1}^{2}\left(\bar{\Psi}i\gamma_{j}\Psi\right),
\\
W_{\Delta_{8z}}^{(2)}
&=&\Delta_{8z}\left(-g_{1}+g_{2}-g_{4}+g_{5}+g_{3z}\right)\nonumber
\\
&&\times\frac{2\Lambda^{\frac{3}{2}}}{5\pi^{2}v^{2}\sqrt{A}}\ell\left(\bar{\Psi}i\gamma_{3}\Psi\right).
\end{eqnarray}

From
\begin{eqnarray}
W_{\Delta_{X}}=W_{\Delta_{X}}^{(1)}+W_{\Delta_{X}}^{(2)},
\end{eqnarray}
we arrive
\begin{eqnarray}
W_{\Delta_{X}}=\delta\Delta_{X}\left(\bar{\Psi}\Gamma_{X}\Psi\right).
\end{eqnarray}
The parameters $\delta\Delta_{X}$ are given by
\begin{eqnarray}
\delta\Delta_{1}
&=&0, \label{Eq:deltaDelta1}
\\
\delta\Delta_{2}
&=&\Delta_{2}\left(-g_{1}+3g_{2}+g_{4}+g_{5}+g_{3z}\right)\nonumber
\\
&&\times\frac{\Lambda^{\frac{3}{2}}}{2\pi^{2}v^{2}\sqrt{A}}\ell, \label{Eq:deltaDelta2}
\\
\delta\Delta_{3\bot}
&=&\Delta_{3\bot}\left(-g_{1}+g_{2}+g_{4}-g_{5}+g_{3z}\right)\nonumber
\\
&&\times\frac{\Lambda^{\frac{3}{2}}}{5\pi^{2}v^{2}\sqrt{A}}\ell, \label{Eq:deltaDelta3bot}
\\
\delta\Delta_{3z}
&=&\Delta_{3z}\left(-g_{1}+g_{2}+g_{4}-g_{5}+3g_{3z}\right)\nonumber
\\
&&\times\frac{\Lambda^{\frac{3}{2}}}{10\pi^{2}v^{2}\sqrt{A}}\ell, \label{Eq:deltaDelta3z}
\\
\delta\Delta_{4}&=&0, \label{Eq:deltaDelta4}
\\
\delta\Delta_{5}
&=&
\Delta_{5}\left(-g_{1}+g_{2}+g_{4}+3g_{5}-g_{3z}\right)\nonumber
\\
&&\times\frac{\Lambda^{\frac{3}{2}}}{2\pi^{2}v^{2}\sqrt{A}}\ell, \label{Eq:deltaDelta5}
\\
\delta\Delta_{6\bot}
&=&\Delta_{6\bot}\left(-g_{1}-g_{2}+g_{4}+g_{5}-g_{3z}\right)\nonumber
\\
&&\times\frac{\Lambda^{\frac{3}{2}}}{5\pi^{2}v^{2}\sqrt{A}}\ell, \label{Eq:deltaDelta6bot}
\\
\delta\Delta_{6z}
&=&\Delta_{6z}\left(-g_{1}-g_{2}+g_{4}+g_{5}+g_{3z}\right)\nonumber
\\
&&\times\frac{\Lambda^{\frac{3}{2}}}{10\pi^{2}v^{2}\sqrt{A}}\ell, \label{Eq:deltaDelta6z}
\\
\delta\Delta_{7\bot}
&=&\Delta_{7\bot}\left(-g_{1}-g_{2}-g_{4}-g_{5}+g_{3z}\right)\nonumber
\\
&&\times\frac{3\Lambda^{\frac{3}{2}}}{10\pi^{2}v^{2}\sqrt{A}}\ell, \label{Eq:deltaDelta7bot}
\\
\delta\Delta_{7z}
&=&\Delta_{7z}\left(-g_{1}-g_{2}-g_{4}-g_{5}-g_{3z}\right)\nonumber
\\
&&\times\frac{2\Lambda^{\frac{3}{2}}}{5\pi^{2}v^{2}\sqrt{A}}\ell,\label{Eq:deltaDelta7z}
\\
\delta\Delta_{8\bot}
&=&\Delta_{8\bot}\left(-g_{1}+g_{2}-g_{4}+g_{5}-g_{3z}\right)\nonumber
\\
&&\times\frac{3\Lambda^{\frac{3}{2}}}{10\pi^{2}v^{2}\sqrt{A}}\ell, \label{Eq:deltaDelta8bot}
\\
\delta\Delta_{8z}
&=&\Delta_{8z}\left(-g_{1}+g_{2}-g_{4}+g_{5}+g_{3z}\right)\nonumber
\\
&&\times\frac{2\Lambda^{\frac{3}{2}}}{5\pi^{2}v^{2}\sqrt{A}}\ell.  \label{Eq:deltaDelta8z}
\end{eqnarray}

\subsection{One-loop order corrections for source terms in particle-particle channels}

In particle-particle channels, to one-loop order, there is one Feynman diagram as shown in Fig.~\ref{Fig:SourceTerm}(c)
resulting in the corrections to source terms.
The correction can be expresses as
\begin{eqnarray}
W_{\Delta_{Y}}&=&2\Delta_{Y}\sum_{a=1,2,4,5,3z}g_{a}
\int_{-\infty}^{+\infty}\frac{d\omega}{2\pi}
\int'\frac{d^3\mathbf{k}}{(2\pi)^{3}}\left(\Psi^{\dag}
\Gamma_{a}^{T}\right.\nonumber
\\
&&\left.\times G_{0}^{T}(i\omega,\mathbf{k})\Gamma_{Y}G_{0}(-i\omega,-\mathbf{k})\Gamma_{a}\Psi^{*}\right), \label{Eq:SourceTermCorrectionPPChannel}
\end{eqnarray}
where $T$ represents transposition.
Substituting Eq.~(\ref{Eq:FermionPropagator}) into Eq.~(\ref{Eq:SourceTermCorrectionPPChannel}), we get
\begin{eqnarray}
W_{\Delta_{Y}}=\delta\Delta_{Y}\left(\Psi^{\dag}\Gamma_{Y}\Psi^{*}\right),
\end{eqnarray}
where
\begin{eqnarray}
\delta\Delta_{S}&=&\Delta_{S}\left(g_{1}-g_{2}+g_{4}+g_{5}-g_{3z}\right)\nonumber
\\
&&\times\frac{2\Lambda^{\frac{3}{2}}}{5\pi^{2}v^{2}\sqrt{A}}\ell,\label{Eq:deltaDeltaS}
\\
\delta\Delta_{op}&=&\Delta_{op}\left(g_{1}+g_{2}+g_{4}-g_{5}+g_{3z}\right)\nonumber
\\
&&\times\frac{2\Lambda^{\frac{3}{2}}}{5\pi^{2}v^{2}\sqrt{A}}\ell, \label{Eq:deltaDeltaop}
\\
\delta\Delta_{V,1}&=&\Delta_{V,1}\left(g_{1}+g_{2}-g_{4}+g_{5}+g_{3z}\right)\nonumber
\\
&&\times\frac{\Lambda^{\frac{3}{2}}}{5\pi^{2}v^{2}\sqrt{A}}\ell, \label{Eq:deltaDeltaV1}
\\
\delta\Delta_{V,2}&=&\Delta_{V,2}\left(g_{1}+g_{2}-g_{4}+g_{5}+g_{3z}\right)\nonumber
\\
&&\times\frac{\Lambda^{\frac{3}{2}}}{5\pi^{2}v^{2}\sqrt{A}}\ell, \label{Eq:deltaDeltaV2}
\\
\delta\Delta_{V,3}&=&\Delta_{V,3}\left(g_{1}+g_{2}-g_{4}+g_{5}-g_{3z}\right)\nonumber
\\
&&\times\frac{\Lambda^{\frac{3}{2}}}{2\pi^{2}v^{2}\sqrt{A}}\ell, \label{Eq:deltaDeltaV3}
\\
\delta\Delta_{V,0}&=&\Delta_{V,0}\left(g_{1}-g_{2}-g_{4}-g_{5}+g_{3z}\right)\nonumber
\\
&&\times\frac{\Lambda^{\frac{3}{2}}}{20\pi^{2}v^{2}\sqrt{A}}\ell. \label{Eq:deltaDeltaV0}
\end{eqnarray}

\subsection{Derivation of the RG equations for source terms}

In particle-hole channels, the bare action for the source terms is
\begin{eqnarray}
S_{s}=\Delta_{X}\int\frac{d\omega}{2\pi}\frac{d^3\mathbf{k}}{(2\pi)^{3}}
\bar{\Psi}(\omega,\mathbf{k})\Gamma_{X}\Psi(\omega,\mathbf{k}).
\end{eqnarray}
Considering the one-loop order corrections, we obtain
\begin{eqnarray}
S_{s}&=&\left(\Delta_{X}+\delta\Delta_{X}\right)
\int\frac{d\omega}{2\pi}\frac{d^3\mathbf{k}}{(2\pi)^{3}}
\bar{\Psi}(\omega,\mathbf{k})\Gamma_{X}\nonumber
\\
&&\times\Psi(\omega,\mathbf{k}).
\end{eqnarray}
Using the transformations Eqs.~(\ref{Eq:Scalingomega})-(\ref{Eq:Scalingk3}) and (\ref{Eq:ScalingPsi}),
we can get
\begin{eqnarray}
S_{s}
&=&\left(\Delta_{X}+\delta\Delta_{X}\right)
e^{\ell}\int\frac{d\omega'}{2\pi}\frac{d^3\mathbf{k}'}{(2\pi)^{3}}
\bar{\Psi}'(\omega',\mathbf{k}')\Gamma_{X}\nonumber
\\
&&\times\Psi'(\omega',\mathbf{k}')\nonumber
\\
&\approx&\left(\Delta_{X}+\Delta_{X}\ell
+\delta\Delta_{X}\right)
\int\frac{d\omega'}{2\pi}\frac{d^3\mathbf{k}'}{(2\pi)^{3}}
\bar{\Psi}'(\omega',\mathbf{k}')\Gamma_{X}\nonumber
\\
&&\times\Psi'(\omega',\mathbf{k}').
\end{eqnarray}
Let
\begin{eqnarray}
\Delta_{X}'=\Delta_{X}+\Delta_{X}\ell
+\delta\Delta_{X},
\end{eqnarray}
the action can be further written as
\begin{eqnarray}
S_{s}&=&\Delta_{X}'\int\frac{d\omega'}{2\pi}\frac{d^3\mathbf{k}'}
{(2\pi)^{3}}\bar{\Psi}'(\omega',\mathbf{k}')\Gamma_{X}\Psi'(\omega',\mathbf{k}'),
\end{eqnarray}
which recovers the form of the original action. We can easily find that the
RG equation for $\Delta_{X}$ is
\begin{eqnarray}
\frac{d\Delta_{X}}{d\ell}&=&\Delta_{X}
+\frac{d\delta\Delta_{X}}{d\ell}. \label{Eq:RGEGeneralDeltaPHChannel}
\end{eqnarray}
Performing similar rescaling transformations, we can get the RG equation for source terms in
particle-particle channels
\begin{eqnarray}
\frac{d\Delta_{Y}}{d\ell}&=&\Delta_{Y}
+\frac{d\delta\Delta_{Y}}{d\ell}. \label{Eq:RGEGeneralDeltaPPChannel}
\end{eqnarray}
Substituting Eqs.~(\ref{Eq:deltaDelta1})-(\ref{Eq:deltaDelta8z}) into Eq.~(\ref{Eq:RGEGeneralDeltaPHChannel}),
and substituting Eqs.~(\ref{Eq:deltaDeltaS})-(\ref{Eq:deltaDeltaV0}) into Eq.~(\ref{Eq:RGEGeneralDeltaPPChannel}), we get the RG equations
\begin{eqnarray}
\bar{\beta}_{1}&=&0, \label{Eq:Beta1}
\\
\bar{\beta}_{2}&=&\frac{1}{2}\left(-g_{1}+3g_{2}+g_{4}+g_{5}+g_{3z}\right), \label{Eq:Beta2}
\\
\bar{\beta}_{3\bot}&=&\frac{1}{5}\left(-g_{1}+g_{2}+g_{4}-g_{5}+g_{3z}\right), \label{Eq:Beta3Bot}
\\
\bar{\beta}_{3z}&=&\frac{1}{10}\left(-g_{1}+g_{2}+g_{4}-g_{5}\right), \label{Eq:Beta3z}
\\
\bar{\beta}_{4}&=&0, \label{Eq:Beta4}
\\
\bar{\beta}_{5}&=&\frac{1}{2}\left(-g_{1}+g_{2}+g_{4}+3g_{5}-g_{3z}\right), \label{Eq:Beta5}
\\
\bar{\beta}_{6\bot}&=&\frac{1}{5}\left(-g_{1}-g_{2}+g_{4}+g_{5}-g_{3z}\right), \label{Eq:Beta6Bot}
\\
\bar{\beta}_{6z}&=&\frac{1}{10}\left(-g_{1}-g_{2}+g_{4}+g_{5}+g_{3z}\right), \label{Eq:Beta6z}
\\
\bar{\beta}_{7\bot}&=&\frac{3}{10}\left(-g_{1}-g_{2}-g_{4}-g_{5}+g_{3z}\right), \label{Eq:Beta7Bot}
\\
\bar{\beta}_{7z}&=&\frac{2}{5}\left(-g_{1}-g_{2}-g_{4}-g_{5}-g_{3z}\right), \label{Eq:Beta7z}
\\
\bar{\beta}_{8\bot}&=&\frac{3}{10}\left(-g_{1}+g_{2}-g_{4}+g_{5}-g_{3z}\right), \label{Eq:Beta8Bot}
\\
\bar{\beta}_{8z}&=&\frac{2}{5}\left(-g_{1}+g_{2}-g_{4}+g_{5}+g_{3z}\right), \label{Eq:Beta8z}
\\
\bar{\beta}_{S}&=&\frac{2}{5}\left(g_{1}-g_{2}+g_{4}+g_{5}-g_{3z}\right), \label{Eq:BetaS}
\\
\bar{\beta}_{op}&=&\frac{2}{5}\left(g_{1}+g_{2}+g_{4}-g_{5}+g_{3z}\right), \label{Eq:Betaop}
\\
\bar{\beta}_{V,1}&=&\frac{1}{5}\left(g_{1}+g_{2}-g_{4}+g_{5}+g_{3z}\right), \label{Eq:BetaV1}
\\
\bar{\beta}_{V,2}&=&\frac{1}{5}\left(g_{1}+g_{2}-g_{4}+g_{5}+g_{3z}\right), \label{Eq:BetaV2}
\\
\bar{\beta}_{V,3}&=&\frac{1}{2}\left(g_{1}+g_{2}-g_{4}+g_{5}-g_{3z}\right), \label{Eq:BetaV3}
\\
\bar{\beta}_{V,0}&=&\frac{1}{20}\left(g_{1}-g_{2}-g_{4}-g_{5}+g_{3z}\right), \label{Eq:BetaV0}
\end{eqnarray}
where
\begin{eqnarray}
\bar{\beta}_{X,Y}=\frac{d\ln(\Delta_{X,Y})}{d\ell}-1.
\end{eqnarray}
For convenience, we show the physical meaning
of different order parameters and corresponding fermion bilinears in Table~\ref{Table:PhysicalMeaning}.

\begin{table*}[htbp]
\caption{ Physical meaning of different order parameters and the corresponding fermion bilinears.
\label{Table:PhysicalMeaning}}
\begin{center}
\begin{ruledtabular}
\begin{tabular}{lll}
\toprule     Order Parameter        & Fermion Bilinear &  Physical meaning
\\ \hline
\midrule $\Delta_{1}$       & $\bar{\Psi}\gamma_{0}\Psi$ & chemical potential
\\
\midrule $\Delta_{2}$       & $\bar{\Psi}\Psi$ & scalar mass
\\
\midrule $\Delta_{3\bot}$       & $\sum\limits_{j=1,2}\bar{\Psi}\gamma_{0}\gamma_{j}\Psi$ & spin-orbit coupling within $xy$ plane
\\
\midrule $\Delta_{3z}$        & $\bar{\Psi}\gamma_{0}\gamma_{3}\Psi$ & spin-orbit coupling along $z$ axis
\\
\midrule $\Delta_{4}$        & $\bar{\Psi}\gamma_{0}\gamma_{5}\Psi$ & axial chemical potential
\\
\midrule $\Delta_{5}$        & $\bar{\Psi}i\gamma_{5}\Psi$ & pseudoscalar mass
\\
\midrule $\Delta_{6\bot}$        & $\bar{\Psi}\left(i\gamma_{2}\gamma_{3}+\gamma_{3}\gamma_{1}\right)\Psi$ & magnetization within $xy$ plane
\\
\midrule $\Delta_{6z}$        & $\bar{\Psi}i\gamma_{1}\gamma_{2}\Psi$ & magnetization along $z$ axis
\\
\midrule $\Delta_{7\bot}$        & $\sum\limits_{j=1,2}\bar{\Psi}i\gamma_{5}\gamma_{j}\Psi$ & axial magnetization within $xy$ plane
\\
\midrule $\Delta_{7z}$        & $\bar{\Psi}i\gamma_{5}\gamma_{3}\Psi$ & axial magnetization along $z$ axis
\\
\midrule $\Delta_{8\bot}$      & $\sum\limits_{j=1,2}\bar{\Psi}i\gamma_{j}\Psi$ & current within $xy$ plane
\\
\midrule $\Delta_{8z}$        & $\bar{\Psi}i\gamma_{3}\Psi$ & current along $z$ axis
\\
\midrule $\Delta_{S}$        & $\Psi^{\dag}i\gamma_{0}\gamma_{5}\gamma_{2}\Psi^{*}$ & $s$-wave paring
\\
\midrule $\Delta_{op}$        & $\Psi^{\dag}i\gamma_{0}\gamma_{2}\Psi^{*}$ & odd-parity pairing
\\
\midrule $\Delta_{V,1}$       & $\Psi^{\dag}\gamma_{3}\Psi^{*}$ & vector pairing along $x$ axis
\\
\midrule $\Delta_{V,2}$        & $\Psi^{\dag}i\gamma_{0}\gamma_{5}\Psi^{*}$ & vector  pairing along $y$ axis
\\
\midrule $\Delta_{V,3}$        & $\Psi^{\dag}\gamma_{1}\Psi^{*}$ & vector pairing along $z$ axis
\\
\midrule $\Delta_{V,0}$        & $\Psi^{\dag}i\gamma_{0}\gamma_{1}\gamma_{3}\Psi^{*}$ & temporal vector paring
\\
\bottomrule
\end{tabular}
\end{ruledtabular}
\end{center}
\end{table*}

\section{Numerical Results\label{App:NumResults}}

\subsection{Fixed points and their properties}

Solving the RG equations for $g_{a}$ as shown in Eqs.~(\ref{Eq:RGEg1})-(\ref{Eq:RGEg3z}), we obtained the real roots as following
\begin{widetext}
\begin{eqnarray}
\mathrm{FP0:}\qquad\left(g_{1}^{*},g_{2}^{*},g_{4}^{*},g_{5}^{*},g_{3z}^{*}\right)&=&\left(0, 0, 0, 0, 0\right)
\\
\mathrm{FP1:}\qquad\left(g_{1}^{*},g_{2}^{*},g_{4}^{*},g_{5}^{*},g_{3z}^{*}\right)&=&\left(0.152019, 1.25444, 0.459247, -0.561711, 0.0551435\right),
\\
\mathrm{FP2:}\qquad\left(g_{1}^{*},g_{2}^{*},g_{4}^{*},g_{5}^{*},g_{3z}^{*}\right)&=&\left(0.140905, -0.585585, 0.418385, 1.34668, 0.06996\right),
\\
\mathrm{FP3:}\qquad\left(g_{1}^{*},g_{2}^{*},g_{4}^{*},g_{5}^{*},g_{3z}^{*}\right)&=&\left(-0.100015, 0.575751, -0.61003, 0.775675, 0.199924\right),
\\
\mathrm{FP4:}\qquad\left(g_{1}^{*},g_{2}^{*},g_{4}^{*},g_{5}^{*},g_{3z}^{*}\right)&=&\left(-2.33263, 0., -0.610178, -1.72246, -1.72246\right),
\\
\mathrm{FP5:}\qquad\left(g_{1}^{*},g_{2}^{*},g_{4}^{*},g_{5}^{*},g_{3z}^{*}\right)&=&\left(0.126936, -0.463077, -0.854005, 0.769245, 1.23232\right),
\\
\mathrm{FP6:}\qquad\left(g_{1}^{*},g_{2}^{*},g_{4}^{*},g_{5}^{*},g_{3z}^{*}\right)&=&\left(0.0860014, 1.37623, 0.236822, -0.304132,0.0941005\right),
\\
\mathrm{FP7:}\qquad\left(g_{1}^{*},g_{2}^{*},g_{4}^{*},g_{5}^{*},g_{3z}^{*}\right)&=&\left(0.103947, -0.465334, 0.29293, 1.43995, 0.0944817\right),
\\
\mathrm{FP8:}\qquad\left(g_{1}^{*},g_{2}^{*},g_{4}^{*},g_{5}^{*},g_{3z}^{*}\right)&=&\left(-3.33745, -1.22097, -1.28072, -1.32395, -0.102973\right),
\\
\mathrm{FP9:}\qquad\left(g_{1}^{*},g_{2}^{*},g_{4}^{*},g_{5}^{*},g_{3z}^{*}\right)&=&\left(-2.68181, 1.06255, -1.97657, -1.35604, -2.41859\right),
\\
\mathrm{FP10:}\qquad\left(g_{1}^{*},g_{2}^{*},g_{4}^{*},g_{5}^{*},g_{3z}^{*}\right)&=&\left(0, 0, -1.25, 1.25, 1.25\right),
\\
\mathrm{FP11:}\qquad\left(g_{1}^{*},g_{2}^{*},g_{4}^{*},g_{5}^{*},g_{3z}^{*}\right)&=&\left(-5.16737, 0, -4.38982, -0.777544, -0.777544\right),
\end{eqnarray}
\end{widetext}
$\mathrm{FP0}$ is the trivial Gaussian fixed point. $\mathrm{FP1}$-$\mathrm{FP11}$ are non-trivial fixed points.

Expanding the RG equations (\ref{Eq:RGEg1})-(\ref{Eq:RGEg3z}) in the vicinity of a fixed point $\left(g_{1}^{*},g_{2}^{*},g_{4}^{*},g_{5}^{*},g_{3z}^{*}\right)$,
we find that
\begin{eqnarray}
\frac{dG}{d\ell}&=&MG,
\end{eqnarray}
where
\begin{eqnarray}
G=\left(
\begin{array}{c}
\delta g_{1}
\\
\delta g_{2}
\\
\delta g_{4}
\\
\delta g_{5}
\\
\delta g_{3z}
\end{array}
\right),
\end{eqnarray}
with $\delta g_{a}=g_{a}-g_{a}^{*}$.
The matrix $M$ is given by
\begin{eqnarray}
M=\left(
\begin{array}{ccccc}
M_{11} & M_{12} & M_{13} & M_{14} & M_{15}
\\
M_{21} & M_{22} & M_{23} & M_{24} & M_{25}
\\
M_{31} & M_{32} & M_{33} & M_{34} & M_{35}
\\
M_{41} & M_{42} & M_{43} & M_{44} & M_{45}
\\
M_{51} & M_{52} & M_{53} & M_{54} & M_{55}
\end{array}
\right),
\end{eqnarray}
where
\begin{eqnarray}
M_{11}&=&-\left(\frac{3}{2}+\frac{2}{5}g_{2}^{*}+\frac{1}{5}g_{4}^{*}+\frac{2}{5}g_{5}^{*}\right),
\\
M_{12}&=&-\frac{2}{5}\left(g_{1}^{*}+g_{5}^{*}\right),
\\
M_{13}&=&-\left(\frac{1}{5}g_{1}^{*}+\frac{2}{5}g_{3z}^{*}\right),
\\
M_{14}&=&-\frac{2}{5}\left(g_{1}^{*}+g_{2}^{*}+g_{3z}^{*}\right),
\\
M_{15}&=&-\frac{2}{5}\left(g_{4}^{*}+g_{5}^{*}\right),
\\
M_{21}&=&-\frac{3}{5}g_{2}^{*}-\frac{2}{5}g_{5}^{*},
\\
M_{22}&=&-\frac{3}{2}+2g_{2}^{*}-\frac{3}{5}g_{1}^{*}+\frac{4}{5}g_{4}^{*}+\frac{3}{5}g_{5}^{*}
+g_{3z}^{*},
\\
M_{23}&=&\frac{4}{5}g_{2}^{*}-g_{5}^{*}+\frac{7}{5}g_{3z}^{*},
\\
M_{24}&=&\frac{3}{5}g_{2}^{*}-\frac{2}{5}g_{1}^{*}-g_{4}^{*}+\frac{2}{5}g_{3z}^{*},
\\
M_{25}&=&g_{2}^{*}+\frac{7}{5}g_{4}^{*}+\frac{2}{5}g_{5}^{*},
\\
M_{31}&=&-\frac{2}{5}g_{1}^{*}+\frac{2}{5}g_{2}^{*}+\frac{2}{5}g_{5}^{*},
\\
M_{32}&=&-\frac{2}{5}g_{2}^{*}+\frac{2}{5}g_{1}^{*}-\frac{7}{5}g_{5}^{*}+g_{3z}^{*},
\\
M_{33}&=&-\frac{3}{2}-\frac{2}{5}g_{4}^{*}+\frac{2}{5}g_{3z}^{*},
\\
M_{34}&=&-\frac{2}{5}g_{5}^{*}+\frac{2}{5}g_{1}^{*}-\frac{7}{5}g_{2}^{*}-\frac{1}{5}g_{3z}^{*},
\\
M_{35}&=&-\frac{2}{5}g_{3z}^{*}+\frac{2}{5}g_{4}^{*}+g_{2}^{*}-\frac{1}{5}g_{5}^{*},
\\
M_{41}&=&\frac{2}{5}g_{1}^{*}-g_{5}^{*}-\frac{4}{5}g_{2}^{*}-\frac{1}{5}g_{4}^{*},
\\
M_{42}&=&\frac{2}{5}g_{2}^{*}+g_{5}^{*}-\frac{4}{5}g_{1}^{*}-\frac{4}{5}g_{4}^{*}-g_{3z}^{*},
\\
M_{43}&=&\frac{2}{5}g_{4}^{*}+g_{5}^{*}-\frac{1}{5}g_{1}^{*}-\frac{4}{5}g_{2}^{*}-\frac{8}{5}g_{3z}^{*},
\\
M_{44}&=&-\frac{3}{2}+\frac{12}{5}g_{5}^{*}-g_{1}^{*}+g_{2}^{*}+g_{4}^{*}-g_{3z}^{*},
\\
M_{45}&=&\frac{2}{5}g_{3z}^{*}-g_{5}^{*}-g_{2}^{*}-\frac{8}{5}g_{4}^{*},
\\
M_{51}&=&\frac{2}{5}g_{1}^{*}-\frac{1}{5}g_{3z}^{*}-\frac{2}{5}g_{2}^{*}-\frac{1}{5}g_{4}^{*}
-\frac{2}{5}g_{5}^{*},
\\
M_{52}&=&\frac{2}{5}g_{2}^{*}-\frac{4}{5}g_{3z}^{*}-\frac{2}{5}g_{1}^{*}
+\frac{1}{5}g_{4}^{*}+\frac{2}{5}g_{5}^{*},
\\
M_{53}&=&\frac{2}{5}g_{4}^{*}-\frac{6}{5}g_{3z}^{*}-\frac{1}{5}g_{1}^{*}
+\frac{1}{5}g_{2}^{*}+\frac{1}{5}g_{5}^{*},
\\
M_{54}&=&\frac{2}{5}g_{5}^{*}-\frac{3}{5}g_{3z}^{*}-\frac{2}{5}g_{1}^{*}+\frac{2}{5}g_{2}^{*}
+\frac{1}{5}g_{4}^{*},
\\
M_{55}&=&-\frac{3}{2}+\frac{4}{5}g_{3z}^{*}
-\frac{1}{5}g_{1}^{*}-\frac{4}{5}g_{2}^{*}-\frac{6}{5}g_{4}^{*}-\frac{3}{5}g_{5}^{*}.
\end{eqnarray}
From eigenvalues of $M$ at a fixed point $(g_{1}^{*}, g_{2}^{*}, g_{4}^{*}, g_{5}^{*},g_{3z}^{*})$, we can get the
properties of the fixed point. A negative (positive) eigenvalue is corresponding to a stable (unstable) eigendirection \cite{Szabo21, Maciejko14}.
For  quantum critical point (QCP),  bicritical point (BCP), and  tricritical point (TCP), there is/are one, two, and three unstable
direction(s) respectively. For a QCP, the correlation length exponent is determined by the inverse of the corresponding positive
eigenvalue.

\begin{table*}[htbp]
\caption{Eigenvalues of matrix $M$ at different fixed points
\label{Table:EigenvaluesM}}
\begin{center}
\begin{ruledtabular}
\begin{tabular}{cccccccccccc}
\toprule            FP0 &  FP1 & FP2 & FP3 & FP4 & FP5 & FP6 & FP7 & FP8 & FP9 & FP10 & FP11
\\ \hline
-1.5 & -3.12277 & -3.05426 & -2.11269  & -4.19874 & -2.50996 & -2.98441 & -2.99224 & -5.70333 & -5.29866 & -2.25 & -9.30126
\\
-1.5 & -2.7634 & -2.48233 & -1.42791  & -2.79748 & -2.05542 & -2.77106 & -2.49091 & -3.26902 & -3.36509 & -2.25 & -1.69424
\\
-1.5 & -1.77432 & -1.76197 & -1.26511  & -2.46863 & -1.41052 & -1.80787 & -1.77301 & -1.46547 & -1.38636 & -1.47474 & 1.5
\\
-1.5 & -0.412398 & -0.231803 & -1.18664  & -0.848268 & -1.06884 & 0.390411 & 0.224733 & 1.5 & 1.5 & 0.974745 & 3.33999
\\
-1.5 & 1.5 & 1.5 & 1.5  & 1.5 & 1.5 & 1.5 & 1.5 & 2.47601 & 2.17557 & 1.5 & 5.46863
\\
\bottomrule
\end{tabular}
\end{ruledtabular}
\end{center}
\end{table*}

Substituting the values of $g_{a}^{*}$ at each fixed point into the expression $M$, we calculate the corresponding eigenvalues
of $M$. The eigenvalues for the fixed points are shown in Table~\ref{Table:EigenvaluesM}.
For FP0, the eigenvalues of $M$ are always negative, thus FP0 is a stable fixed point. We can find that there is one positive
eigenvalue for FP1, FP2, FP3, FP4, and FP5, and there are two positive eigenvalues for FP6, FP7, FP8, FP9, and FP10, and three positive eigenvalues for
FP11. Thus, FP1, FP2, FP3, FP4, and FP5 are QCPs, FP6, FP7, FP8, FP9, and FP10 are BCPs, and FP11 is a TCP.

It is easy to find that the correlation length exponent at the QCPs FP1, FP2, FP3, FP4 and FP5 all
satisfy
\begin{eqnarray}
\nu^{-1}&=&1.5.
\end{eqnarray}

Substituting the values of
$g_{a}^{*}$ with $i=1,2,4,5,3z$ into Eqs.~(\ref{Eq:Beta1})-(\ref{Eq:BetaV0}), we can get values of $\bar{\beta}_{X,Y}$ for different
$\Delta_{X,Y}$, which are shown in Tabel~\ref{Table:BetaDeltaXFPs}.  For a QCP, the largest value of $\beta_{X,Y}$ is marked by
the bold style. It represents that the fixed point is a QCP to the new state in which $\Delta_{X,Y}$ acquires finite value.
FP1, FP2, FP4, and FP5 are corresponding to QCPs to a state in which $\Delta_{2}$, $\Delta_{5}$, $\Delta_{7z}$ and $\Delta_{8z}$ acquire finite
value respectively. For FP3, it stands for a QCP to a state in which both $\Delta_{2}$ and $\Delta_{5}$ become finite generally.
This state represents an axionic insulator whose order parameter can be written as
$\langle\bar{\Psi}\left(\cos(\theta)+i\gamma_{5}\sin(\theta)\right)\Psi\rangle$ \cite{Roy16}.

\begin{table*}[htbp]
\caption{$\beta_{X,Y}$ at different fixed points. The largest value at a QCP is marked by the bold style. Notice that FP1, FP2, FP3, FP4, FP5 are QCPs.
\label{Table:BetaDeltaXFPs}}
\begin{center}
\begin{ruledtabular}
\begin{tabular}{lccccccccccc}
\toprule             & FP1 & FP2 & FP3 & FP4 & FP5 & FP6 & FP7 & FP8 & FP9 & FP10 & FP11
\\ \hline
\midrule $\bar{\beta}_{1}$        & 0 & 0 & 0  & 0 & 0 & 0 & 0 & 0 & 0 & 0 & 0
\\
\midrule $\bar{\beta}_{2}$        & \textbf{1.78199} & -0.0313166 & \textbf{1.09642}  & -0.861228 & -0.184302 & 2.03474 & 0.163708 & -1.51656 & 0.0591369 & 0.625 & -0.388772
\\
\midrule $\bar{\beta}_{3\bot}$    & 0.435705 & -0.316965 & -0.102003  & 0.344491 & -0.196188 & 0.385057 & -0.324365 & 0.411346 & 0.141049 & -0.25 & 0.155509
\\
\midrule $\bar{\beta}_{3z}$       & 0.212338 & -0.165478 & -0.070994  & 0.344491 & -0.221326 & 0.183119 & -0.171631 & 0.21597 & 0.312383 & -0.25 & 0.155509
\\
\midrule $\bar{\beta}_{4}$        & 0 & 0 & 0  & 0 & 0 & 0 & 0 & 0 & 0 & 0 & 0
\\
\midrule $\bar{\beta}_{5}$        & -0.0893033 & \textbf{1.83099} & \textbf{1.09642}  & -0.861228 & -0.184302 & 0.260278 & 1.97451 & -1.51656 & 0.0591369 & 0.625 & -0.388772
\\
\midrule $\bar{\beta}_{6\bot}$    & -0.312814 & 0.427957 & -0.102003  & 0.344491 & -0.196188 & -0.324729 & 0.399958 & 0.411346 & 0.141049 & -0.25 & 0.155509
\\
\midrule $\bar{\beta}_{6z}$       & -0.145378 & 0.227971 & -0.0110167  & -0.172246 & 0.14837 & -0.143544 & 0.218875 & 0.185078 & -0.413193 & 0.125 & -0.0777544
\\
\midrule $\bar{\beta}_{7\bot}$    & -0.374656 & -0.375128 & -0.132437  & 0.882843 & 0.495967 & -0.390247 & -0.383104 & 2.11804 & 0.759982 & 0.375 & 2.86716
\\
\midrule $\bar{\beta}_{7z}$       & -0.543656 & -0.556138 & -0.336522  & \textbf{2.55509} & -0.324568 & -0.59561 & -0.586391 & 2.90643 & 2.94818 & -0.5 & 4.44491
\\
\midrule $\bar{\beta}_{8\bot}$    & 0.00789679 & 0.0395537 & 0.558464  & 0.882843 & -0.0597252 & 0.196553 & 0.144978 & 0.652867 & 2.03504 & 0.375 & 2.86716
\\
\midrule $\bar{\beta}_{8z}$       & 0.0546439 & 0.108706 & 0.904558 & -0.20084 & \textbf{0.906224} & 0.337352 & 0.26889 & 0.788112 & 0.778521 & 1.5 & 3.20084
\\
\midrule $\bar{\beta}_{S}$        & -0.504013 & 0.968638 & -0.284018  & -1.17712 & -0.290828 & -0.580657 & 0.883073 & -1.84727 & -1.86335 & -0.5 & -3.82288
\\
\midrule $\bar{\beta}_{op}$       & 0.993025 & -0.521206 & -0.284018  & -1.17712 & -0.290828 & 0.838916 & -0.565572 & -1.84727 & -1.86335 & -0.5 & -3.82288
\\
\midrule $\bar{\beta}_{V,1}$      & 0.0881294 & 0.110715 & 0.412273  & -1.03347 & 0.503886 & 0.203076 & 0.176024 & -0.940925 & -0.683464 & 0.75 & -0.466527
\\
\midrule $\bar{\beta}_{V,2}$      & 0.0881294 & 0.110715 & 0.412273  & -1.03347& 0.503886 & 0.203076 & 0.176024 & -0.940925 & -0.683464 & 0.75 & -0.466527
\\
\midrule $\bar{\beta}_{V,3}$      & 0.16518 & 0.206828 & 0.830758  & -0.861228& 0.027394 & 0.41359 & 0.345578 & -2.24934 & 0.709928 & 0.625 & -0.388772
\\
\midrule $\bar{\beta}_{V,0}$      & -0.0472408 & -0.0484308 & -0.0320743  & -0.0861228 & 0.0953547 & -0.0564411 & -0.053456 & 0.019261 & -0.141517 & 0.0625 & -0.0388772
\\
\bottomrule
\end{tabular}
\end{ruledtabular}
\end{center}
\end{table*}

\section{Interplay of four-fermion interaction and long-range Coulomb interaction \label{App:InterplayCoulomb}}

The Coulomb interaction between fermions can be described by the coupling between fermion field $\Psi$ and boson field $\phi$
as the following action
\begin{eqnarray}
S_{\psi\phi}
&=&i\lambda\int\frac{d\omega_{1}}{2\pi}\frac{d^3\mathbf{k}_{1}}{(2\pi)^{3}}\frac{d\omega_{2}}{2\pi}
\frac{d^3\mathbf{k}_{2}}{(2\pi)^{3}}\bar{\Psi}(\omega_{1},\mathbf{k}_{1})\gamma_{0}\Psi(\omega_{2},\mathbf{k}_{2})\nonumber
\\
&&\times\phi(\omega_{1}-\omega_{2},\mathbf{k}_{1}
-\mathbf{k}_{2}),
\end{eqnarray}
where $\lambda=\frac{e}{\sqrt{\epsilon}}$ with $e$ the elementary charge and $\epsilon$ the dielectric constant.
The free action of $\phi$ is given by
\begin{eqnarray}
S_{\phi}^{0}=\int\frac{d\omega}{2\pi}\frac{d^{3}\mathbf{k}}{(2\pi)^{3}}\phi(\omega,\mathbf{k})
\left(\frac{1}{\sqrt{\eta}}k_{\bot}^{2}+\sqrt{\eta} k_{z}^{2}
\right)\phi(\omega,\mathbf{k}).
\end{eqnarray}

\begin{figure}[htbp]
\center
\includegraphics[width=3.1in]{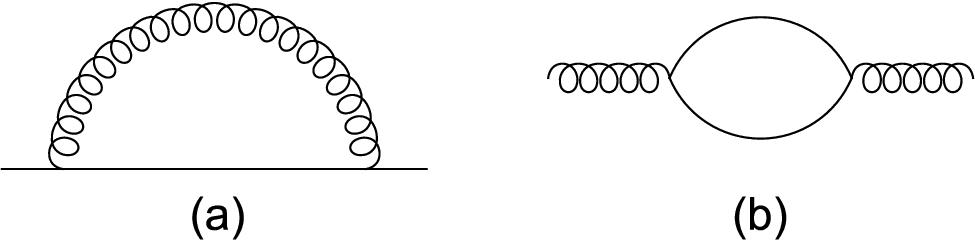}
\caption{(a) Feynman diagram for the self-energy  of fermions induced by long-range Coulomb interaction; (b) Feynman diagram for
self-energy of boson field. The solid line represents the fermion propagator, and spiral line stands for the boson field which is equivalent to
the long-range Coulomb interaction.\label{Fig:FermionBosonSelfEnergyCoulomb}}
\end{figure}

\subsection{Interaction Corrections related to Coulomb interaction}

\subsubsection{Fermion self-energy induced by Coulomb interaction}

As the shown in Fig.~\ref{Fig:FermionBosonSelfEnergyCoulomb}(a), the fermion self-energy induced by long-range Coulomb interaction is given by
\begin{eqnarray}
\Sigma_{C}(i\omega,\mathbf{k})&=&-\lambda^2\int_{-\infty}^{+\infty}\frac{d\Omega}{2\pi}\int'\frac{d^3\mathbf{q}}{(2\pi)^3}
\gamma_{0}G_{0}(i\Omega,\mathbf{q})\gamma_{0}\nonumber
\\
&&\times D_{0}(i\omega-i\Omega,\mathbf{k}-\mathbf{q}), \label{Eq:FermionSelfEnergyCoulomb}
\end{eqnarray}
where
\begin{eqnarray}
D_{0}(i\Omega,\mathbf{q})&=&\frac{\sqrt{\eta}}{q_{\bot}^{2}+\eta q_{3}^{2}}. \label{Eq:BosonPropagator}
\end{eqnarray}
Substituting Eqs.~(\ref{Eq:FermionPropagator}) and (\ref{Eq:BosonPropagator}) into Eq.~(\ref{Eq:FermionSelfEnergyCoulomb}), we obtain
\begin{eqnarray}
\Sigma_{C}(i\omega,\mathbf{k})
&=&-iv\left(k_{1}\gamma_{1}+k_{2}\gamma_{2}\right)\Sigma_{C,\bot}\nonumber
\\
&&-iAk_{3}^{2}\gamma_{3}\Sigma_{C,3},
\end{eqnarray}
where
\begin{eqnarray}
\Sigma_{C,\bot}&=&\frac{\lambda^2\sqrt{\eta}}{4\pi^{2}}\int'dq_{\bot}d|q_{3}|q_{\bot}
\frac{q_{\bot}^{2}}{\sqrt{v^{2}q_{\bot}^{2}
+A^{2}q_{3}^{4}}}\nonumber
\\
&&\times\frac{1}{\left(q_{\bot}^{2}+\eta
q_{3}^{2}\right)^{2}},
\\
\Sigma_{C,3}&=&\frac{\lambda^2\eta^{\frac{3}{2}}}{4\pi^{2}}\int'dq_{\bot}d|q_{3}|q_{\bot}
\frac{q_{3}^{2}\left(-q_{\bot}^{2}+3\eta q_{3}^{2}\right)}{\sqrt{v^{2}q_{\bot}^{2}
+A^{2}q_{3}^{4}}}\nonumber
\\
&&\times\frac{1}{\left(q_{\bot}^{2}+\eta
q_{3}^{2}\right)^{3}}.
\end{eqnarray}
A constant term that does not depend on energy and momenta has been discarded.

Utilizing the transformations Eqs.~(\ref{Eq:TransformationA})-(\ref{Eq:TransformationC}) and carrying out the
integrations of $E$ and $\delta$ within the ranges $b\Lambda<E<\Lambda$ and $0<\delta<+\infty$, we get
\begin{eqnarray}
\Sigma_{C,\bot}\approx C_{1}\ell,\qquad
\Sigma_{C,3}=C_{2}\ell,
\end{eqnarray}
where
\begin{eqnarray}
C_{1}&=&\frac{\lambda^2\zeta^{\frac{3}{2}}}{8\pi^{2}v}
\int_{0}^{+\infty}d\delta\frac{1}{\sqrt{\delta}\left(1+\delta^2\right)^{\frac{1}{4}}}\nonumber
\\
&&\times\frac{1}{\left(\zeta+
\delta \left(1+\delta^2\right)^{\frac{1}{2}}\right)^{2}},
\\
C_{2}&=&\frac{\lambda^2\zeta^{\frac{1}{2}}}{8\pi^{2}v}
\int_{0}^{+\infty}d\delta\sqrt{\delta}\left(1+\delta^2\right)^{\frac{1}{4}}\nonumber
\\
&&\times\frac{\left(-\zeta+3\delta \left(1+\delta^2\right)^{\frac{1}{2}}\right)}{\left(\zeta+
\delta\left(1+\delta^2\right)^{\frac{1}{2}}\right)^{3}},
\end{eqnarray}
with  $\zeta=\frac{A\Lambda}{v^{2}\eta}$.

\subsubsection{Boson self-energy}

As depicted in Fig.~\ref{Fig:FermionBosonSelfEnergyCoulomb}(b), the boson self-energy is given by
\begin{eqnarray}
\Pi(i\Omega,\mathbf{q})&=&-\lambda^{2}\int\frac{d\omega}{2\pi}\int'\frac{d^3\mathbf{k}}{(2\pi)^{3}}
\mathrm{Tr}\left[\gamma_{0}G_{0}(i\omega,\mathbf{k})\gamma_{0}\right.\nonumber
\\
&&\left.\times G_{0}(i\omega+i\Omega,\mathbf{k}+\mathbf{q})\right]. \label{Eq:BosonSelfEnergy}
\end{eqnarray}
Substituting Eq.~(\ref{Eq:FermionPropagator})  into Eq.~(\ref{Eq:BosonSelfEnergy}) and expanding to quadratic order of $\Omega$ and $q_{i}$,
we arrive
\begin{eqnarray}
\Pi(i\Omega,\mathbf{q})
&=&\lambda^{2}v^{2}q_{\bot}^{2}\frac{1}{8\pi^{2}}\int' dk_{\bot}d|k_{3}|k_{\bot}\left(\frac{2}{E_{\mathbf{k}}^{3}}-\frac{v^{2}k_{\bot}^{2}}{E_{\mathbf{k}}^{5}}\right)\nonumber
\\
&&+\lambda^{2}v^{2}A^{2}q_{3}^{2}\frac{1}{\pi^{2}}\int' dk_{\bot}d|k_{3}|k_{\bot}\frac{k_{3}^{2}k_{\bot}^{2}}{E_{\mathbf{k}}^{5}}.
\end{eqnarray}
Employing the transformations Eqs.~(\ref{Eq:TransformationA})-(\ref{Eq:TransformationC}) and performing the
integrations of $E$ and $\delta$, $\Pi$ can be expressed as
\begin{eqnarray}
\Pi(i\Omega,\mathbf{q})
&=&C_{\bot}q_{\bot}^{2}\ell+C_{z}q_{3}^{2}\ell,
\end{eqnarray}
where
\begin{eqnarray}
C_{\bot}&=&\frac{3\lambda^{2}}{20\pi^{2} \sqrt{A}\sqrt{\Lambda}},
\\
C_{z}&=&\frac{4\lambda^{2}\sqrt{A}\Lambda^{\frac{1}{2}}}{21\pi^{2} v^{2}}.
\end{eqnarray}

\subsubsection{Corrections to fermion-boson coupling}

\begin{figure}[htbp]
\center
\includegraphics[width=3.1in]{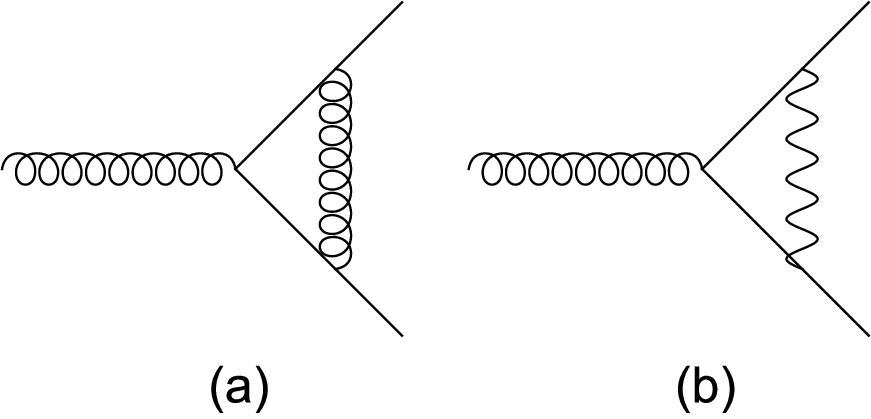}
\caption{Feynman diagrams for the vertex corrections to fermion-boson coupling
 due to (a) long-range Coulomb interaction and (b) four-fermion interaction.\label{Fig:FermionBosonVertexCorrection}}
\end{figure}

As displayed in Fig.~\ref{Fig:FermionBosonVertexCorrection}(a), the correction to fermion-boson coupling induced by Coulomb interaction takes the form
\begin{eqnarray}
V_{C}^{(1)}
&=&-i\lambda^{3}\int'\frac{d\Omega}{2\pi}\frac{d^3\mathbf{q}}{(2\pi)^{3}}\gamma_{0}
G_{0}(i\Omega,\mathbf{q})\gamma_{0}G_{0}(i\Omega,\mathbf{q})\gamma_{0}\nonumber
\\
&&\times D_{0}(i\Omega,\mathbf{q}). \label{Eq:CoulombVertexCorrectionA}
\end{eqnarray}
Substituting Eqs.~(\ref{Eq:FermionPropagator}) and (\ref{Eq:BosonPropagator}) into Eq.~(\ref{Eq:CoulombVertexCorrectionA}), we find
\begin{eqnarray}
V_{C}^{(1)}
&=&-i\lambda^{3}\gamma_{0}\int'\frac{d^3\mathbf{q}}{(2\pi)^{3}}\int_{-\infty}^{+\infty}
\frac{d\Omega}{2\pi}
\frac{-\Omega^{2}+E_{\mathbf{q}}^{2}}{\left(\Omega^2+E_{\mathbf{q}}^{2}\right)^{2}}\nonumber
\\
&&\times\frac{\sqrt{\eta}}{q_{\bot}^{2}+\eta q_{z}^{2}}\nonumber
\\
&=&0,
\end{eqnarray}
which means
\begin{eqnarray}
\delta\lambda^{(1)}=0.
\end{eqnarray}
As presented in Fig.~\ref{Fig:FermionBosonVertexCorrection}(b), the correction to fermion-boson coupling generated  by four-fermion interactions can be written as
\begin{eqnarray}
V_{C}^{(2)}
&=&i\lambda^{3}\sum_{a=1,2,4,5,3z}g_{a}^{2}\int'\frac{d\Omega}{2\pi}\frac{d^3\mathbf{q}}{(2\pi)^{3}}\Gamma_{a}
G_{0}(i\Omega,\mathbf{q})\gamma_{0}\nonumber
\\
&&\times G_{0}(i\Omega,\mathbf{q})\Gamma_{a}. \label{Eq:CoulombVertexCorrectionB}
\end{eqnarray}
Substituting Eq.~(\ref{Eq:FermionPropagator}) into Eq.~(\ref{Eq:CoulombVertexCorrectionB}), one can obtain
\begin{eqnarray}
V_{C}^{(2)}
&=&i\lambda^{3}\sum_{a=1,2,4,5,3z}g_{a}^{2}\int'\frac{d^3\mathbf{q}}{(2\pi)^{3}}\int_{-\infty}^{+\infty}
\frac{d\Omega}{2\pi}
\Gamma_{a}\gamma_{0}\nonumber
\\
&&\times\frac{-\Omega^{2}+E_{\mathbf{q}}^{2}}{\left(\Omega^2+E_{\mathbf{q}}^{2}\right)^{2}}\Gamma_{a}
=0.
\end{eqnarray}
Thus, $\delta\lambda^{(2)}$ is given by
\begin{eqnarray}
\delta\lambda^{(2)}=0.
\end{eqnarray}
The total correction to fermion-boson coupling is
\begin{eqnarray}
\delta\lambda&=&\delta\lambda^{(1)}+\delta\lambda^{(2)}=0.
\end{eqnarray}

\subsubsection{Corrections to four-fermion couplings induced by long-range Coulomb interaction}

\begin{figure}[htbp]
\center
\includegraphics[width=3.3in]{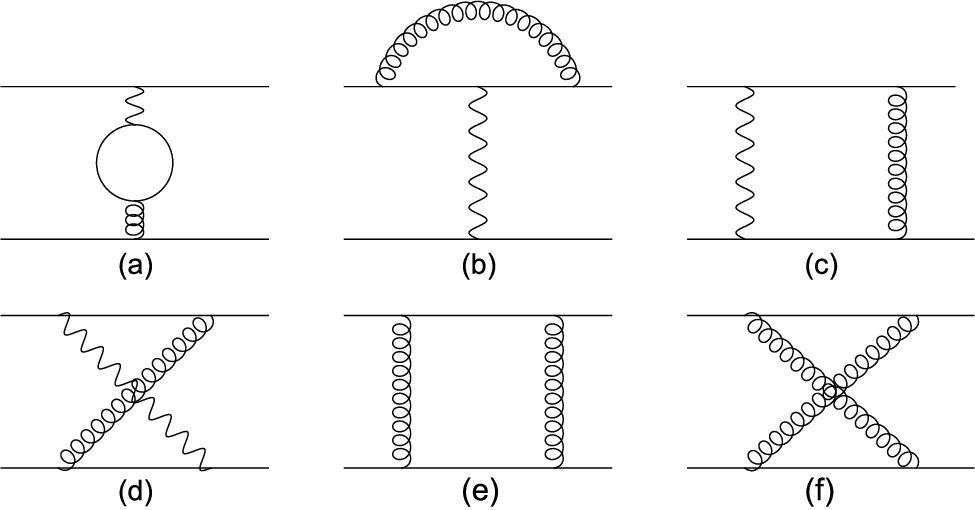}
\caption{Feynman diagrams for the vertex corrections to four-fermion interaction
 induced by long-range Coulomb interaction.\label{Fig:VertexCorrectionDueToCoulomb}}
\end{figure}

The correction from Fig.~\ref{Fig:VertexCorrectionDueToCoulomb}(a) is
\begin{eqnarray}
V_{a}^{(5)}&=&2\lambda^{2}g_{a}\left(\bar{\Psi}\Gamma_{a}\Psi\right)^{2}\int'\frac{d\omega}{2\pi}\frac{d^3\mathbf{k}}{(2\pi)^{3}}
\mathrm{Tr}\left[\gamma_{0}G_{0}(i\omega,\mathbf{k})\Gamma_{a}\right.\nonumber
\\
&&\left.\times G_{0}(i\omega+i\Omega,\mathbf{k}+\mathbf{q})\right]D_{0}(i\Omega,\mathbf{q}). \label{Eq:FifthDiagramExpression}
\end{eqnarray}
Fig.~\ref{Fig:VertexCorrectionDueToCoulomb}(b) leads to the correction
\begin{eqnarray}
V_{a}^{(6)}
&=&-4\lambda^{2}g_{a}\left(\bar{\Psi}\Gamma_{a}\Psi\right)\int'\frac{d\Omega}{2\pi}\frac{d^3\mathbf{q}}{(2\pi)^{3}}
\left(\bar{\Psi}\gamma_{0}G_{0}(i\Omega,\mathbf{q})\Gamma_{a}\right.\nonumber
\\
&&\left.\times G_{0}(i\Omega,\mathbf{q})\gamma_{0}\Psi\right)
D_{0}(i\Omega,\mathbf{q}). \label{Eq:SixthDiagramExpression}
\end{eqnarray}
The correction from Figs.~\ref{Fig:VertexCorrectionDueToCoulomb}(c) and \ref{Fig:VertexCorrectionDueToCoulomb}(d) takes the form
\begin{eqnarray}
V_{a}^{(7)+(8)}
&=&-4\lambda^{2}g_{a}\int'\frac{d\Omega}{2\pi}\frac{d^3\mathbf{q}}{(2\pi)^{3}}\left(\bar{\Psi}\Gamma_{a}
G_{0}(i\Omega,\mathbf{q})\gamma_{0}\Psi\right)\nonumber
\\
&&\times\left\{\bar{\Psi}\left[\gamma_{0}G_{0}(i\Omega,\mathbf{q})\Gamma_{a}\right.\right.\nonumber
\\
&&\left.\left.+\Gamma_{a}G_{0}(-i\Omega,-\mathbf{q})\gamma_{0}\right]
\Psi\right\}D_{0}(i\Omega,\mathbf{q}). \label{Eq:SeventhEighthDiagramExpression}
\end{eqnarray}
Figs.~\ref{Fig:VertexCorrectionDueToCoulomb}(e) and \ref{Fig:VertexCorrectionDueToCoulomb}(f) generate the correction
\begin{eqnarray}
V^{(9)+(10)}
&=&4\lambda^{4}\int'\frac{d\Omega}{2\pi}\frac{d^3\mathbf{q}}{(2\pi)^{3}}\left(\bar{\Psi}\gamma_{0}G_{0}\left(i\Omega,\mathbf{q}\right)\gamma_{0}
\Psi\right)\nonumber
\\
&&\times D_{0}(i\Omega,\mathbf{q})\left\{\bar{\Psi}\left[\gamma_{0}G_{0}\left(i\Omega,\mathbf{q}\right)\gamma_{0}\right.\right.\nonumber
\\
&&\left.\left.+\gamma_{0}G_{0}(-i\Omega,-\mathbf{q})\right]\gamma_{0}\Psi\right\}
D_{0}(i\Omega,\mathbf{q}). \label{Eq:NinethTenthDiagramExpression}
\end{eqnarray}

Substituting Eqs.~(\ref{Eq:FermionPropagator}) and (\ref{Eq:BosonPropagator}) into Eq.~(\ref{Eq:FifthDiagramExpression}), we get
\begin{eqnarray}
V_{a}^{(5)}&=&\delta g_{a}^{(5)}\left(\bar{\Psi}\Gamma_{a}\Psi\right)^{2},
\end{eqnarray}
where
\begin{eqnarray}
\delta g_{1}^{(5)}&=&-2g_{1}
\left(\sqrt{\eta}C_{\bot}+\frac{C_{z}}{\sqrt{\eta}}\right)\ell,
\\
\delta g_{2}^{(5)}&=&0,
\\
\delta g_{4}^{(5)}&=&0,
\\
\delta g_{5}^{(5)}&=&0,
\\
\delta g_{3z}^{(5)}&=&0.
\end{eqnarray}

Substituting Eqs.~(\ref{Eq:FermionPropagator}) and (\ref{Eq:BosonPropagator}) into Eq.~(\ref{Eq:SixthDiagramExpression}), we arrive
\begin{eqnarray}
V_{a}^{(6)}&=&\delta g_{a}^{(6)}\left(\bar{\Psi}\Gamma_{a}\Psi\right)^{2},
\end{eqnarray}
where
\begin{eqnarray}
\delta g_{1}^{(6)}
&=&0,
\\
\delta g_{2}^{(6)}
&=&g_{2}C_{3}\ell,
\\
\delta g_{4}^{(6)}
&=&0,
\\
\delta g_{5}^{(6)}
&=&g_{5}C_{3}\ell,
\\
\delta g_{3z}^{(6)}
&=&-
g_{3z}C_{4}\ell,
\end{eqnarray}
with
\begin{eqnarray}
C_{3}&=&\frac{\lambda^{2}\sqrt{\zeta}}{2\pi^{2}v}
\int_{0}^{+\infty}d\delta\nonumber
\\
&&\times\frac{1}{\sqrt{\delta}\left(1+\delta^2\right)^{\frac{1}{4}}}
\frac{1}{\zeta
+\delta \left(1+\delta^2\right)^{\frac{1}{2}}}, \label{Eq:CoffC3}
\\
C_{4}&=&\frac{\lambda^{2}\sqrt{\zeta}}{2\pi^{2}v}
\int_{0}^{+\infty}d\delta\frac{\delta^{\frac{3}{2}}}{\left(1+\delta^2\right)^{\frac{5}{4}}}
\frac{1}{\zeta+\delta \left(1+\delta^2\right)^{\frac{1}{2}}}. \label{Eq:CoffC4}
\end{eqnarray}

Substituting Eqs.~(\ref{Eq:FermionPropagator}) and (\ref{Eq:BosonPropagator}) into Eq.~(\ref{Eq:SeventhEighthDiagramExpression}), $V_{a}^{(7)+(8)}$ can
be written as
\begin{eqnarray}
V_{1}^{(7)+(8)}
&=&-\left(\bar{\Psi}i\gamma_{3}\Psi\right)^{2}
g_{1}C_{4}\ell,
\\
V_{2}^{(7)+(8)}
&=&\sum_{j=1}^{2}\left(\bar{\Psi}
\gamma_{0}\gamma_{j}\Psi\right)^{2}g_{2}C_{5}\ell,
\\
V_{4}^{(7)+(8)}
&=&-\left(\bar{\Psi}i\gamma_{5}\gamma_{3}\Psi\right)^{2}
g_{4}C_{4}\ell,
\\
V_{5}^{(7)+(8)}
&=&-\sum_{<<lk>>}\left(\bar{\Psi}
\gamma_{l}\gamma_{k}\Psi\right)^{2}g_{5}C_{5}\ell,
\\
V_{3z}^{(7)+(8)}
&=&0,
\end{eqnarray}
where
\begin{eqnarray}
C_{5}&=&\frac{\lambda^{2}\sqrt{\zeta}}{4\pi^{2}v}
\int_{0}^{+\infty}d\delta\frac{1}{\sqrt{\delta}\left(1+\delta^2\right)^{\frac{5}{4}}}\nonumber
\\
&&\times\frac{1}{\zeta+\delta\left(1+\delta^2\right)^{\frac{1}{2}} }.
\label{Eq:CoffC5}
\end{eqnarray}
Through the relations Eqs.~(\ref{Eq:semiDSMFFCouplingRelation1}), (\ref{Eq:semiDSMFFCouplingRelation2}), (\ref{Eq:semiDSMFFCouplingRelation5}),
and (\ref{Eq:semiDSMFFCouplingRelation7}), we obtain
\begin{eqnarray}
V_{1}^{(7)+(8)}
&=&\left[\left(\bar{\Psi}\gamma_{0}\gamma_{5}\Psi\right)^{2}
-\left(\bar{\Psi}i\gamma_{5}\Psi\right)^{2}
-\left(\bar{\Psi}\gamma_{0}\gamma_{3}\Psi\right)^{2}\right]\nonumber
\\
&&\times g_{1}C_{4}\ell,
\\
V_{2}^{(7)+(8)}
&=&\left[-\left(\bar{\Psi}\gamma_{0}\Psi\right)^{2}
+\left(\bar{\Psi}\Psi\right)^{2}
+\left(\bar{\Psi}\gamma_{0}\gamma_{5}\Psi\right)^{2}\right.\nonumber
\\
&&\left.-2\left(\bar{\Psi}i\gamma_{5}\Psi\right)^{2}
-\left(\bar{\Psi}\gamma_{0}\gamma_{3}\Psi\right)^{2}\right]g_{2}C_{5}\ell,
\\
V_{4}^{(7)+(8)}
&=&\left[\left(\bar{\Psi}\gamma_{0}\Psi\right)^{2}
+\left(\bar{\Psi}i\gamma_{5}\Psi\right)^{2}
+\left(\bar{\Psi}\gamma_{0}\gamma_{3}\Psi\right)^{2}\right]\nonumber
\\
&&\times g_{4}C_{4}\ell,
\\
V_{5}^{(7)+(8)}
&=&\left[\left(\bar{\Psi}\gamma_{0}\Psi\right)^{2}
+\left(\bar{\Psi}\Psi\right)^{2}
-\left(\bar{\Psi}\gamma_{0}\gamma_{5}\Psi\right)^{2}\right.\nonumber
\\
&&\left.+\left(\bar{\Psi}\gamma_{0}\gamma_{3}\Psi\right)^{2}\right]g_{5}C_{5}\ell.
\end{eqnarray}
Thus, the total correction from Figs.~\ref{Fig:VertexCorrectionDueToCoulomb}(c) and  \ref{Fig:VertexCorrectionDueToCoulomb}(d) can be expressed as
\begin{eqnarray}
V^{(7)+(8)}&=&\sum_{a=1,2,4,5,3z}V_{a}^{(7)+(8)}\nonumber
\\
&=&\sum_{a=1,2,4,5,3z}\delta g_{a}^{(7)+(8)}\left(\bar{\Psi}\Gamma_{a}\Psi\right)^{2},
\end{eqnarray}
where
\begin{eqnarray}
\delta g_{1}^{(7)+(8)}&=&\left(-g_{2}C_{5}+g_{2}C_{4}+g_{5}C_{5}\right)\ell,
\\
\delta g_{2}^{(7)+(8)}&=&\left(g_{2}C_{5}+g_{5}C_{5}\right)\ell,
\\
\delta g_{4}^{(7)+(8)}&=&\left(g_{1}C_{4}+g_{2}C_{5}-g_{5}C_{5}\right)\ell,
\\
\delta g_{5}^{(7)+(8)}&=&\left(-g_{1}C_{4}-2g_{2}C_{5}+g_{4}C_{4}\right)\ell,
\\
\delta g_{3z}^{(7)+(8)}&=&\left(-g_{1}C_{4}-g_{2}C_{5}+g_{4}C_{4}+g_{5}C_{5}\right)\ell.
\end{eqnarray}

Substituting Eqs.~(\ref{Eq:FermionPropagator}) and (\ref{Eq:BosonPropagator}) into Eq.~(\ref{Eq:NinethTenthDiagramExpression}), one can get
\begin{eqnarray}
V^{(9)+(10)}
&=&\left(\bar{\Psi}
i\gamma_{3}\Psi\right)^{2}\frac{\pi^{2} v^{2}A^{\frac{1}{2}}}{\Lambda^{\frac{3}{2}}}C_{6}\ell,
\end{eqnarray}
where
\begin{eqnarray}
C_{6}&=&\frac{\lambda^{4} \zeta}{2\pi^{4}v^{2}}
\int_{0}^{+\infty}d\delta\frac{\delta^{\frac{3}{2}}}{\left(1+\delta^2\right)^{\frac{1}{4}}}\nonumber
\\
&&\times\frac{1}{\left[\zeta+ \delta\left(1+\delta^2\right)^{\frac{1}{2}} \right]^{2}}. \label{Eq:CoffC6}
\end{eqnarray}
Using Eq.~(\ref{Eq:semiDSMFFCouplingRelation7}),
we can get
\begin{eqnarray}
V_{3z}^{(9)+(10)}
&=&\left[-\left(\bar{\Psi}\gamma_{0}\gamma_{5}\Psi\right)^{2}
+\left(\bar{\Psi}i\gamma_{5}\Psi\right)^{2}\right.\nonumber
\\
&&\left.+\left(\bar{\Psi}\gamma_{0}\gamma_{3}\Psi\right)^{2}\right]\frac{\pi^{2} v^{2}A^{\frac{1}{2}}}{\Lambda^{\frac{3}{2}}}C_{6}\ell.
\end{eqnarray}
It indicates that
\begin{eqnarray}
\delta g_{1}^{(9)+(10)}&=&0,
\\
\delta g_{2}^{(9)+(10)}&=&0,
\\
\delta g_{4}^{(9)+(10)}&=&-\frac{\pi^{2}v^{2}A^{\frac{1}{2}}}{\Lambda^{\frac{3}{2}}}C_{6}\ell,
\\
\delta g_{5}^{(9)+(10)}&=&\frac{\pi^{2}v^{2}A^{\frac{1}{2}}}{\Lambda^{\frac{3}{2}}}C_{6}\ell,
\\
\delta g_{3z}^{(9)+(10)}&=&\frac{\pi^{2}v^{2}A^{\frac{1}{2}}}{\Lambda^{\frac{3}{2}}}C_{6}\ell.
\end{eqnarray}

\subsection{RG equations}

Considering the correction of interactions, the action of fermions becomes
\begin{eqnarray}
S_{\Psi}
&=&\int\frac{d\omega}{2\pi}\frac{d^{3}\mathbf{k}}{(2\pi)^{3}}\bar{\Psi}(\omega,\mathbf{k})
\left(i\omega\gamma_{0}+ivk_{1}\gamma_{1}+ivk_{2}\gamma_{2}\right.\nonumber
\\
&&\left.+iAk_{3}^{2}\gamma_{3}-\Sigma_{C}(i\omega,\mathbf{k})
\right)\Psi(\omega,\mathbf{k})\nonumber
\\
&\approx&\int\frac{d\omega}{2\pi}\frac{d^{3}\mathbf{k}}{(2\pi)^{3}}\bar{\Psi}(\omega,\mathbf{k})
\left(i\omega\gamma_{0}+ivk_{1}\gamma_{1}+ivk_{2}\gamma_{2}\right.\nonumber
\\
&&\left.\times e^{C_{1}\ell}+iAk_{3}^{2}\gamma_{3}e^{C_{2}\ell}
\right)\Psi(\omega,\mathbf{k}).
\end{eqnarray}
Employing the transformations Eqs.~(\ref{Eq:Scalingomega})-(\ref{Eq:Scalingk3}), (\ref{Eq:ScalingPsi}), and
\begin{eqnarray}
v&=&v'e^{-C_{1}\ell}, \label{Eq:ScalingvNew}
\\
A&=&A'e^{-C_{2}\ell}, \label{Eq:ScalingANew}
\end{eqnarray}
the action becomes
\begin{eqnarray}
S_{\Psi'}&=&\int\frac{d\omega'}{2\pi}\frac{d^{3}\mathbf{k}'}{(2\pi)^{3}}\bar{\Psi}'(\omega',\mathbf{k}')
\left(i\omega'\gamma_{0}+iv'k_{1}'\gamma_{1}+iv'k_{2}'\gamma_{2}\right.\nonumber
\\
&&\left.+iA'k_{3}'^{2}\gamma_{3}
\right)\Psi'(\omega',\mathbf{k}'),
\end{eqnarray}
which recovers the original form of the fermion action.

Including the correction of boson self-energy, the action of $\phi$ can be written as
\begin{eqnarray}
S_{\phi}&=&\int\frac{d\omega}{2\pi}\frac{d^{3}\mathbf{k}}{(2\pi)^{3}}\phi(\omega,\mathbf{k})
\left(\frac{1}{\sqrt{\eta}}k_{\bot}^{2}+\sqrt{\eta} k_{z}^{2}+\Pi(\mathbf{k})
\right)\nonumber
\\
&&\times\phi(\omega,\mathbf{k})\nonumber
\\
&\approx&\int\frac{d\omega}{2\pi}\frac{d^{3}\mathbf{k}}{(2\pi)^{3}}\phi(\omega,\mathbf{k})
\left(\frac{1}{\sqrt{\eta}}k_{\bot}^{2}e^{\sqrt{\eta}C_{\bot}\ell}\right.\nonumber
\\
&&\left.+\sqrt{\eta} k_{z}^{2}e^{\frac{C_{z}}{\sqrt{\eta}}\ell}
\right)\phi(\omega,\mathbf{k}).
\end{eqnarray}
Utilizing the transformations Eqs.~(\ref{Eq:Scalingomega})-(\ref{Eq:Scalingk3}), and
\begin{eqnarray}
\phi&=&\phi'e^{\left(\frac{5}{2}-\frac{\sqrt{\eta}C_{\bot}+\frac{C_{z}}{\sqrt{\eta}}}{4}\right)\ell}, \label{Eq:Scalingphi}
\\
\eta&=&\eta'e^{\left(-1+\sqrt{\eta}C_{\bot}-\frac{C_{z}}{\sqrt{\eta}}\right)\ell},  \label{Eq:Scalingeta}
\end{eqnarray}
the action can be expressed as
\begin{eqnarray}
S_{\phi'}
&=&\int\frac{d\omega'}{2\pi}\frac{d^{3}\mathbf{k}'}{(2\pi^{3})}\phi'(\omega',\mathbf{k}')
\left(\frac{1}{\sqrt{\eta'}}k_{\bot}'^{2}+\sqrt{\eta'} k_{z}'^{2}
\right)\nonumber
\\
&&\times\phi'(\omega',\mathbf{k}'),
\end{eqnarray}
which has the same form as the original action of boson.

Including the correction of one-loop Feynaman diagrams, the action of fermion-boson couplings can be written as
\begin{eqnarray}
S_{\psi\phi}&=&i\left(\lambda+\delta \lambda\right)\int\frac{d\omega_{1}}{2\pi}\frac{d^3\mathbf{k}_{1}}{(2\pi)^{3}}\frac{d\omega_{2}}{2\pi}
\frac{d^3\mathbf{k}_{2}}{(2\pi)^{3}}\bar{\Psi}(\omega_{1},\mathbf{k}_{1})\gamma_{0}\nonumber
\\
&&\times\Psi(\omega_{2},\mathbf{k}_{2})\phi(\omega_{1}-\omega_{2},\mathbf{k}_{1}
-\mathbf{k}_{2})\nonumber
\\
&=&i\lambda\int\frac{d\omega_{1}}{2\pi}\frac{d^3\mathbf{k}_{1}}{(2\pi)^{3}}\frac{d\omega_{2}}{2\pi}
\frac{d^3\mathbf{k}_{2}}{(2\pi)^{3}}\bar{\Psi}(\omega_{1},\mathbf{k}_{1})\gamma_{0}\Psi(\omega_{2},\mathbf{k}_{2})\nonumber
\\
&&\times\phi(\omega_{1}-\omega_{2},\mathbf{k}_{1}
-\mathbf{k}_{2}),
\end{eqnarray}
since $\delta\lambda=0$. Employing the transformations Eqs.~(\ref{Eq:Scalingomega})-(\ref{Eq:Scalingk3}), (\ref{Eq:ScalingPsi}), (\ref{Eq:Scalingphi}), and
\begin{eqnarray}
\lambda=\lambda' e^{\left(\frac{\sqrt{\eta}C_{\bot}+\frac{C_{z}}{\sqrt{\eta}}}{4}\right)\ell}, \label{Eq:Scalinglambda}
\end{eqnarray}
the action becomes
\begin{eqnarray}
S_{\psi'\phi'}
&=&i\lambda'\int\frac{d\omega_{1}'}{2\pi}\frac{d^3\mathbf{k}_{1}'}{(2\pi)^{3}}\frac{d\omega_{2}'}{2\pi}
\frac{d^3\mathbf{k}_{2}'}{(2\pi)^{3}}\bar{\Psi}'(\omega_{1}',\mathbf{k}_{1}')\gamma_{0}\nonumber
\\
&&\times\Psi'(\omega_{2}',\mathbf{k}_{2}')\phi'(\omega_{1}'-\omega_{2}',\mathbf{k}_{1}'
-\mathbf{k}_{2}'),
\end{eqnarray}
which recovers the original form the action of fermion-boson coupling.

Including the corrections of one-loop Feynman diagrams, the action of four-fermion interaction becomes
\begin{eqnarray}
S_{\Psi^{4}}&=&\sum_{a=1,2,4,5,3z}\left(g_{a}+\delta g_{a}\right)\int\frac{d\omega_{1}}{2\pi}\frac{d^{3}\mathbf{k}_{1}}{(2\pi)^{3}}
\frac{d\omega_{2}}{2\pi}\frac{d^{3}\mathbf{k}_{2}}{(2\pi)^{3}}\nonumber
\\
&&\times\frac{d\omega_{3}}{2\pi}\frac{d^{3}\mathbf{k}_{3}}{(2\pi)^{3}}\bar{\Psi}(\omega_{1},\mathbf{k}_{1})
\Gamma_{a}\Psi(\omega_{2},\mathbf{k}_{2})\bar{\Psi}(\omega_{3},k_{3})\Gamma_{a}\nonumber
\\
&&\times\Psi(\omega_{1}-\omega_{2}+\omega_{3},
\mathbf{k}_{1}-\mathbf{k}_{2}+\mathbf{k}_{3}).
\end{eqnarray}
Using the transformations Eqs.~(\ref{Eq:Scalingomega})-(\ref{Eq:Scalingk3}), (\ref{Eq:ScalingPsi}), and
\begin{eqnarray}
g_{a}'&=&\left(g_{a}+\delta g_{a}\right)e^{-\frac{3}{2}\ell}\approx g_{a}-\frac{3}{2}g_{a}\ell+\delta g_{a},\label{Eq:GeneralRescalinggaNew}
\end{eqnarray}
we get
\begin{eqnarray}
S_{\Psi'^{4}}
&=&\sum_{a=1,2,4,5,3z}g_{a}'\int\frac{d\omega_{1}'}{2\pi}\frac{d^{3}\mathbf{k}_{1}'}{(2\pi)^{3}}
\frac{d\omega_{2}'}{2\pi}\frac{d^{3}\mathbf{k}_{2}'}{(2\pi)^{3}}
\frac{d\omega_{3}'}{2\pi}\frac{d^{3}\mathbf{k}_{3}'}{(2\pi)^{3}}\nonumber
\\
&&\times\bar{\Psi}'(\omega_{1}',\mathbf{k}_{1}')
\Gamma_{a}\Psi'(\omega_{2}',\mathbf{k}_{2}')\bar{\Psi}'(\omega_{3}',k_{3}')\Gamma_{a}\nonumber
\\
&&\times\Psi'(\omega_{1}'-\omega_{2}'+\omega_{3}',
\mathbf{k}_{1}'-\mathbf{k}_{2}'+\mathbf{k}_{3}'),
\end{eqnarray}
which recovers the original form of the action.

From the transformations as shown in Eqs.~(\ref{Eq:ScalingvNew}), (\ref{Eq:ScalingANew}), (\ref{Eq:Scalingeta}), (\ref{Eq:Scalinglambda}),
(\ref{Eq:GeneralRescalinggaNew}), we can get the RG equations
\begin{eqnarray}
\frac{dv}{d\ell}&=&C_{1}v,
\\
\frac{dA}{d\ell}&=&C_{2}A,
\\
\frac{d\eta}{d\ell}&=&\left(1-\beta+\gamma\right)\eta,
\\
\frac{dg}{d\ell}&=&-\frac{\beta+\gamma}{4}g,
\\
\frac{d\bar{A}}{d\ell}
&=&\left(-\frac{1}{2}+\frac{1}{2}C_{2}-C_{1}+\frac{1}{2}\beta-\frac{1}{2}\gamma
\right)\bar{A},
\\
\frac{d\alpha}{d\ell}
&=&\left(-C_{1}-\frac{1}{2}\beta-\frac{1}{2}\gamma
\right)\alpha,
\\
\frac{d\beta}{d\ell}
&=&\left(\frac{1}{2}-\frac{1}{2}C_{2}-\beta
\right)\beta,
\\
\frac{d\gamma}{d\ell}
&=&\left(-\frac{1}{2}+\frac{1}{2}C_{2}-2C_{1}-\gamma
\right)\gamma.
\\
\frac{dg_{1}}{d\ell}
&=&-\frac{3}{2}g_{1}-\frac{2}{5}g_{1}\left(g_{2}+\frac{1}{2}g_{4}+g_{5}\right)
-\frac{2}{5}\left(g_{2}g_{5}\right.\nonumber
\\
&&\left.+g_{4}g_{3z}+g_{5}g_{3z}\right)-2g_{1}
\left(\beta+\gamma\right)+\bigg(-2g_{1}C_{1}\nonumber
\\
&&-\frac{1}{2}g_{1}C_{2}-g_{2}C_{5}+g_{2}C_{4}+g_{5}C_{5}\bigg),
\\
\frac{dg_{2}}{d\ell}
&=&-\frac{3}{2}g_{2}+g_{2}^{2}+g_{2}\left(-\frac{3}{5}g_{1}+\frac{4}{5}g_{4}+\frac{3}{5}g_{5}+g_{3z}\right)\nonumber
\\
&&-\frac{2}{5}g_{1}g_{5}+g_{4}\left(-g_{5}+\frac{7}{5}g_{3z}\right)+\frac{2}{5}g_{5}g_{3z}\nonumber
\\
&&+\bigg(-2g_{2}C_{1}-\frac{1}{2}g_{2}C_{2}+g_{2}C_{3}+g_{2}C_{5}\nonumber
\\
&&+g_{5}C_{5}\bigg),
\\
\frac{dg_{4}}{d\ell}
&=&-\frac{3}{2}g_{4}-\frac{1}{5}g_{4}^{2}-\frac{1}{5}\left(g_{1}^{2}+g_{2}^{2}+g_{5}^{2}+g_{3z}^{2}\right)
\nonumber
\\
&&+\frac{2}{5}g_{4}g_{3z}+\frac{2}{5}g_{1}\left(g_{2}+g_{5}\right)+g_{2}\left(-\frac{7}{5}g_{5}+g_{3z}\right)
\nonumber
\\
&&-\frac{1}{5}g_{5}g_{3z}+\bigg(g_{1}C_{4}+g_{2}C_{5}-2g_{4}C_{1}-\frac{1}{2}g_{4}C_{2}\nonumber
\\
&&-g_{5}C_{5}\bigg)-\frac{2}{5}C_{6},
\\
\frac{ dg_{5}}{d\ell}
&=&-\frac{3}{2}g_{5}+\frac{6}{5}g_{5}^{2}+\frac{1}{5}\left(g_{1}^{2}+g_{2}^{2}+g_{4}^{2}+g_{3z}^{2}\right)\nonumber
\\
&&+g_{5}\left(-g_{1}+g_{2}+g_{4}-g_{3z}\right)-\frac{2}{5}g_{1}\left(2g_{2}+\frac{1}{2}g_{4}\right)
\nonumber
\\
&&-g_{2}\left(\frac{4}{5}g_{4}+g_{3z}\right)-\frac{8}{5}g_{4}g_{3z}+\bigg(-g_{1}C_{4}-2g_{2}C_{5}\nonumber
\\
&&+g_{4}C_{4}-2g_{5}C_{1}-\frac{1}{2}g_{5}C_{2}+g_{5}C_{3}\bigg)
+\frac{2}{5}C_{6},
\\
\frac{dg_{3z}}{d\ell}
&=&-\frac{3}{2}g_{3z}+\frac{2}{5}g_{3z}^{2}+\frac{1}{5}\left(g_{1}^{2}+g_{2}^{2}+g_{4}^{2}+g_{5}^{2}\right)\nonumber
\\
&&-\frac{2}{5}g_{3z}\left(\frac{1}{2}g_{1}+2g_{2}+3g_{4}+\frac{3}{2}g_{5}\right)\nonumber
\\
&&-\frac{2}{5}g_{1}\left(g_{2}+\frac{1}{2}g_{4}+g_{5}\right)+\frac{2}{5}g_{2}\left(\frac{1}{2}g_{4}+g_{5}\right)
\nonumber
\\
&&+\frac{1}{5}g_{4}g_{5}+\bigg(-g_{1}C_{4}-g_{2}C_{5}+g_{4}C_{4}+g_{5}C_{5}\nonumber
\\
&&-2g_{3z}C_{1}-\frac{1}{2}g_{3z}C_{2}-
g_{3z}C_{4}\bigg)
+\frac{2}{5}C_{6},
\end{eqnarray}
where
\begin{eqnarray}
\alpha&=&\frac{\lambda^{2}}{4\pi v},
\\
\bar{A}&=&\frac{\sqrt{A}\sqrt{\Lambda}}{v\sqrt{\eta}},
\\
\beta&=&\sqrt{\eta}C_{\bot}=\frac{3}{5\pi}\frac{\alpha}{\bar{A}}, \label{Eq:CoffBeta}
\\
\gamma&=&\frac{C_{z}}{\sqrt{\eta}}=\frac{16}{21\pi}\alpha\bar{A}, \label{Eq:CoffGamma}
\end{eqnarray}
and redefinition
\begin{eqnarray}
\frac{\Lambda^{\frac{3}{2}}g_{a}}{\pi^{2}v^{2}\sqrt{A}}\rightarrow g_{a}
\end{eqnarray}
has been employed.

\begin{figure}[htbp]
\center
\includegraphics[width=3.1in]{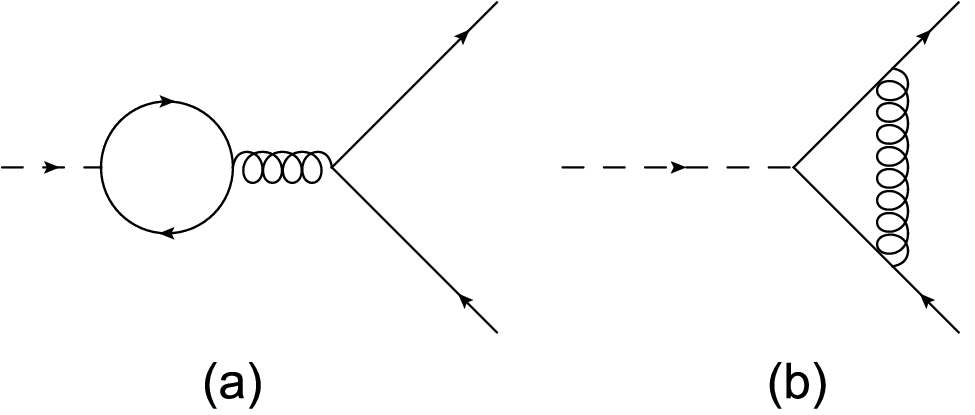}
\caption{One-loop Feynman diagrams for the corrections to the
source terms in particle-hole channels induced by long-range Coulomb interaction.
\label{Fig:SourceTermCoulomb}}
\end{figure}

\begin{figure}[htbp]
\center
\includegraphics[width=3.3in]{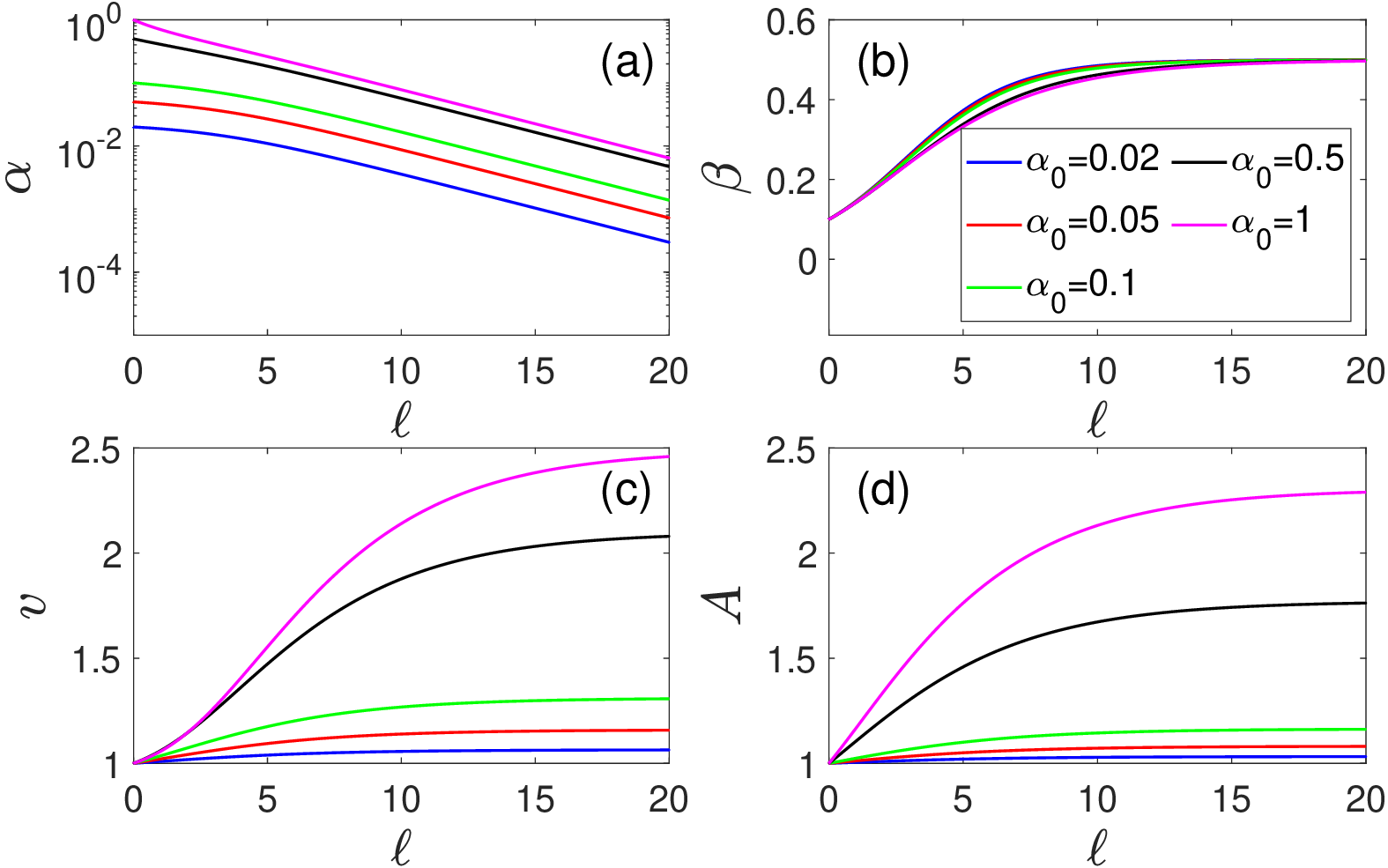}
\caption{Flows of $\alpha$, $\beta$, $v$, and $A$ with different initial values of Coulomb strength. $\beta_{0}=0.1$ is taken.
\label{Fig:OnlyCoulomb}}
\end{figure}

\begin{figure}[htbp]
\center
\includegraphics[width=3.3in]{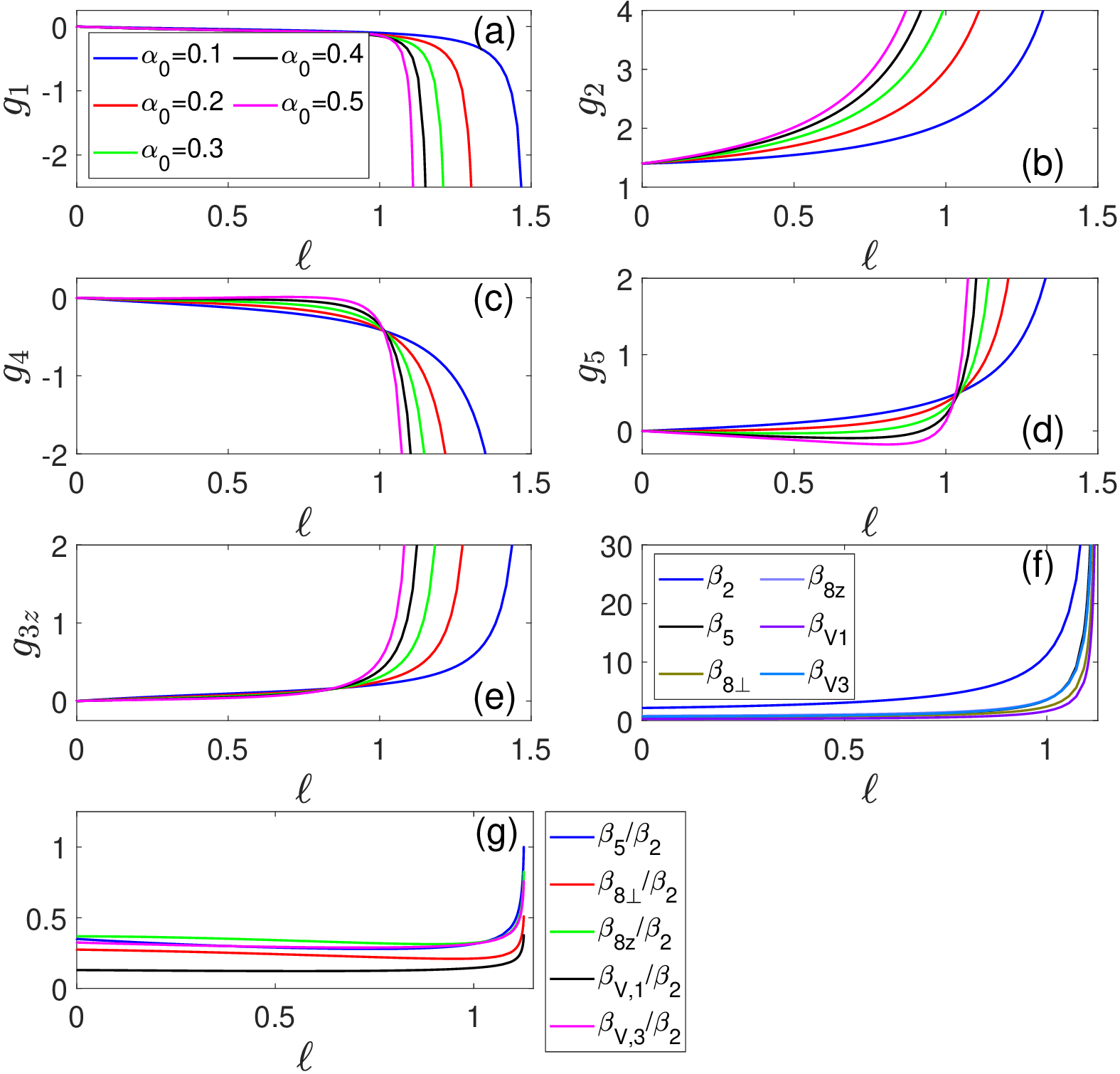}
\caption{Flows of $g_{1}$, $g_{2}$, $g_{4}$, $g_{5}$, and $g_{3z}$ are shown in (a)-(e). (f) and (g):
Flows of $\bar{\beta}_{X,Y}$ which approach to positive infinity and  ratios between $\bar{\beta}_{X,Y}$.
In (a)-(g) $g_{1,0}=0$, $g_{2,0}=1.4$, $g_{4,0}=0$, $g_{5,0}=0$, $g_{3z,0}=0$, and $\beta_{0}=0.1$ are taken. $\alpha_{0}=0.5 $ is
taken in (f) and (g).
\label{Fig:InterplyCoulombA}}
\end{figure}

\subsection{Source terms}

The one loop correction for the source term $\Delta_{X}$ in particle-hole channels induced by long-range Coulomb interaction as shown in
Fig.~\ref{Fig:SourceTermCoulomb}(a) can be written as
\begin{eqnarray}
W_{\Delta_{X}}^{(3)}&=&2\Delta_{X}\lambda^{2}\left(\bar{\Psi}\Gamma_{X}\Psi\right)\int_{-\infty}^{+\infty}d\omega
\int'\frac{d^3\mathbf{k}}{(2\pi)^{3}}\nonumber
\\
&&\times\mathrm{Tr}\left[\Gamma_{X}G_{0}(i\omega+i\Omega,\mathbf{k}+\mathbf{q})\gamma_{0}G_{0}(i\omega,\mathbf{k})\right]\nonumber
\\
&&\times D_{0}(i\Omega,\mathbf{q}). \label{Eq:SourceTermCorrectionCoulombA}
\end{eqnarray}
In particle-hole channels, the one-loop correction for the source term $\Delta_{X}$ from Fig.~\ref{Fig:SourceTermCoulomb}(b) is given by
\begin{eqnarray}
W_{\Delta_{X}}^{(4)}&=&-2\Delta_{X}\lambda^{2}
\int_{-\infty}^{+\infty}\frac{d\Omega}{2\pi}
\int'\frac{d^3\mathbf{q}}{(2\pi)^{3}}\left(\bar{\Psi}
\gamma_{0}G_{0}(i\Omega,\mathbf{q})\Gamma_{X}\right.\nonumber
\\
&&\left.\times G_{0}(i\Omega,\mathbf{q})\gamma_{0}\Psi\right)D_{0}(i\Omega,\mathbf{q}).
\label{Eq:SourceTermCorrectionCoulombB}
\end{eqnarray}
It should be noticed that long-range Coulomb interaction does not induce correction for the source terms in particle-particle channels.

\begin{figure}[htbp]
\center
\includegraphics[width=3.3in]{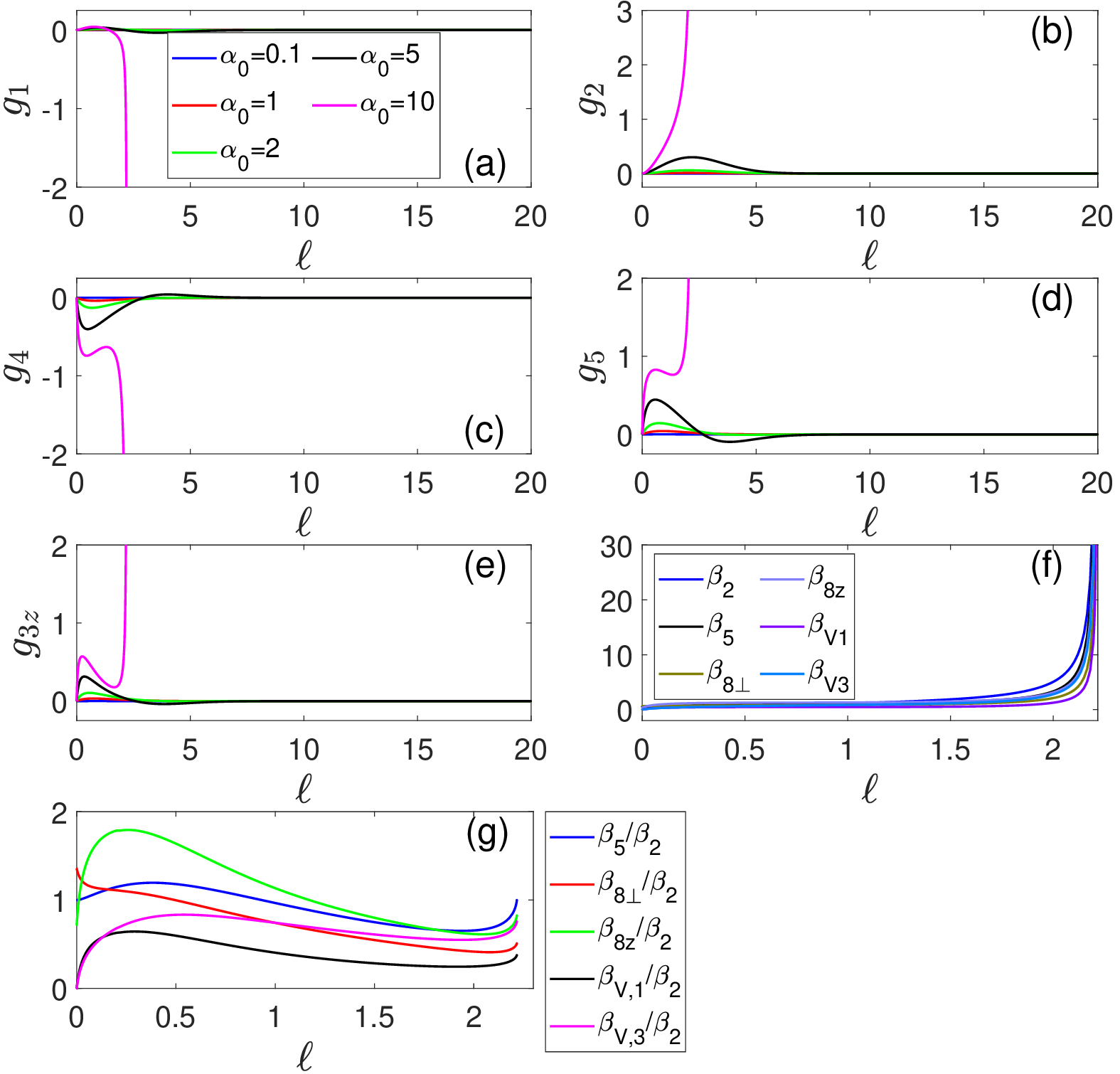}
\caption{Flows of $g_{1}$, $g_{2}$, $g_{4}$, $g_{5}$, and $g_{3z}$ are shown in (a)-(e). (f) and (g):
Flows of $\bar{\beta}_{X,Y}$ which approach to positive infinity and  ratios between $\bar{\beta}_{X,Y}$.
In (a)-(g) $g_{1,0}=0$, $g_{2,0}=0$, $g_{4,0}=0$, $g_{5,0}=0$, $g_{3z,0}=0$, and $\beta_{0}=0.1$ are taken. $\alpha_{0}=10$ is
taken in (f) and (g).
\label{Fig:InterplyCoulombB}}
\end{figure}

Calculating the corrections for source terms in particle-hole channels induced by long-range Coulomb interaction through
Eqs.~(\ref{Eq:SourceTermCorrectionCoulombA}) and (\ref{Eq:SourceTermCorrectionCoulombB}), and re-deriving the RG equations
for $\Delta_{X}$, we finally obtain
\begin{eqnarray}
\bar{\beta}_{\Delta_{1}}&=&-2\left(\beta+\gamma\right),
\\
\bar{\beta}_{\Delta_{2}}&=&\frac{1}{2}\left(-g_{1}+3g_{2}+g_{4}+g_{5}+g_{3z}\right)\nonumber
\\
&&+\frac{1}{2}C_{3},
\\
\bar{\beta}_{\Delta_{3\bot}}&=&\frac{1}{5}\left(-g_{1}+g_{2}+g_{4}-g_{5}+g_{3z}\right)\nonumber
\\
&&+\frac{1}{2}C_{5},
\\
\bar{\beta}_{\Delta_{3z}}&=&\frac{1}{10}\left(-g_{1}+g_{2}+g_{4}-g_{5}+3g_{3z}\right)\nonumber
\\
&&+\frac{1}{2}C_{4},
\\
\bar{\beta}_{\Delta_{4}}&=&0,
\\
\bar{\beta}_{\Delta_{5}}&=&\frac{1}{2}\left(-g_{1}+g_{2}+g_{4}+3g_{5}-g_{3z}\right)\nonumber
\\
&&+\frac{1}{2}C_{3},
\\
\bar{\beta}_{\Delta_{6\bot}}&=&\frac{1}{5}\left(-g_{1}-g_{2}+g_{4}+g_{5}-g_{3z}\right)\nonumber
\\
&&+\frac{1}{2}C_{5}\ell,
\\
\bar{\beta}_{\Delta_{6z}}&=&\frac{1}{10}\left(-g_{1}-g_{2}+g_{4}+g_{5}+g_{3z}\right)\nonumber
\\
&&+\frac{1}{2}C_{4}\ell,
\\
\bar{\beta}_{\Delta_{7\bot}}&=&\frac{3}{10}\left(-g_{1}-g_{2}-g_{4}-g_{5}+g_{3z}\right)\nonumber
\\
&&+\frac{1}{2}\left(C_{4}+C_{5}
\right),
\\
\bar{\beta}_{\Delta_{7z}}&=&\frac{2}{5}\left(-g_{1}-g_{2}-g_{4}-g_{5}-g_{3z}\right)+C_{5},
\\
\bar{\beta}_{\Delta_{8\bot}}&=&\frac{3}{10}\left(-g_{1}+g_{2}-g_{4}+g_{5}-g_{3z}\right)\nonumber
\\
&&+\frac{1}{2}\left(C_{4}+C_{5}\right),
\\
\bar{\beta}_{\Delta_{8z}}&=&\frac{2}{5}\left(-g_{1}+g_{2}-g_{4}+g_{5}+g_{3z}\right)+C_{5},
\end{eqnarray}
where $\beta$ and $\gamma$ are given by Eqs.~(\ref{Eq:CoffBeta}) and (\ref{Eq:CoffGamma}), and $C_{3}$, $C_{4}$ and $C_{5}$ are given by Eq.~(\ref{Eq:CoffC3}), (\ref{Eq:CoffC4}), and (\ref{Eq:CoffC5}) respectively.
The RG equations for source terms in particle-particle channels are still given by Eqs.~(\ref{Eq:BetaS})-(\ref{Eq:BetaV0}).

\subsection{Numerical Results}

The flows of $\alpha$, $\beta$, $v$, and $A$ are shown in Figs.~\ref{Fig:OnlyCoulomb}(a)-\ref{Fig:OnlyCoulomb}(d) respectively. We can find that $\alpha$ approaches to
zero quickly in the lowest energy limit. It represents that long-range Coulomb interaction becomes irrelevant in the lowest energy
regime. As shown in Fig.\ref{Fig:OnlyCoulomb}~(b), $\beta\rightarrow\frac{1}{2}$, which indicates the anisotropic screening of Coulomb interaction.
According to Figs.~\ref{Fig:OnlyCoulomb}(c) and \ref{Fig:OnlyCoulomb}(d), $v$ and $A$ approach to constant values in the
lowest energy limit. Thus, the fermion dispersion is not changed qualitatively by long-range Coulomb interaction.

According to Fig.~\ref{Fig:InterplyCoulombA}, the four-fermion interactions become divergent more quickly with increasing of initial value of
Coulomb strength. This result reveals that the long-range Coulomb interaction  can enhance the instabilities in particle-hole
channels although it becomes irrelevant in the low energy regime,
As shown in Figs.~\ref{Fig:InterplyCoulombB}(a)-\ref{Fig:InterplyCoulombB}(e), if the initial value of the Coulomb strength is large enough,
we can find that even if the initial values of the four-fermion interactions all vanish, the four-fermion interactions can
be generated and become divergent finally at a finite energy scale. According to Figs.~\ref{Fig:InterplyCoulombB}(f) and \ref{Fig:InterplyCoulombB}(g), axionic insulating phase
is generated if the initial value of Coulomb interaction is strong enough.

\end{document}